\documentclass[12pt,a4paper]{article}
\usepackage{bbm}
\usepackage{amssymb,amsmath}
\usepackage{rotating}
\usepackage[centerlast]{caption}
\usepackage{graphicx}
\usepackage{verbatim}
\usepackage{enumerate}
\usepackage{mathrsfs}
\usepackage{amsfonts}
\usepackage{multirow}
\usepackage{amsthm,amscd,graphics,latexsym,color}
\usepackage{booktabs,tabularx}
\usepackage{caption}
\usepackage{subcaption}
\usepackage[title,titletoc]{appendix}
\usepackage[authoryear]{natbib}
\usepackage{amsmath}
\usepackage{amssymb}

\usepackage[a4paper,text={18.0cm,26.2cm},centering]{geometry}

\def\be{\begin{equation}}
\def\ee{\end{equation}}
\def\bestar{\begin{eqnarray*}}
\def\eestar{\end{eqnarray*}}
\def\bea{\begin{eqnarray}}
\def\eea{\end{eqnarray}}
\theoremstyle{remark}

\theoremstyle{definition}
\newtheorem{defn}{Definition}
\theoremstyle{plain}
\newtheorem{theo}{Theorem}
\theoremstyle{plain}

\theoremstyle{plain}
\newtheorem{cor}{Corollary}
\theoremstyle{plain}

\theoremstyle{plain}
\newtheorem{lemma}{Lemma}

\numberwithin{equation}{section} \numberwithin{theo}{section}
\numberwithin{defn}{section} \numberwithin{rem}{section}
\numberwithin{cor}{section} \numberwithin{lemma}{section}
\numberwithin{prop}{section} \numberwithin{assumption}{section}
\allowdisplaybreaks[4]

\newcommand{\D}{\displaystyle}
\newcommand{\DF}[2]{\frac{\D#1}{\D#2}}
\newcommand{\e}{{\mathbb E}}
\newcommand{\p}{{\mathbb P}}
\newcommand{\ra}{{\cal F}}

\begin{document}

\title{Robust M--Estimation for Additive Single--Index \\Cointegrating Time Series Models}

\author{Chaohua Dong$^{a}$, Jiti Gao$^{b}$, Bin Peng$^{b}$ and Yundong Tu$^{c}$\footnote{Corresponding author: Guanghua
School of Management and Center for Statistical Science, Peking University,
Beijing, 100871, China. E-mail: yundong.tu@gsm.pku.edu.cn.}\\
$^{a}$Zhongnan University of Economics and Law, China\\ $^{b}$Monash University, Australia, and $^{c}$Peking University, China}

\date{\today}
\maketitle

\begin{abstract}
Robust M--estimation uses loss functions, such as least absolute deviation (LAD), quantile loss and Huber's loss, to construct its objective function, in order to for example eschew the impact of outliers, whereas the difficulty in analysing the resultant estimators rests on the nonsmoothness of these losses.

Generalized functions have advantages over ordinary functions in several aspects, especially generalized functions possess derivatives of any order. Generalized functions incorporate local integrable functions, the so-called regular generalized functions, while the so-called singular generalized functions (e.g. Dirac delta function) can be obtained as the limits of a sequence of sufficient smooth functions, so-called regular sequence in generalized function context. This makes it possible to use these singular generalized functions through approximation. Nevertheless, a significant contribution of this paper is to establish the convergence rate of regular sequence to nonsmooth loss that answers a call from the relevant literature.

For parameter estimation where objective function may be nonsmooth, this paper first shows as a general paradigm that how generalized function approach can be used to tackle the nonsmooth loss functions in Section two using a very simple model. This approach is of general interest and applicability. We further use the approach in robust M--estimation for additive single--index cointegrating time series models; the asymptotic theory is established for the proposed estimators. We evaluate the finite--sample performance of the proposed estimation method and theory by both simulated data and an empirical analysis of predictive regression of stock returns.
\smallskip

\smallskip


\end{abstract}

\section{Introduction}

Robust model building and determination of good models for estimation and prediction is a very important part of much empirical research in economics and finance. Since real world data often display characteristics, such as nonlinearity, nonstationarity and spatiality, an important aspect of model building and determination is how to take such characteristics into account. Meanwhile, the increasing availability of economic and financial data in recent years has been accompanied by increasing interest in both theoretical research and empirical data analysis in diverse fields of sciences as well as more broadly within social and medical sciences.

One challenge is how to select the most relevant variables when there are many variables available from the data. One initial step is to explore a simple method associated with principal component analysis. Different from dealing with high-dimensional data issues in medical sciences for example, we do need to pay attention to many econometric issues, such as multicollinearity when aggregating real datasets for model building. Meanwhile, one additional complexity is that many original data appear to be highly dependent and even nonstationary. Simply using a transformed version of the original data may completely ignore some key features involved in the original data, such as trending behaviours, which can be a key interest in modelling time series data in climatology, energy, epidemiology, health and macroeconomics. More importantly, transformed versions of two different datasets may have similar features, but the original datasets can have very different structures. Another challenge is the development of new models and estimation methods that are not only robust in theory, but also capable of offering user-friendly tools for both the computational and implementational procedures of  empirical data analysis in practice.

There are several levels of robustness involved in the discussion of this paper. First, the proposed model building and estimation methods are robust to data with outliers, spatiality and temporality, different types of nonstationarity. Second, the class of models we propose are applicable to several types of data, such as stationary time series data with deterministic trends, highly dependent and nonstationary time series data. Third, the proposed robust estimation procedures are iterative and computationally tractable. Fourth, the proposed models and robust estimation methods as well as the associated computational algorithms are user-friendly and easily implementable for empirical researchers and practitioners. While the relevant literature accounts for one or more of the model building and estimation issues we will address in this paper, the novelty of this paper is addressing them all simultaneously to produce new models and feasible estimation methods with computational tractability as well as empirical relevance and applicability.

We therefore propose a parametric additive single-index model of the form: 
\begin{align}
y_t=&\sum_{j=1}^{p_1}\gamma_{1j}^{0}\,g_{1j}(x_{t}^\top\theta_{1j}^{0})+ \sum_{j=1}^{p_2}\gamma_{2j}^{0}\,g_{2j}(z_{t}^\top\theta_{2j}^{0})+e_t
\label{m1}
\end{align}
for $1\leq t\leq n$, where $\gamma_{ij}^{0}\in \mathbb{R}^{1}$ and $\theta_{ij}^{0}\in \mathbb{R}^{d_i}$, for $1\leq j\leq p_i$ and $1\leq i\leq 2$, are all unknown parameters, $g_{ij}(\cdot)$, for $1\leq j\leq p_i$ and $1\leq i\leq 2$, are all known functions, $p_1$ and $p_2$ are known positive integers, $x_t$ is a $d_1$--dimensional vector of nonstationary time series, $z_t$ is a $d_2$--dimensional vector of trending stationary time series, and $e_t$ is the error term.
\smallskip

Before we discuss about how to estimate model (\ref{m1}) and then summarize the main contributions of this paper, we would like to first point out some novel features of model (\ref{m1}) and explain why we start with model (\ref{m1}) in this paper.

(i) Model (\ref{m1}) offers a simple way of dimension reduction by involving a partial sum of a number of linear combinations of the form: $\left(x_t^{\top} \theta_{11}^{0}, \cdots, x_t^{\top} \theta_{1, p_1}^{0}; z_t^{\top} \theta_{21}^{0}, \cdots, z_t^{\top} \theta_{2, p_2}^{0}\right)$, which naturally use a principal component analysis approach to obtaining the first $p_1$ leading components from $x_t$, and also the first $p_2$ leading components from $z_t$. Even though in this paper we focus on the case where $d_1, p_1, d_2$ and $p_2$ do not diverge with $n$, the actual values of $d_1$ and $d_2$ can be very large in both theory and practice, and much larger than $p_1$ and $p_2$, respectively.

(ii) Model (\ref{m1}) may be considered as a parametric neural network (NN) model with one hidden layer. In this case, $\{g_{ij}(\cdot): 1\leq j\leq p_i, 1\leq i\leq 2\}$ may be generated by a sequence of activation functions of the form: $g_{ij}(\cdot)= \sigma_i(\cdot + c_{ij})$, in which each $\sigma_i(\cdot)$ may be chosen as a commonly used activation function, and each $c_{ij}$ is the location parameter. The relevant NN literature includes \cite{cg1989}, \cite{cs1998}, \cite{cw1999}, \cite{crs2001}, \cite{chen2007}, \cite{kk2017}, \cite{bk2019}, and \cite{sh2020}.  

(iii) Note that model (\ref{m1}) covers a wide class of linear combinations of single--index parameters and basis functions commonly used in the NN literature for approximating unknown nonparametric functions by such linear combinations, as discussed in the NN references cited above. Note also that the choice of $\{g_{ij}(\cdot): 1\leq j\leq p_i, 1\leq i\leq 2\}$ allows for many commonly used polynomials, such as Hermite polynomials used in the empirical analysis in Section 5.2 below. 

(iv) The functional forms in model (\ref{m1}) considerably extend those studied in the relevant literature by \cite{dgp2015}, \cite{dgd2016}, \cite{degui2016}, \cite{degui2021} and others. As discussed below, moreover, we are able to consistently estimate both the single--index parameters, $\{\theta_{ij}^{0}\}$, and their coefficients, $\{\gamma_{ij}^{0}\}$, under the data structure outlined in model (\ref{m0}) below.

(v) The parametric form: $m(x, z; \theta_0, \gamma_0) = \sum_{j=1}^{p_1}\gamma_{1j}^{0}\,g_{1j}(x^\top\theta_{1j}^{0})+ \sum_{j=1}^{p_2}\gamma_{2j}^{0}\,g_{2j}(z^\top\theta_{2j}^{0})$ represents the true conditional mean when the null hypothesis $H_0: \, E\left[y_t|(x_t=x, z_t=z)\right] = m(x, z; \theta_0, \gamma_0)$ is true, as studied in \cite{dgdy2017} for a special case of model (\ref{m1}).  We therefore think that it is both important and necessary to investigate model (\ref{m1}) before we may study estimation and inferential problems for various semi--parametric forms of model (\ref{m1}).
\medskip

We now come back to depict time series sequence $\{x_t, z_t, \tau_t, 1\leq t \leq n\}$ in model \eqref{m1}. We specify the respective structures of $x_t$ and $z_t$ by
\begin{align}\label{m0}
\begin{split}
&x_t=x_{t-1}+w_t \ \ \ \mbox{and} \ \ \ z_t=h(\tau_t,v_t),
\end{split}
\end{align}
where $\tau_t=\frac{t}{n}$, as defined later, vector $w_t$ is a linear process, and $h(\cdot)$ is a known vector of functions driven by the time trend $\tau_t$ and stationary $v_t$.

In the formulation of \eqref{m0}, $x_t$ is a vector of linearly integrated processes of order one. It is known that $x_t$ may capture `large' trending behaviours. Meanwhile, $z_t$ is a function in ($\tau_t, v_t$) that can capture `small' fluctuations.  The general form of $z_t$ accommodates possible nonlinearity, as it is quite common in nonstationary time series settings. Such general structure covers some important special cases, such as (1) $z_t=h(\tau_t)+v_t$, a stationary vector superimposed with a vector of time trends; (2) $z_t=\sigma(\tau_t)^\top v_t$, a combination of $v_t$ with varying coefficients ($\sigma(\cdot)$ may be a matrix); and (3) $z_t = h(\tau_t) + \sigma(\tau_t)^\top v_t$, which accommodates both deterministic and stochastic trending behaviours in the respective mean and volatility components.

Different from existing settings, model (\ref{m1}) takes into account not only the `large' stochastic trending component driven by $x_t^{\top} \theta_{1j}^{0}$, but also the `small' slowly varying component $z_t^{\top} \theta_{2j}^{0}$ mainly represented by the deterministic time trend $\tau_t$.

Note that the products of $\gamma_{ij}$ and $g_{ij}(\cdot)$ may give raise to an identification issue. For example, if $g_{11}(u)=u$, $\gamma_{11}g_{11}(x_{t}^\top\theta_{11}^0)=\gamma_{11}x_{t}^\top\theta_{11}^0$. For the sake of identification, we shall assume that $\|\theta_{ij}^0\|=1$ for all $1\leq j \leq p_i$ and $1\leq i\leq 2$, and possibly the first elements of $\theta_{ij}^0$ are positive that is not necessary in the case that $g_{ij}(\cdot)$ are nonlinear.

We shall adopt robust M--estimation (robust estimation hereafter for simplicity) method to estimate all unknown parameters. With regard to robust estimation, there is an extensive literature. Starting from the seminal work of \citet{huber1964} with Huber's loss, robust estimation methods gradually extend to some general loss functions extensively studied in the literature, such as \citet{bai1992}, \citet{fanjq1994}, \citet{fan2018} and \citet{degui2021}.


By robust estimation, the unknown parameters $\theta, \gamma$ are estimated by minimizing
\begin{equation}\label{objfun}
L_n(\theta,\gamma)=\sum_{t=1}^n\rho\left(y_t- \sum_{j=1}^{p_1} \gamma_{1j}g_{1j}(x_{t}^\top\theta_{1j})- \sum_{j=1}^{p_2} \gamma_{2j} g_{2j}(z_{t}^\top\theta_{2j})\right),
\end{equation}
where $\rho(\cdot)$ is some loss function, $\theta$ and $\gamma$ stand for generic vectors in a compact sets including the true vectors $\gamma_i^{0}$ and $\theta_{ij}^{0}$, $i=1,2$, $j=1,\cdots,p_i$, as interior points.

Though there are several loss functions in the robust estimation literature, the most important ones are: (1) Least absolute deviation (LAD), $\rho(u)=|u|$; (2) Quantile loss \citep{koenker1978}, $\rho_{\tau}(u)=u(\tau-I(u<0))$ for $\tau\in (0,1)$; and (3) Huber's loss \citep{huber1964}, i.e. for fixed $c>0$,
\begin{align*}
\rho_{c}(u)=\begin{cases}
\frac{1}{2}u^2, & |u|\le c,\\
c |u|-\frac{1}{2}c^2, & |u|>c.
\end{cases}
\end{align*}
The first two loss functions are not smooth while the third does not have second derivative at $\pm c$. This feature makes aymptotic analysis of the estimators much difficult than usual where the loss is sufficiently smooth.

We shall adopt generalized function approach to dealing with the nonsmooth loss function $\rho(\cdot)$. Generalized functions started with the use in physics, and the most famous one is Dirac delta function that is actually not a function in ordinary sense. Mathematicians make the concept of generalized functions rigorous by defining them as functionals acting on tempered test function space $S$ (another test space is $D$ but $D\subset S$), introduced in the supplementary material of the paper. Because of the property of functions in $S$, any local integrable function with no faster than a polynomial increase at infinity is a generalized function, and all loss functions in the literature belong to the category. See, for example, \citet{gelfand1964} and \citet{kanwal1983}. In particular, LAD and quantile losses have Dirac delta function $\delta(\cdot)$ as their second-order generalized derivatives.

As is well known, generalized functions can not be regarded as ordinary functions because they might not be well-defined at each point. Nonetheless, their operation is valid under integration with certain conditions. For example, for any random variable $e$ with density function $f_e(x)$,
\begin{equation*}
\e[\delta(e)]=\int \delta(x)f_e(x)dx=f_e(0),
\end{equation*}
provided $f_e(x)$ is continuous at $x=0$. Thus, in our proposed mechanism we use generalized functions only in expectations, whereas if we have to consider the derivatives of loss functions, we use their regular sequences defined later.

The focus of this paper is on non-differentiable convex loss functions $\rho(\cdot)$. It is well known that the subgradient $\psi(\cdot)$ of convext $\rho(\cdot)$ always exists, see \citet{rockafellar1970}; subgradient will coincide with derivative wherever it is differentiable. The step stone in dealing with non-differentiable convex loss $\rho(\cdot)$ is using a so-called regular sequence $\{\rho_m(\cdot), m=1,2, \cdots\}$ that approximates to $\rho(\cdot)$, as $m\to\infty$; and $\rho_m(\cdot)$ are infinitely differentiable, $\rho_m^{(k)}(u)\to \rho^{(k)}(u)$ for any $k$ in generalized function sense. More importantly, we establish the rates of $\rho_m(u)-\rho(u)$, $\rho_m'(e_t)-\psi(e_t)$ in probability and $\e[\rho_m''(e_t)-\rho''(e_t)]$ in Theorem 2.1 below that are crucial for our theory, where $\rho''(\cdot)$ is understood in generalized function sense such that $\e[\rho''(e_t)]$ exists. As far as we are aware, this is the debut of these rates in the literature.

Closely related to our generalized function approach, \citet{phillips1991, phillips1995} use generalized function approach to studying asymptotic properties for LAD and M-estimators in linear models. From then on, to the best of our knowledge, there has been no attempts for using generalized functions to tackle robust estimation. The Achilles heel to the approach is that the approximation rate of regular sequence to nonsmooth loss functions is not established in \citet{phillips1991, phillips1995}, and this approximation rate nevertheless is crucial in the theoretical side that, as what aforementioned, we shall establish in Section two below. Notice also that the regular sequence (also delta-convergent sequence) approach already has been used in earlier studies in statistics to estimate density function, such as \citet{walter1979} and \citet{walter1981}. Thus, we believe that the rationale of the estimation method we propose considerably enriches the relevant literature.

Our methodology is to approximate $L_n$ by a quadratic form $Q_{n}$ of parameters whose minimizer has an explicit expression, and we show that the difference between the minimizers of $L_n$ and $Q_{n}$ is negligible, so that the limit of $L_{n}$'s minimizer is the same as that of $Q_{n}$'s. This mechanism is much simpler and more generally applicable than those developed in the existing literature, such as \citet{knight1989}, \citet{phillips1991, phillips1995}, \citet{pollard1991}, \citet{bai1992}, \citet{fanjq1994}, \citet{vaart1996}, \cite{gao2009}, \citet{fan2018} and \citet{degui2021}. In addition, the estimator's explicit expression derived in our paper is comparable with Bahadur representation in the relevant literature; under the same setting, our remainder term is $O_P(n^{-1/2+\lambda})$ for any $\lambda\in (0,1/2)$, whereas \citet{bahadur1966} has the remainder $O_{a.s.}(n^{-1/4}\log(n))$.

To illustrate our methodology of the proposed generalized function approach, we use a linear model as an exemplar in Section 2 to show the essence of the approach; in Section 3 we present all assumptions. In Section 4 we establish the corresponding asymptotic distributions of the proposed estimators of all the parameters in model \eqref{m1} in two scenarios where $g_{1j}(\cdot)$ are $H$-regular and $I$-regular, respectively. Section 5.1 gives Monte Carlo simulation results to illustrate the performance of our estimators in finite sample situations; an empirical study about stock returns is shown in Section 5.2. Technical lemmas are given in Appendix A, whose proofs however are shown in supplementary materials; all main results are proven in Appendix B.

Let us finally introduce the following notation.  $\|\cdot\|$ is either Euclidean norm for vectors or element-wise norm for matrices; $\equiv$ means equal by definition; $\int$ means an integration on $\mathbb{R}$; $C$, $C_1$ and so on represent absolute constants that may be different at each appearance; $\dot{g}, \ddot{g}$ signify the first and second derivatives of $g$, respectively.

\section{Generalized functions and robust estimation}

This section will show the rationale of robust estimation based on generalized function approach, temporarily letting model \eqref{m1} alone. To begin with, we study the convergence rate of regular sequence of loss functions where regular sequence is a tool by which generalized functions are defined. See Chapter 5 of \citet{milne1980}.

The loss functions $\rho(\cdot)$ we use satisfy the following assumption. Recall that $\psi(\cdot)$ is the subgradient of $\rho(\cdot)$. Though the value of $\psi(\cdot)$ at some points may not be unique, we allow $\psi(\cdot)$ to take any of them while the analysis in the sequel remains unchanged. The focus then rests on the second derivative of $\rho(\cdot)$ that may not exist in ordinary sense.

\medskip

\noindent \textbf{Assumption 2.1}.
\begin{enumerate}[(a)]
\item \emph{Suppose that $\rho(\cdot)$ is positive convex, and has its only minimum at zero; moreover, it is locally integrable function on the real line and increase at infinity no faster than a polynomial.}
\item \emph{$\rho(\cdot)$ satisfies Lipschitz condition $|\rho(x+u)-\rho(u)|\le C |x|$ for any $x, u\in \mathbb{R}$ where $C$ is an absolute constant}.
\end{enumerate}

By convexity, $\rho(\cdot)$ is a continuous function. See, Corollary 10.1.1 of \citet[p 83]{rockafellar1970}. The other conditions in (a) (local integrability and no increasing faster than a polynomial) make operation of generalized function in the sequel valid, and these are certainly fulfilled by all loss functions encountered in the literature.

The Lipschitz condition in (b) of the assumption is the center that depicts a technical requirement in the following analysis, and it is guaranteed by the boundedness of subgradient. Notice also that it is satisfied by several most popular loss functions. For example, when $\rho(u)=|u|$, we have $C=1$; when $\rho(u)$ is the check function with parameter $\tau\in (0,1)$, we have $C=\max(\tau, 1-\tau)$; when $\rho(u)$ is Huber's loss with parameter $c$, we have $C=c$.

We next give crucial results about regular sequences of loss functions that is used as a bridge between nonsmooth loss and its approximation counterpart in generalized function context.

\begin{theo}\label{th21}
Let $\rho(u)$ satisfy Assumption 2.1. Define a regular sequence of $\rho(u)$,
\begin{equation}\label{regularseq2a}
\rho_m(u)=\int \rho(x)\phi_m(x-u)dx, \ \ \phi_m(x)=\sqrt{m/\pi} \exp(-mx^2),
\end{equation}
for $m=1,2, \ldots$.

Then, (1) $\rho_m(u), m=1,2,\cdots,$ are differentiable with any order, and in particular,
\begin{align}\label{derivatives2a}
  \rho_m'(u)=&-\int \rho(x)\phi_m'(x-u)dx, & \rho_m''(u)=&\int \rho(x)\phi_m''(x-u)dx,
\end{align}
and $\rho_m'(x)\to \psi(x)$ and $\rho_m''(x)\to \rho''(x)$ as $m\to\infty$ except on a set of measure zero.

(2) $\sup_{u\in \mathbb{R}}|\rho_m(u)-\rho(u)|\le Cm^{-1/2}$ where absolute constant $C$ may be different from Lipschitz constant in Assumption 2.1. Consequently, $\rho_m(e)-\rho(e)=O(m^{-1/2})$ almost surely for any random variable $e$.

(3) Let $f(u)$ be the density of $e$. Suppose that $f(u)$ is differentiable, $\int |f'(u)|du <\infty$, and $\rho(u)f(u)\to 0$ as $|u|\to\infty$, then $\rho_m'(e)-\psi(e)=O_P(m^{-1/2})$.

(4) Suppose $\e[\rho''(e)]$ exists where $\rho''(u)$ may be considered as a generalized function; suppose also $\int |f''(u)|du <\infty$, $\rho(u)f(u)\to 0$ and $\psi(u)f'(u)\to 0$ when $|u|\to\infty$. Then $\e[\rho_m''(e)-\rho''(e)]=O(m^{-1/2})$.

(5) $\e[\rho_m''(e+\epsilon)-\rho''_m(e)]=O(m^{-1/2}+\epsilon)$ for any given small $\epsilon$.
\end{theo}

The proof is given in Appendix B. Note that each $\phi_m(x)$ is the density of variables $N(0, (2m)^{-1})$ which form a delta-convergent sequence, i.e. $\phi_m(x)\to \delta(x)$ as $m\to \infty$ in generalized function sense; see \citet{kanwal1983} and \citet{stein2003}. Note also that in the last assertion, because of convexity of $\rho(u)$ and $\rho_m''(u)\to \rho''(u)$ in generalized function sense, $\e[\rho_m''(e)]>0$ for large $m$.

\begin{figure}
\begin{subfigure}{.32\textwidth}
  \centering
    \includegraphics[width=0.85\linewidth]{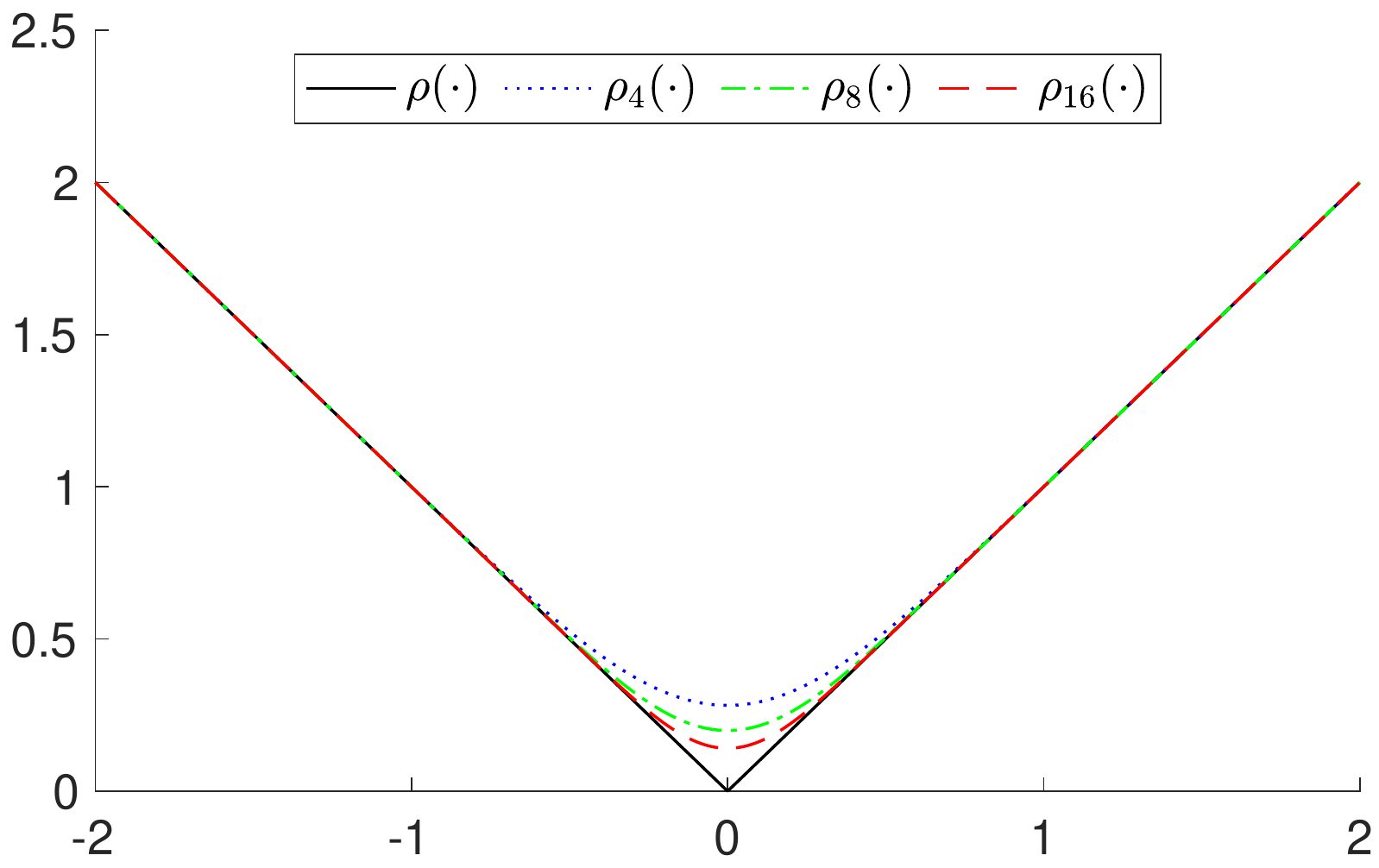}\\
  \caption{$\rho(u)=|u|$}\label{absolute}
  \end{subfigure}
\begin{subfigure}{.32\textwidth}
\centering
    \includegraphics[width=0.85\linewidth]{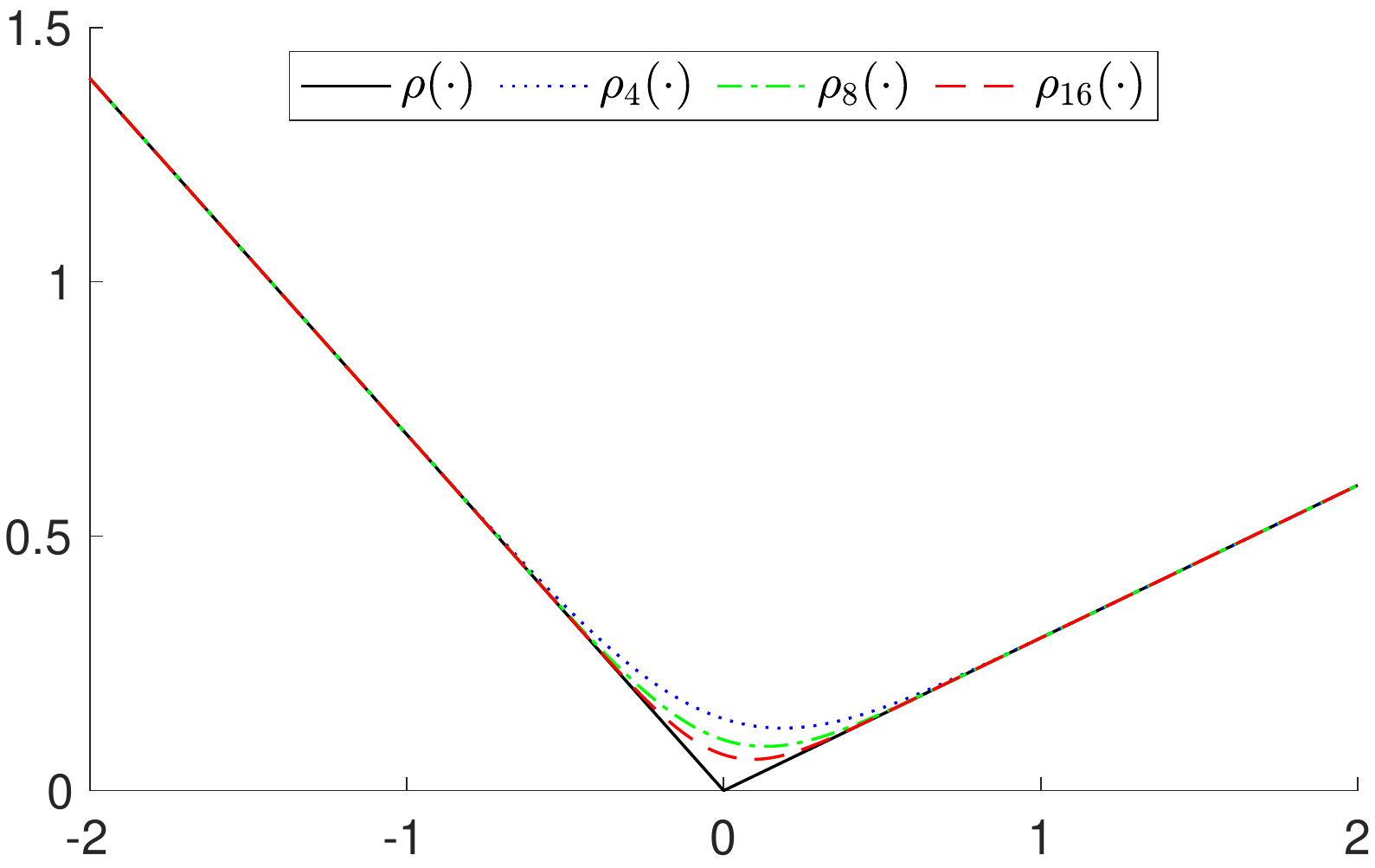}\\
  \caption{$\rho(u)=u(0.3-I(u<0))$}\label{quantile}
  \end{subfigure}
  \begin{subfigure}{.32\textwidth}
\centering
    \includegraphics[width=0.85\linewidth]{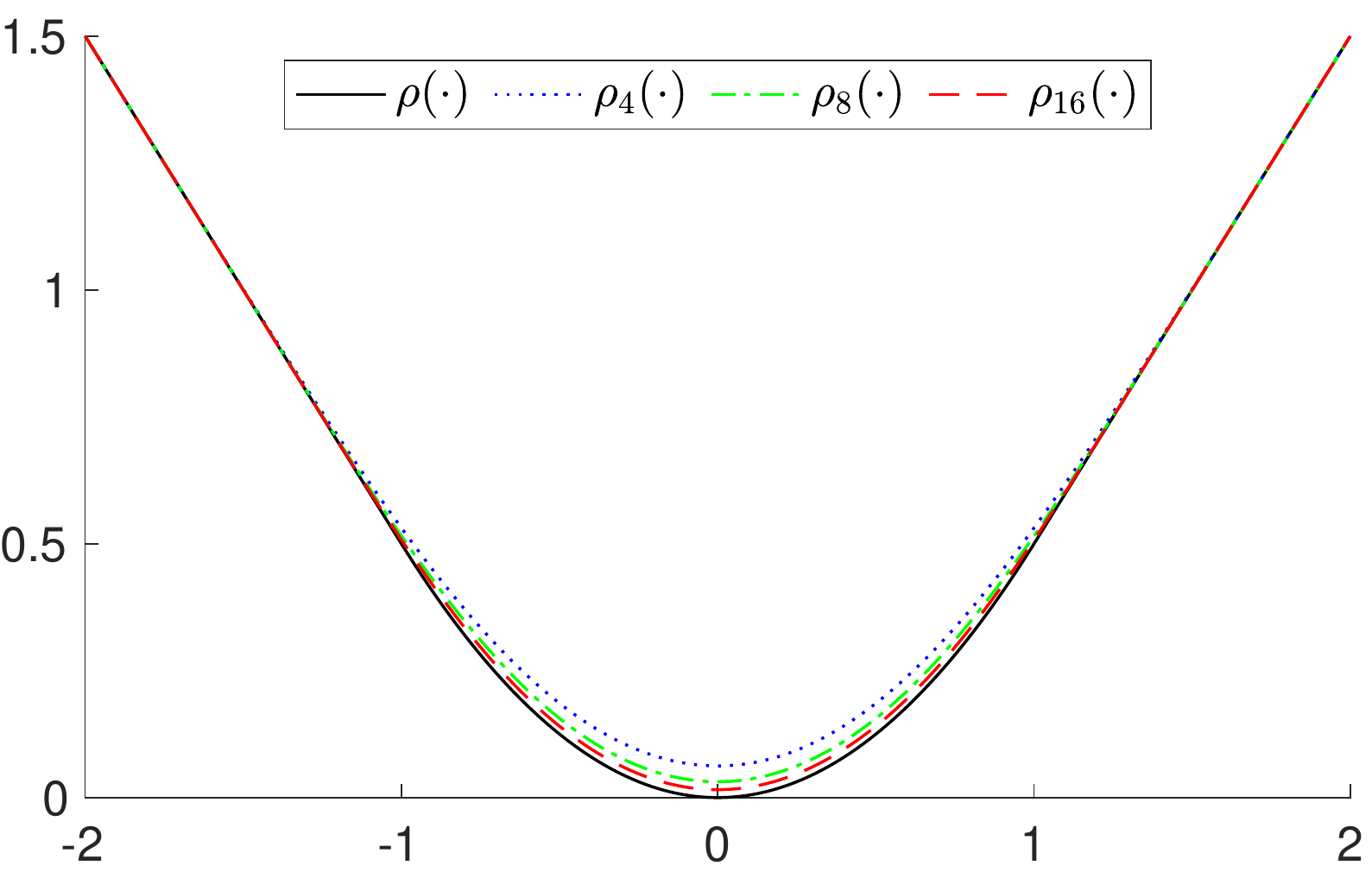}\\
  \caption{Huber loss with $c=1$}\label{huber}
  \end{subfigure}
  \caption{Loss functions with their regular sequences}
  \label{losses}
\end{figure}

{\bf Remark 2.1}. \ Theorem \ref{th21} is of independent interest. The importance of Theorem \ref{th21} is the convergence rate $\sup_{u\in \mathbb{R}}|\rho_m(u)-\rho(u)|\le Cm^{-1/2}$. To the best of our knowledge, this is first shown in the literature although regular sequences are discussed in several papers, such as \citet{phillips1991, phillips1995}. It is due to this rate that our analysis below is on a solid and rigorous ground. In Figure \ref{losses} above three losses and their regular sequences are plotted to visualize the approximations established in Theorem \ref{th21}. Note from the proof that the Lipschitz condition in Assumption 2.1 can be relaxed as $|\rho(x)-\rho(y)|\le C|x-y|^\alpha$ for $\alpha\in (0,1]$. All the results below under the relaxation hold with a change on the choice of $m$. \qed

\medskip

To show the essence of our generalized function approach, in the sequel we shall illustrate through a linear parametric regression. To do so, let us consider a very simple regression model of the form:
\begin{align}
y_t=&x_t'\theta_0+e_t, \ \ \ t=1, 2, \cdots, n, \label{2a}
\end{align}
where unknown parameter $\theta_0\in \Theta$, a compact subset of $\mathbb{R}^d$. Notice that the notation in this section has a different meaning from other sections.

The estimator of $\theta_0$ in the regression \eqref{2a} is defined by
\begin{equation}\label{2b}
\widehat{\theta}=\underset{\theta\in \Theta}{\arg\min}\ \Pi_n(\theta)=\sum_{t=1}^n\rho(y_t- x_{t}'\theta),
\end{equation}
where $\rho(\cdot)$ satisfies Assumption 2.1.

To analyse $\widehat{\theta}$, instead of considering minimization over $\theta\in\Theta$, we shall follow the literature to consider vectors $\beta$ in the tangent cone $T_{\Theta}(\theta_0)$ of $\Theta$ at $\theta_0$, that is, $\beta=\sqrt{n}(\theta-\theta_0)$ where $\theta\in\Theta$. See among others, \citet{bickel1974}, \citet{badu1989}, \citet{davis1992}, \citet{phillips1995} and \cite{gao2009}. In particular, \citet{charles1994} argues that $\beta=\sqrt{n}(\theta-\theta_0)\in T_{\Theta}(\theta_0)$, and shows the convergence of $\widehat\theta$ through that of $\widehat\beta$.

Therefore, we focus on the minimization of $\Pi_n(\theta)-\Pi_n(\theta_0)$ instead of $\Pi_n(\theta)$ in \eqref{2b}, and write
\begin{align*}
\Pi_n(\theta)-\Pi_n(\theta_0)=&\sum_{t=1}^n[\rho(y_t- x_t'\theta)-\rho(y_t- x_t'\theta_0)]\\
=&\sum_{t=1}^n[\rho(e_t-n^{-1/2}x_t'\beta)-\rho(e_t)]
\equiv\tilde{\Pi}_n(\beta).
\end{align*}
Hereby, the objective function is reparametrized in $\beta\in \mathbb{R}^d$, and when $\widehat\beta$ minimizes $\tilde{\Pi}_n(\beta)$, $\widehat\theta$ minimizes $\Pi_n(\theta)$ where $\widehat\beta=\sqrt{n}(\widehat\theta -\theta_0)$.

\begin{theo}\label{th22}
Suppose that $(e_t, x_t)$ is an independent and identically distributed sequence, $e_t$ and $x_t$ are mutually independent, and the fourth moment of $x_t$ exists. Define
\begin{equation}\label{th22a}
Q_n(\beta)=\left(-\DF{1}{\sqrt{n}}\sum_{t=1}^n\psi(e_t)x_t'\right)\beta
+\DF{a_2}{2}\beta' \left(\DF{1}{n} \sum_{t=1}^nx_tx_t'\right)\beta,
\end{equation}
where $a_2=\e[\rho''(e_t)]>0$. Then,
\begin{equation}\label{th22b}
  |\tilde{\Pi}_{n}(\beta)-Q_n(\beta)|=O_P(n^{-1/2}),
\end{equation}
for each $\beta$; and
\begin{equation}\label{th22c}
  \sup_{\|\beta\|\le c}|\tilde{\Pi}_{n}(\beta)-Q_n(\beta)|=O_P(n^{-1/2}),
\end{equation}
for any $c>0$; moreover,
\begin{equation}\label{th22d}
  \sup_{\|\beta\|\le c_n}|\tilde{\Pi}_{n}(\beta)-Q_n(\beta)|=o_P(1),
\end{equation}
for any $c_n=o(n^{1/4})$.

Furthermore, define $\widehat\beta_Q$ as the minimizer of $Q_n(\beta)$, then as $n\to\infty$ we have
\begin{equation}\label{th22e}
\widehat{\beta}-\widehat\beta_Q=o_P(n^{-1/2+\lambda}), \ \ \forall\ \lambda\in (0, 1/2).
\end{equation}
\end{theo}

Notice that, to illustrate our method manifestly, we impose very strong conditions on the regressors and error term. Such strong conditions are used only in this section. It is clear that
\begin{align*}
\widehat{\beta}_Q=\left(\DF{a_2}{n} \sum_{t=1}^nx_tx_t'\right)^{-1} \left(\DF{1}{\sqrt{n}}\sum_{t=1}^n\psi(e_t)x_t\right),
\end{align*}
and it is the unique minimizer of $Q_n(\beta)$.

{\bf Remark 2.2}. \ Note that \citet[p187]{pollard1991} establishes `Convexity Lemma' and points that if convex sequence $\lambda_n(\theta)$ converges to $\lambda(\theta)$ in probability for each $\theta\in \Theta$ where $\Theta$ is convex and open subset of $\mathbb{R}^d$, then $\sup_{\theta\in K}|\lambda_n(\theta)- \lambda(\theta)|\to 0$ in probability for any compact $K\subset \Theta$. This is much weaker than our results \eqref{th22b} and \eqref{th22c}. In addition, \citet[p308]{fan2003} use quadratic approximation for smooth objective function and derive limit theory from this quadratic function. By contrast, the generalized function approach we propose mainly deals with nonsmooth loss functions although it can also be used for smooth losses. \qed

{\bf Remark 2.3}. \ Interestingly, equation \eqref{th22e} gives $\widehat{\beta}=\widehat\beta_Q+o_P(n^{-1/2+\lambda})$ for any $\lambda\in (0, 1/2)$ that is better than Bahadur representation in some sense. See, for example, \citet{bahadur1966} where the reminder term is of order $O_{a.s.}(n^{-1/4}\log(n))$. \qed

{\bf Remark 2.4}. \ As argued in  \citet[p187]{pollard1991}, quadratic approximation to objective function of M estimation avoids technical difficulty such as stochastic equicontinuity in an asymptotic proof. Indeed, once we establish \eqref{th22e}, the limit of $\widehat{\beta}$ is the same as $\widehat\beta_Q$ that is quite easily derived from its explicit expression and some regular conditions such as $\e[\psi(e_t)]=0$. \qed

\medskip

We shall use this mechanism for model \eqref{m1} in the following section.

\section{Assumptions in model \eqref{m1}}

Now we turn back to model \eqref{m1}, and we shall use the generalized function approach for the establishment of the asymptotic limits of the proposed robust estimators. We give the following assumptions.

\medskip

\noindent {\bf Assumption 3.1}
\begin{enumerate}[(i)]
\item \emph{Let $\{\eta_j, j\in \mathbb{Z}\}$ be a sequence of $d_1$-vector independent and identically distributed (i.i.d.) continuous random variables satisfying $E\eta_0=0$, $E\eta_0\eta_0'=I_{d_1}$ and $E\|\eta_0\|^4<\infty$. Let $\varphi(u)$ be the characteristic function of $\eta_0$ satisfying $\int \|u\|\, |\varphi(u)| du <\infty$.}
   \item \emph{Let $\{w_t\}$ be a linear process defined by $w_t=\sum_{j=0}^\infty A_j \eta_{t-j}$, where $\{A_j\}$ is a sequence of square matrices, $A_0=I_{d_1}$, and $\sum_{j=0}^\infty j \|A_j\| <\infty$. Moreover, suppose that $A\equiv\sum_{j=0}^\infty A_j$ has full rank.}
  \item \emph{Let $x_{t}$ follow a linearly integrated process of the form: $x_t=x_{t-1}+w_{t}$ for $t\geq 1$ with $x_{0}=O_P(1)$. Furthermore, we assume that $x_t^{\top} \theta_{1j}^{0}$ remains linearly integrated for each given $1\leq j\leq p_1$.}
\end{enumerate}

The structure of the unit root process $x_t$ is commonly imposed in the literature since $w_t$ has a considerably general form that covers many special cases. See \citet{phillips1999,phillips2001} and \citet{dgd2016}. It is straightforward to calculate that $d_t^2\equiv\mathbb{E} (x_t x_t^\top)= AA^\top t(1+o(1))$, and it is well known that, for $r\in [0,1]$, $n^{-1/2}x_{[nr]}\to_DB(r)$ as $n\to\infty$ where $B(r)$ is a standard Brownian motion on $[0,1]$ with covariance $AA^\top$.

In the case where some of $x_t^{\top} \theta_{1j}^{0}$ or even all of them reduce to stationary processes, the estimation method of $\theta_{1j}^{0}$, for $1\leq j \leq p_1$, and the other parameters remains the same. The corresponding asymptotic properties, however, become different. In such cases, the corresponding assumptions on $g_{1j}(\cdot)$ are roughly the same as those on $g_{2j}(\cdot)$. Since theoretical developments for such cases require substantially different technologies, we wish to leave this for future research.
\medskip

\noindent \textbf{Assumption 3.2}
\begin{enumerate}[(i)]
\item \emph{Let $h(r, v)$ be a $d_2$-dimensional vector of continuous functions in $r\in [0,1]$.}

\item \emph{Suppose that $v_t=q(\eta_t,\cdots, \eta_{t-d_0+1}; \tilde{v}_t)$ for fixed $d_0\geq 1$, where $\{\tilde{v}_t, 1\le t\le n\}$ is also a vector sequence of strictly stationary and $\alpha$-mixing time series, and independent of $\eta_j$'s introduced in Assumption 3.1, and the vector function $q(\cdots)$ is measurable such that
    \begin{equation*}
    \sup_{r\in [0,1], j=1,\cdots,p_2}\e\|\dot{g}_{2j}^2(\theta_{2j}^{0\top} h(r, v_1))\; h(r, v_1)h(r, v_1)^\top\|^4<\infty.
    \end{equation*}
    Moreover, the $\alpha$-mixing coefficient $\alpha(k)$ of $\{\tilde{v}_t, 1\le t\le n\}$ satisfies $\sum_{k=1}^\infty \alpha^{1/2} (k) <\infty$.}
\end{enumerate}

Basically, $\{v_t\}$ is a vector sequence of strictly stationary time series variables and $v_t$ can be correlated with $x_t$ through a finite number of innovations of $\eta$'s. The existence of the fourth moment and the condition on the mixing coefficients guarantee the convergence of the average of $\dot{g}_{2j}^2(\theta_{2j}^{0\top} z_t)z_t z_t^\top$ in probability by virtue of Davydov's inequality, as shown in the supplementary material of the paper. Since this relies on a known function $h(\cdot, \cdot)$, so we impose in 3.2(i) the continuity on $h(\cdot, v)$ to make sure that $h(r,v)$ is bounded in $r\in [0,1]$ for each $v$.

Recall that $\psi(\cdot)$ is the subgradient of $\rho(\cdot)$ that satisfies Assumption 2.1.

\medskip

\noindent \textbf{Assumption 3.3}
\begin{enumerate}[(i)]
\item \emph{Suppose that $(\psi(e_t),\mathcal{F}_{t})$ is a martingale difference sequence where $\mathcal{F}_t$ is a $\sigma$-field such that $x_t$ and $v_t$ are adapted to $\mathcal{F}_{t-1}$. Moreover, $\mathbb{E}[\psi^2(e_t)|\mathcal{F}_{t-1}]=a_1$ and $\mathbb{E}[\rho''(e_t)|\mathcal{F}_{t-1}]=a_2$ almost surely for some $0<a_1,a_2<\infty$, where $\rho''(\cdot)$ may be understood as a generalized function when it does not exist in ordinary sense. In addition, $\max_{1\le t\le n} \mathbb{E}[\psi^4(e_t)| \mathcal{F}_{t-1}]<\infty$ uniformly in $n\geq 1$.}

\item \emph{Denote by $f_{t,e}(u)$ the density of $e_t$ given $\ra_{t-1}$ ($t=1,\cdots,n$). For all $n$, $\max_{1\le t\le n} f_{t,e}(u)\le f(u)$ where $f(u)$ belongs to $S$ defined in Appendix C in the supplementary material of the paper}.

\item \emph{Suppose that, as $n\to\infty$, for $r\in [0,1]$,}
\begin{equation*}
  \left(\frac{1}{\sqrt{n}}\sum_{t=1}^{[nr]} \psi(e_t), \frac{1}{\sqrt{n}}\sum_{t=1}^{[nr]}\eta_t\right)\Rightarrow (U_\rho(r), W(r)),
\end{equation*}
\end{enumerate}
\emph{which is a vector of standard Brownian motion processes of $(d_1+1)$ dimension, where $U_\rho(r)$ and $W(r)$ have variance and covariance $a_1$ and $I_{d_1}$, respectively}.

\medskip

In Assumption 3.3(i), $\psi(\cdot)$ and $\rho''(\cdot)$ are the first two derivatives of $\rho(\cdot)$ possibly in the generalized function sense, while they are the same as the derivatives in ordinary sense whenever exist.  The filtration in 3.3(i) can be taken as $\mathcal{F}_{t}=\sigma(e_1,\cdots, e_t; x_s, z_s, s\le t+1)$ that is less restrictive than mutual independence between $e_t$ and $(x_t, z_t)$. The martingale difference sequence condition renders $\mathbb{E}[\psi(e_t)|\mathcal{F}_{t-1}]=0$ that is easily fulfilled for the loss functions of LAD and Huber's when the conditional distribution of $e_t$ given $\mathcal{F}_{t-1}$ is symmetric, because these functions $\psi(\cdot)$ are odd. Also, this condition is equivalent to $F_e(0|\mathcal{F}_{t-1}) =\tau$ for quantile loss (see, for example, Assumption 2 in \citealp{degui2019}). Some authors obtain this condition by adjusting the intercept that we also have to deal with when it is violated, like \citet[p. 124]{hexuming2000} and \citet[p. 251]{xiao2009a}.

When $\rho''(\cdot)$ in 3.3(i) is genuinely a generalized function, the conditional expectations can be defined as
\begin{equation*}
\e[\rho''(e_t)|\mathcal{F}_{t-1}]=\int \rho(u)f''_{t,e}(u)du
\end{equation*}
under some conditions on the conditional density $f_{t,e}(u)$ of $e_t$ given $\mathcal{F}_{t-1}$. Though $\rho(\cdot)$ may not be differentiable at some points, $\mathbb{E}[\psi(e_t)|\mathcal{F}_{t-1}]$ and $\mathbb{E}[\rho''(e_t)|\mathcal{F}_{t-1}]$ are well-defined given that the conditional density is smooth and is a member in $S$ (the rapid decay test function space, see Appendix C in the supplementary material); this is due to the use of generalized functions that extend the ordinary derivatives. The condition D.2 in \citet[p. 125]{hexuming2000} defines $\mathbb{E}[\rho''(e_t)]$ in the same fashion as for non-differentiable loss functions.

In addition, the positiveness $\mathbb{E}[\rho''(e_t)|\mathcal{F}_{t-1}] =a_2>0$ is due to the convexity of the loss functions under consideration. In particular, $a_2=f(0)>0$ for both check loss and LAD loss with $f(\cdot)$ being the density of $e_t$, while for Huber's loss, $a_2= P(|e_t|\le c)>0$ in exogenous situation.

Condition 3.3(ii) is used as a technical requirement in related calculations. See Lemma \ref{lemma3}. This holds in particular when the regressors are independent of the error terms.

Note that 3.3(iii) is quite commonly encountered in the cointegrating regression literature when there is no additional stationary regressor involved (see \citealp{phillips2001}, \citealp{xiao2009a} and \citealp{qiying2018}). Nevertheless, here the subscript $\rho$ indicates that the joint convergence relies on the loss function. For example, equation (3) of \citet{phillips1995} shows the functional invariance principle for the derivative of LAD loss, i.e. sign$(\cdot)$. For simplicity of notation we suppress the subscript, viz., denote $U_\rho(r)$ by $U(r)$, if there is no confusion raised.

Note that 3.3(ii), in particular, ensures that as $n\rightarrow \infty$
\begin{equation}\label{bmotion}
 \left(\frac{1}{\sqrt{n}}\sum_{t=1}^{[nr]} \psi(e_t), \frac{1}{\sqrt{n}}x_{[nr]}\right)\Rightarrow (U(r), B(r)),
\end{equation}
where $B(r)$ is a vector of Brownian motion with covariance matrix $AA^\top$.

\medskip

Because of the involvement of the unit root process $x_t$, we need a similar classification on the function classes, such as $H$-regular and $I$-regular in \citet{phillips1999,phillips2001}. However, in our paper we only consider a simpler version of $H$-regular class, that is, power functions, because our model incorporates an additive form of $H$-regular functions.

\begin{defn}
\emph{A function $f(\cdot)$ defined on $\mathbb{R}$ is called $H$-regular with homogeneity order $\nu(\cdot)$ if $f(\lambda x)=\nu(\lambda)f(x)$ for all $\lambda>0$ and any $x$. A function $f(\cdot)$ defined on $\mathbb{R}$ is called $I$-regular if $\int |f(x)|dx<\infty$.}
\end{defn}

The class of $H$-regular functions mainly contains power functions while the class of $I$-regular functions has all integrable functions on the entire real line as its member, such as probability density functions. More detailed discussion on these definitions can be found in \citet{phillips1999, phillips2001}.

\medskip

\noindent \textbf{Assumption 3.4} \ \ \emph{Let all $g_{ij}(\cdot)$, for $1\leq  j \leq p_i$ and $1\leq i\leq 2$, be continuously differentiable up to the second order. Suppose further that}
\begin{enumerate}[(i)]
  \item \emph{All $g_{1j}(u)$, $\dot{g}_{1j}(u)$ and $\ddot{g}_{1j}(u)$ are $H$-regular with homogeneity order $\nu_j(\cdot)$, $\dot{\nu}_j(\cdot)$ and $\ddot{\nu}_j(\cdot)$, respectively, such that $\lim_{\lambda\to +\infty} \dot{\nu}_j(\lambda)/\nu_j(\lambda) =0$, and $\lim_{\lambda\to +\infty} \ddot{\nu}_j(\lambda)/\dot{\nu}_j (\lambda) =0$, $j=1,\ldots, p_1$.}

    \item \emph{(a) $\theta_{1j}^0=\theta_1^0$, $j=1,\ldots, p_1$. (b) Suppose that $g_{1j}(u)$, $g_{1j}^2(u)$, $\dot{g}_{1j}(u)$, $\dot{g}_{1j}^2(u)$, $u\dot{g}_{1j}(u)$, and $[u\dot{g}_{1j}(u)]^2$ are all $I$-regular for $j=1,\ldots, p_1$.}
\end{enumerate}

\medskip

Assumption 3.4 requires all $g_{ij}(\cdot)$, for $1\leq  j \leq p_i$ and $1\leq i\leq 2$ to be continuously differentiable up to the second order, while considering two classes of functions for all $g_{1j}(u)$, $j=1,\ldots,p_1,$ that is, $H$-regular and $I$-regular. This is because, when $\theta'x_t$ is a unit root process, the behaviour of $g(\theta'x_t)$ depends heavily on whether $g(\cdot)$ is $H$-regular or $I$-regular (see, for example, \cite{dgd2016}). More importantly, in $I$-regular case we require that all index vectors be the same $\theta_{1j}^0=\theta_1^0$, $j=1,\ldots, p_1$. That is, all integrable functions share the common stochastic trend $x_t'\theta_1^0$, since otherwise there would be a technical bottle neck. Under this setting, the first part of the regression function can be rewritten as
\begin{equation*}
\sum_{j=1}^{p_1}\gamma_{1j}^0g_{1j}(x_{t}^\top \theta_{1}^0)\equiv g_1(x_{t}^\top \theta_{1}^0; \gamma_{1}^0), \ \ \ \text{with}\ \gamma_{1}^0=( \gamma_{11}^0, \cdots, \gamma_{1p_1}^0),
\end{equation*}
where $g_1(x^\top \theta_{1}^0; \gamma_{1}^0)$ is known up to $\theta_{1}^0$ and $\gamma_{1}^0$.

\section{Asymptotic theory}

The asymptotic behaviours of the proposed estimators of the parameters associated with the nonstationary part in our regression equation depend heavily on the functional functions of $g_{1j}(\cdot)$, $j=1,\cdots, p_1$. We therefore establish our asymptotic theory according to $H$-regular and $I$-regular functions, respectively.

\subsection{$H$-regular regression functions}

We firstly consider the case of $H$-regular functions, that is, Assumption 3.4(i) holds. For better exposition, denote $x_{nt}=n^{-1/2}x_t$ and two vectors associated with the regression functions and regressors by
\begin{eqnarray}\label{41a}
\begin{split}
Z_{1}(x_{nt})&\equiv&
(-\gamma_{11}^0\dot{g}_{11}(x_{nt}^\top \theta_{11}^0)x_{nt}^\top, \ldots, -\gamma_{1p_1}^0\dot{g}_{1p_1}(x_{nt}^\top \theta_{1p_1}^0)x_{nt}^\top, g_{11}(x_{nt}^\top \theta_{11}^0),\ldots, g_{1p_1}(x_{nt}^\top \theta_{1p_1}^0) )^\top, \\
Z_{2}(z_t) &\equiv& (-\gamma_{21}^0\dot{g}_{21}(z_{t}^\top \theta_{21}^0)z_{t}^\top, \ldots, -\gamma_{2p_2}^0\dot{g}_{2p_2}(z_{t}^\top \theta_{2p_2}^0)z_{t}^\top, g_{21}(z_{t}^\top \theta_{21}^0),\ldots, g_{2p_2}(z_{t}^\top \theta_{2p_2}^0) )^\top.
\end{split}
\end{eqnarray}
In addition, we define a symmetric matrix $B$ and a vector $B_1$ given by
\begin{align}\label{41b}
\begin{split}
B\equiv&a_{2}\begin{pmatrix}\int_0^1Z_1(B(r))Z_1(B(r))^\top dr & \int_0^1Z_1(B(r))[\mathbb{E}Z_2(h(r, v_1))]^\top dr\\ \int_0^1[\mathbb{E}Z_2(h(r, v_1))]Z_1(B(r))^\top dr
 & \int_0^1\mathbb{E}[Z_2(h(r, v_1))Z_2(h(r, v_1))]^\top dr \end{pmatrix},\\
B_1\equiv&\begin{pmatrix}\int_0^1Z_1(B(r))dU(r)\\ N(0, a_1\Sigma)\end{pmatrix}, \ \ \ \text{where}\;  \Sigma=\int_0^1\mathbb{E}[Z_2(h(r, v_1))Z_2(h(r, v_1))]^\top dr,
\end{split}
\end{align}
where the positive constants $a_1, a_2$ are given by Assumption 3.3 and processes $B(r)$ and $U(r)$ are defined by \eqref{bmotion}. Also, let
\begin{align}\label{41b1}
\begin{split}
 \widehat{\Lambda}_1=&\Big[\dot{\nu}_1(\sqrt{n})n(\widehat\theta_{11}-
 \theta_{11}^0),
 \cdots, \dot{\nu}_{p_1}(\sqrt{n})n(\widehat\theta_{1p_1}-\theta_{1p_1}^0),\\ &\quad
 \nu_1(\sqrt{n})\sqrt{n}(\widehat\gamma_{11}-\gamma_{11}^0), \cdots,  \nu_{p_1}(\sqrt{n})\sqrt{n}(\widehat\gamma_{1p_1}-\gamma_{1p_1}^0)\Big]^\top, \\
 \widehat{\Lambda}_2=&\Big[\sqrt{n}(\widehat\theta_{21}-
 \theta_{21}^0),
 \cdots, \sqrt{n}(\widehat\theta_{2p_2}-\theta_{2p_2}^0),
 \sqrt{n}(\widehat\gamma_{21}-\gamma_{21}^0), \cdots,  \sqrt{n}(\widehat\gamma_{2p_2}-\gamma_{2p_2}^0)\Big]^\top,
 \end{split}
\end{align}
where $\widehat\theta_{1j}$, $\widehat\theta_{2j}$, $\widehat\gamma_{1j}$ and $\widehat\gamma_{2j}$ are the minimizers of the objective function $L_n(\theta, \gamma)$ defined by \eqref{objfun}.

\begin{theo}\label{th41}
Under Assumptions 2.1, 3.1-3.3 and 3.4(i), as $n\to\infty$ we have $\widehat{\Lambda}=O_P(1)$ where $\widehat{\Lambda}=(\widehat{\Lambda}_1^\top, \widehat{\Lambda}_2^\top)^\top$, and moreover,
\begin{equation}\label{th41a}
\widehat{\Lambda}\to_DB^{-1}B_1.
\end{equation}
\end{theo}

The proof of the theorem is given in Appendix B. Note that in the proof we make use of the regular sequence of $\rho(\cdot)$ defined in Theorem 2.1, and then similar to Theorem 2.2, the objective function can be approximated by a quadratic function. Hence, the consistency and convergence rate of all estimators are implied by $\widehat{\Lambda}=O_P(1)$.

As can be seen from the asymptotics in \eqref{th41a}, all convergence rates of $\widehat\theta_{1j}$ and $\widehat\gamma_{1j}$ are affected by the $H$-regular property of $g_{1j}(\cdot)$, $j=1,\cdots, p_1$. Precisely, $\widehat\theta_{1j}$ has convergence rate $[\dot{\nu}_j(\sqrt{n})n]^{-1}$ where $\nu_j(\cdot)$ is the asymptotic order of $g_{1j}(\cdot)$. For example, if $g_{11}(u)=u$, $\nu_1(\lambda)=\lambda$, then $\widehat\theta_{11}$ has rate $n^{-1}$, so-called super rate in nonstationary context; if $g_{12}(u)=u^2$, $\nu_1(\lambda)=\lambda^2$, then $\widehat\theta_{12}$ has rate $n^{-3/2}$.

Notice that, given the identification condition $\|\theta_{1j}^0\|=1$, we are able to identify $\gamma_{1j}$. Moreover, the rate of $\widehat\gamma_{1j}$ is affected by $g_{1j}(\cdot)$ too, as it converges at rate of $[\nu_j(\sqrt{n})\sqrt{n}]^{-1}$. Since $x_t^\top \theta_{1j}^0$ is still a unit root process, $g_{1j}(x_t^\top \theta_{1j}^0)$ has asymptotic order $\nu_j(\sqrt{n})$ that results in a rapid rate for $\widehat\gamma_{1j}$ by a factor $[\nu_j(\sqrt{n})]^{-1}$, comparing with stationary case.

By contrast, all the estimators $\widehat\theta_{2j}$ and $\widehat\gamma_{2j}$, $j=1,\cdots, p_2$, have usual square-root-$n$ convergence rate, although these parameters are involved in the regression accommodating both stationary and nonstationary variables. This is simply because the regression function takes additive form so that the stationary part and the nonstationary part are separated, and thus the parameters in the stationary part retain the usual square-root-$n$. As can be seen below, however, the limits of all estimators in both parts include all ingredients in the regression. This was also found in \citet{DL2018}.

In order to obtain the asymptotic properties of $\widehat{\Lambda}_1$ and $\widehat{\Lambda}_2$ separately, we partition the matrix $B=(B_{ij})_{2\times 2}$ and vector $B_1=(b_{11}^\top, b_{12}^\top)^\top$ conformably,
\begin{align*}
B_{11}=&a_2\int_0^1Z_1(B(r))Z_1(B(r))^\top dr, & B_{12}=&a_2\int_0^1Z_1(B(r))[\mathbb{E}Z_2(h(r, z_1))]^\top dr \\
B_{21}= & B_{12}^\top, & B_{22}=&a_2\Sigma,\\
b_{11}=&\int_0^1Z_1(B(r))dU(r), & b_{12}=&\sqrt{a_1}\;N(0,\Sigma).
\end{align*}
Then, \eqref{th41a} implies that
\begin{align}\label{41c}
\begin{split}
\widehat{\Lambda}_1\to_D&(B_{11}-B_{12}B_{22}^{-1}B_{21})^{-1}b_{11}
-B_{11}^{-1}B_{12}
(B_{22}-B_{21}B_{11}^{-1}B_{12})^{-1}b_{12},\\
\widehat{\Lambda}_2\to_D&(B_{22}-B_{21}B_{11}^{-1}B_{12})^{-1}b_{12}-
(B_{22}-B_{21}B_{11}^{-1}B_{12})^{-1}B_{21}B_{11}^{-1}b_{11}.
\end{split}
\end{align}

To understand this, take $p_1=p_2=1$ as example. Consider
\begin{equation*}
  y_t=\gamma_1^0g_1(x_t^\top \theta_1^0)+\gamma_2^0g_2(z_t^\top \theta_2^0)+e_t, \ \ t=1,\cdots, n,
\end{equation*}
where $g_1(\cdot)$ is $H$-regular with asymptotic order $\nu_1(\cdot)$, $\mathbb{E}z_t=0$ that implies $B_{12}=B_{21}^\top =0$. Theorem \ref{th41}, in particular \eqref{41c}, gives
\begin{align}\label{41d}
\begin{split}
\begin{pmatrix}\dot{\nu}_1(\sqrt{n})n(\widehat\theta_{1}-
\theta_1^0)\\ \nu_1(\sqrt{n})\sqrt{n}(\widehat\gamma_1-\gamma_1^0)
\end{pmatrix}
\to_D&B_{11}^{-1}b_{11},\\
\begin{pmatrix}\sqrt{n}(\widehat\theta_{2}-
\theta_2^0)\\ \sqrt{n}(\widehat\gamma_2-\gamma_2^0)
\end{pmatrix}\to_D&B_{22}^{-1}b_{12},
\end{split}
\end{align}
where all notation $B_{11}, B_{22}, b_{11}, b_{12}$ can be explicitly given from the above discussion.

It is noteworthy that the impact of the loss function to estimation is through the constants $a_1$ and $a_2$ that are easily separated from the matrices and vectors, so the limits of $\widehat{\Lambda}_1$ and $\widehat{\Lambda}_2$ have factors $1/a_2$ and $\sqrt{a_1}/a_2$, respectively.

In order to make a comparison with the relevant literature, here we simply suppose the model is exogenous. (1) When $\rho(u)=|u|$, $\rho'(u)=-H(-u)+H(u)$ with $H(\cdot)$ being Heaviside function and $\rho''(u)=2\delta(u)$. We have $a_1=\mathbb{E}[\rho'(e_t)]^2=1$ and $a_2=\mathbb{E}[\rho''(e_t)] =2f_e(0)>0$ where $f_e(\cdot)$ is the density of $e_t$. Moreover, if $g_1(u)=u$, $\gamma_1^0=1$ needed not to estimate, and $g_2(v)\equiv 0$, the model reduces to a usual linear cointegrating model. Our result implies that
$$n(\widehat \theta_1-\theta_1^0)\to_D [2f_e(0)]^{-1}\left[\int_0^1B(r)B(r)^\top dr\right]^{-1} \int_0^1 B(r)dU(r),$$
which is the same as \citet{phillips1995} without correlation between $x_t$ and $e_t$; if $g_2(v)=v$, $\gamma_2^0=1$ needed not to estimate, and $g_1(u)\equiv 0$, $h(r, v)\equiv v$, the model reduces to a linear model with stationary regressors. Our result also naturally covers that $\sqrt{n}(\widehat \theta_2-\theta_2^0)\to_D[2f_e(0)]^{-1} [\mathbb{E}(v_1v_1^\top)]^{-1/2} N(0,I_{d_2})$ as shown in \citet{pollard1991}.

(2) When $\rho(u)$ is the quantile loss, $\rho'(u)=\tau H(u)+(\tau-1)H(-u)$ and $\rho''(u)=\delta(u)$. We have $a_1=\mathbb{E}[\rho'(e_t)]^2 =\tau(1-\tau)$ and $a_2=\mathbb{E}[\rho''(e_t)]=f_e(0)>0$ where $f_e(\cdot)$ is the density of $e_t$. Further, if $g_1(u)=u$, $\gamma_1^0=1$ needed not to estimate and $g_2(v)\equiv 0$, the model reduces to a linear cointegrating model. Our result implies that $n(\widehat \theta_1-\theta_1^0)\to_D [f_e(0)]^{-1} [\int_0^1B(r)B(r)^\top dr]^{-1} \int_0^1 B(r)dU(r)$ that is the same result as in \citet{xiao2009a} without correlation between $x_t$ and $e_t$; if $g_2(v)=v$, $\gamma_2^0=1$ need not to be estimated and $g_1(u)\equiv 0$, $h(r,v)\equiv v$, the model reduces to a linear model with one stationary regressor. Our result implies that $\sqrt{n}(\widehat \theta_2- \theta_2^0) \to_D\sqrt{\tau(1-\tau)}[f_e(0)]^{-1} [\mathbb{E}(v_1v_1^\top)]^{-1/2} N(0,I_{d_2})$, which is the result of quantile regression for linear model. See \citet{koenker1978}.

\medskip

While $\Sigma$ can easily be consistently estimated by the function $h$ and observations of $v_t$, we need to construct consistent estimators for both $a_1$ and $a_2$ in order to use the asymptotic limits for statistical inferences. These estimators are given by
\begin{align}\label{41f}
\widehat{a}_1=\frac{1}{n}\sum_{t=1}^n [\psi(\widehat{e}_t)]^2 \ \ \ \mbox{and} \ \ \ \widehat{a}_2=\frac{1}{n}\sum_{t=1}^n \rho''_m(\widehat{e}_t),
\end{align}
where $\widehat{e}_t=y_t-\sum_{j=1}^{p_1}\widehat\gamma_{1j}g_{1j}
(\widehat{\theta}_{1j}^{\,\top} x_t) -\sum_{j=1}^{p_2}\widehat\gamma_{2j}g_{2j}(\widehat{\theta}_{2j}^{\,\top} z_t)$ for $1\leq t\leq n$, and $m=[n^{2+\varepsilon}]$ for some $\varepsilon>0$. Although both $a_1$ and $a_2$ are defined in terms of conditional expectation, they are the corresponding expectations as well, so we simply define their estimators as the corresponding sample averages. We then have the following corollary.

\begin{cor}\label{cor1}
Let the conditions of Theorem \ref{th41} hold. Then we have $\widehat{a}_1\to_Pa_1$. If, in addition, suppose that $\{\rho_m''(e_t), t=1, \cdots, n\}$ is ergodic, we have $\widehat{a}_2\to_Pa_2$ as $n\to\infty$.
\end{cor}

In the proof of $\widehat{a}_1\to_Pa_1$ we use the ergodicity of $\{\psi(e_t)\}$ that is implied from its martingale difference sequence property; similarly in the proof of $\widehat{a}_2\to_Pa_2$, we need the ergodicity of $\{\rho_m''(e_t)\}$. Certainly, the ergodicity of the two process is ensured by that of $\{e_t\}$, but we do not impose this condition in order to keep the condition as weak as possible.

\subsection{$I$-regular regression functions}

In this subsection, we shall consider the case where all $g_{1j}(\cdot)$ are $I$-regular satisfying Assumption 3.4(ii). This assumption will change the asymptotic theory considerably. Note that in 3.4(ii), $\theta_{1j}^0= \theta_1^0, j=1, \cdots, p_1$, so all $I$-regular functions have the same stochastic trend $x_t^\top\theta_{1}^0$ as their argument. Moreover, the limit theory depends on the local time defined for scalar Brownian motion $W(r)$ as
\begin{equation}\label{42a}
  L_W(p,s)=\lim_{\epsilon\to 0}\frac{1}{2\epsilon} \int_0^pI(|W(r)-s|<\epsilon) dr,
\end{equation}
which measures the sojourning time of $W(r)$ around $s$ during the time period $(0,p)$. In our context, the scalar Brownian motion is $B_{1}(r)\equiv (\theta_{1}^0)^\top B(r)$ where $B(r)$ is defined by \eqref{bmotion}, and we denote its local time process by $L_1(p,s)$.

Another feature of the limit theory about this kind of regression functions is that we derive the asymptotics in a new coordinate systems first, then recover the limits in the original coordinate. This is the same as \citet{phillips2000}, and \citet{dgd2016}. To do so, using $\theta_{1}^0$ that is a unit vector, we construct an orthogonal matrix $P=(\theta_{1}^0, P_{2})_{d_1\times d_1}$ to rotate all the vectors of interest, including $\theta_{1}^0$, $x_t$ and $B(r)$. Hence, $P^\top \theta_{1}^0 =(1,0,\cdots,0)^\top$; other relevant rotated vectors are $P^\top x_t$ that have the first element $x_{1,t}\equiv (\theta_{1}^0)^\top x_t$ and sub-vector $x_{2,t}\equiv P_{2}^\top x_t$; due to continuous mapping theorem, after normalized by $\sqrt{n}$ they converge jointly as follows:
\begin{equation}\label{42b}
\frac{1}{\sqrt{n}}x_{1,[nr]}\to_DB_{1}(r)\equiv (\theta_{1}^0)^\top B(r) \ \ \ \mbox{and} \ \ \ \frac{1}{\sqrt{n}}x_{2,[nr]}\to_DB_{2}(r)\equiv P_{2}^\top B(r).
\end{equation}

To proceed, denote $D_n=\text{diag}(\sqrt[4]{n}, \sqrt[4]{n}^3I_{d_1-1})$, $p_n=\sqrt{n}$ and $q_n=\sqrt[4]{n}$, and define
\begin{align}\label{42c}
\begin{split}
\alpha_{1}=&D_nP^\top (\theta_{1}-\theta_{1}^0),\ \beta_{1j}=q_n (\gamma_{1j}-\gamma_{1j}^0), \ \ j=1,\cdots, p_1, \\ \alpha_{2j}=&p_n(\theta_{2j}-\theta_{2j}^0),\ \ \ \beta_{2j}=p_n (\gamma_{2j}-\gamma_{2j}^0),\ j=1,\cdots, p_2.
\end{split}
\end{align}
Here, $\theta_1, \theta_{2j}, \gamma_{1j}, \gamma_{2j}$ are generic parameters, from which $\alpha_1, \beta_{1j}, \alpha_{2j}, \beta_{2j}$ are centered and rescaled, and in particular $\alpha_1$ is the centered vector $\theta_{1}-\theta_{1}^0$ rotated by $P$ and rescaled by $D_n$.

The role that the rotation takes is for asymptotic analysis only, while in Monte Carlo experiments and empirical study the rotation is not necessary; more details can be seen in the relevant literature such as \citet{phillips2000}, and \citet{dgd2016}.

To make the notation compact, we also define $\Lambda\equiv(\Lambda_1^\top, \Lambda_2^\top)^\top$ where
\begin{align*}
\Lambda_1\equiv &(\alpha_{1}^\top, \beta_{11}, \cdots, \beta_{1p_1})^\top, &
\Lambda_2\equiv &(\alpha_{21}^\top, \cdots, \alpha_{2p_2}^\top, \beta_{21}, \cdots, \beta_{2p_2})^\top.
\end{align*}
We shall give the asymptotics of $\widehat\Lambda=(\widehat\Lambda_1^\top, \widehat\Lambda_2^\top)^\top$ where
\begin{align*}
\widehat\Lambda_1= &(\widehat\alpha_{1}^\top, \widehat\beta_{11}, \cdots, \widehat\beta_{1p_1})^\top, &
\widehat\Lambda_2= &(\widehat\alpha_{21}^\top, \cdots, \widehat\alpha_{2p_2}^\top, \widehat\beta_{21}, \cdots, \widehat\beta_{2p_2})^\top,
\end{align*}
that are defined in terms of $\widehat\theta_1, \widehat\theta_{2j}, \widehat\gamma_{1j}, \widehat\gamma_{2j}$ through the relationship in \eqref{42c}.

To state the asymptotic properties of these estimators, we need to define several much more complicated vectors and matrices. First, let $x_{nt}\equiv D_n^{-1}P^\top x_t$, $\dot{g}(u)\equiv\sum_{j=1}^{p_1} \gamma_{1j}^0 \dot{g}_{1j}(u)$ and $Z_t\equiv(Z_1(x_{nt})^\top, Z_2(z_t)^\top)^\top$, where
\begin{align}\label{42c1}
\begin{split}
Z_1(x_{nt})\equiv &\left(-\dot{g}(x_t^\top \theta_{1}^0)x_{nt}^\top,\ \ q_n^{-1}g_{11}(x_t^\top \theta_{1}^0), \cdots, q_n^{-1}g_{1p_1}(x_t^\top \theta_{1}^0)\right)^\top, \\
Z_2(z_t)\equiv &p_n^{-1}(-\gamma_{21}^0\dot{g}_{21}(z_t^\top \theta_{21}^0)z_{t}^\top, \cdots, -\gamma_{2p_2}^0\dot{g}_{2p_2}(z_t^\top \theta_{2p_2}^0)z_{t}^\top, g_{21}(z_t^\top \theta_{21}^0), \cdots, g_{2p_2}(z_t^\top \theta_{2p_2}^0))^\top,
\end{split}
\end{align}
where $p_n$ and $q_n$ are the same as in \eqref{42c}.

Second, to show the limits of $\sum_{t=1}^nZ_tZ_t^\top$ and $\sum_{t=1}^n\psi(e_t)Z_t$ we need to define a matrix. Let $\mathcal{R}$ be a square matrix of order $(d_1+p_1)+p_2(d_2+1)$, standing for the limit of $\sum_{t=1}^nZ_tZ_t^\top$, that we divide into blocks $\mathcal{R}=(\mathcal{R}_{ij})_{2\times 2}$, conformably with the blocks in $Z_tZ_t^\top$, i.e. $Z_1Z_1^\top$, $Z_1Z_2^\top$, $Z_2Z_1^\top$ and $Z_2Z_2^\top$. Thus, $\mathcal{R}_{11}=(R_{ij}^{11})$ is a symmetric matrix of order $d_1+p_1$ that has elements,
\begin{align*}
R_{11}^{11}=& L_1(1,0)\int  [\dot{g}(u)]^2u^2du,  &
R_{1,2:d_1}^{11}=&\int_0^1B_2(r)^\top dL_1(r,0)\int [\dot{g}(u)]^2udu, \\
R_{1, d_1+1}^{11}=&-L_1(1,0)\int ug_{11}(u)\dot{g}(u)du, \cdots&
R_{1, d_1+p_1}^{11}=&-L_1(1,0)\int ug_{1p_1}(u)\dot{g}(u)du \\
R_{2:d_1, 2:d_1}^{11}=& \int_0^1B_2(r)B_2(r)^\top dL_1(r,0)\int [\dot{g}(u)]^2du, & &\\
R_{2:d_1, d_1+1}^{11}=&-\int_0^1B_2(r) dL_1(r,0)\int g_{11}(u)\dot{g}(u)du, & &\cdots\\
R_{2:d_1, d_1+p_1}^{11}=&-\int_0^1B_2(r) dL_1(r,0)\int g_{1p_1}(u)\dot{g}(u)du,& &\\
R_{d_1+1,d_1+1}^{11}=&L_1(1,0)\int [g_{11}(u)]^2du, \cdots & R_{d_1+1,d_1+p_1}^{11}=&L_1(1,0)\int g_{11}(u)g_{1p_1}(u)du\\
\vdots & & &\vdots \\
R_{d_1+p_1,d_1+1}^{11}=&L_1(1,0)\int g_{11}(u)g_{1p_1}(u)du, \cdots & R_{d_1+p_1,d_1+p_1}^{11}=&L_1(1,0)\int [g_{1p_1}(u)]^2du,
\end{align*}
where $R_{2:d_1, 2:d_1}^{11}$ denotes all elements $R_{i, j}^{11}$ with $i,j =2, \cdots, d_1$; other symbols, such as $R_{2:d_1, d_1+1}^{11}, \cdots$ and $R_{2:d_1, d_1+p_1}^{11}$, are explained similarly.

Note that $\mathcal{R}_{22}=\Sigma$, the same as given in Theorem 4.1, while $\mathcal{R}_{12}=\mathcal{R}_{12}^\top=0$ a zero matrix of order $(d_1+p_1)\times (d_2p_2+p_2)$. Thus, $\mathcal{R}$ is a diagonal block matrix.

It can be seen that there may be many zeros even in the block $\mathcal{R}_{11}$ if $g_{11}(u), \cdots, g_{1p_1}(u)$ are taken from an orthogonal sequence such that $\int g_{1j}(u)g_{1k}(u)du=0$ for $j\ne k$. This is most likely to be true because the orthogonality among $g_{11}(u), \cdots, g_{1p_1}(u)$ is the primary motivation for their choice.

\begin{theo}\label{th42}
Under Assumptions 2.1, 3.1-3.3 and 3.4(ii), as $n\to\infty$ we have $\widehat{\Lambda}=O_P(1)$ and moreover,
\begin{equation}\label{th42a}
\widehat{\Lambda}\to_D\DF{\sqrt{a_1}}{a_2}\mathcal{R}^{-1/2}N(0, I),
\end{equation}
where $I$ is an identity matrix of order $d_1+p_1+p_2d_2+p_2$, the positive constants $a_1$ and $a_2$ are given by Assumption 3.3.
\end{theo}

The proof of the theorem is given in Appendix B. Similar to the comment right below Theorem 4.1, the consistency and convergence order of all estimators are implied by $\widehat{\Lambda}=O_P(1)$. 

Due to diagonal block structure of $\mathcal{R}$, the assertion \eqref{th42a} implies
\begin{align}\label{42d}
\widehat{\Lambda}_1\to_D \DF{\sqrt{a_1}}{a_2}\mathcal{R}^{-1/2}_{11}N(0, I) \ \ \ \mbox{and} \ \ \ \widehat{\Lambda}_2\to_D \DF{\sqrt{a_1}}{a_2}\mathcal{R}^{-1/2}_{22}N(0, I)
\end{align}
as $n\to\infty$. It follows from \eqref{42c} that all estimators $\widehat\theta_{2j}$ and $\widehat\gamma_{2j}$, $j=1,\cdots, p_2$, have conventional $\sqrt{n}$-convergent rate, whereas the rates of $\widehat\theta_{1}$ and $\widehat\gamma_{1j}$, $j=1,\cdots, p_1$, affected by the unit root property of $x_t$ and $I$-regularity of functions $f_{1j}$, are distorted, similar to the results in \citet{dgd2016}. This feature coincides with \citet{DL2018} since these models have additive form where these nonstationary and stationary variables are separated in different components.

In particular, $\widehat\gamma_{1j}$ have a very slow rate $n^{-1/4}$, because they are the coefficients of $g_{1j}(x_t^\top \theta_1^0)$, but $g_{1j}(x_t^\top \theta_1^0)$ attenuates to zero when $t$ gets large. This rate is the same as classical papers such as \citet{phillips1999, phillips2001}. However, the rate of $\widehat\theta_{1}$ is blurred due to the involvement of $D_n$ and $P$. To find out its exact rate, denote by $r_{11}$ the left-top $d_1\times d_1$ submatrix of $\mathcal{R}^{-1}_{11}$. Thus,
\begin{equation}\label{42e}
D_nP^\top (\widehat\theta_{1}-\theta_{1}^0)\to_D
\DF{\sqrt{a_1}}{a_2}N(0, r_{11}).
\end{equation}

The following corollary recovers the asymptotic distribution of $\widehat\theta$ from the above limit of its rotation.

\begin{cor}\label{cor2}
Under the conditions in Theorem \ref{th42}, we have as $n\to\infty$
\begin{equation*}
\sqrt[4]{n}(\widehat\theta_1-\theta_1^0)\to_D\DF{\sqrt{a_1}}{a_2}N(0, r_{11}^{11}\theta_1^0(\theta_1^0)^\top),
\end{equation*}
where $r_{11}^{11}$ is the left-top element of $r_{11}$.
\end{cor}

Though indicated by \eqref{42e} that the coordinates of $\widehat\theta_1$ on $\theta_1^0$ and $P_2$ have rates $n^{-1/4}$ and $n^{-3/4}$, respectively, the estimator $\widehat\theta_1$ eventually has slower rate $n^{-1/4}$. This is the same as \citet{dgd2016} and one may find more detailed explanation therein. In addition, if we normalize $\widehat\theta_1$ to be $\widehat\theta_{1,unit}=\widehat\theta_1/\|\widehat\theta_1\|$, one may find $\widehat\theta_{1,unit}$ converges to $\theta_1^0$ with a quicker rate $n^{-3/4}$. Since this is exactly the same as \citet{dgd2016}, we omit the derivation and refer the readers to the reference paper.

Similar to Corollary \ref{cor1}, we may define $\widehat{a}_1$ and $\widehat{a}_2$ and then establish their consistency. This however is omitted due to similarity.

\section{Numerical results}
\subsection{Simulations}

This subsection presents simulation experiments to evaluate the finite sample performances of the robust M estimators in the multiple index cointegration model. For space consideration, we only entertain two examples. The first example considers homogeneous cointegrating functions, while the second example studies the case with an integrable cointegrating function.

\medskip

The time series $\{x_t, z_t\}$ used in the following two examples are generated as follows. $x_t=\rho_{1} x_{t-1}+\sigma_1 w_{t}$, $z_t=h(t/n)+v_t$, $h(\tau)=(\tau, \tau)^\top$, $v_t=\rho_{2} v_{t-1}+\sigma_2 \epsilon_{t}$, both $x_t$ and $v_t$ are bivariate autoregressive vector processes with $x_0=0$,  $v_0=0$,  $\rho_{1}=I_2$, $\rho_{2}=0.5\times I_2$, $\sigma_1={\rm diag}\{0.2,0.5\}$, $\sigma_2=I_2$, and $(w _{t}^\top, \epsilon_{t}^\top)$ is a vector of independent and normally distributed random variables with zero mean and identity covariance matrix.

\medskip

{\noindent \bf Example 5.1}. We first consider the following model:
\begin{equation*}
  y_t=\gamma_1 (x_{t}^\top {\bf\theta_1})+\gamma_2(x_{t}^\top {\bf\theta_2})^2+ \gamma_3(z_{t}^\top {\bf\theta_3})+e_t,
\end{equation*}
where  ${\bf\theta}^\top_1= (\theta_{11},\theta_{12})=(1,1)/\sqrt{2}$,
${\bf\theta}^\top_2=(\theta_{21},\theta_{22})=(1, 1)/\sqrt{2})$, ${\bf\theta}^\top_3=(\theta_{31},\theta_{32})=(1,1)/\sqrt{2}$,
$\gamma_1=\gamma_2=2$, $\gamma_3=1$. The regression residual is set as $e_t=0.5\cdot  u_t$ with $u_t$ independently generated according to four distributions: (D1) standard normal, $N(0,1)$;  (D2) mixed normal, $0.9\cdot N(0,1)+0.1\cdot N(0,4)$; (D3) $t(2)$, $t$ distribution with 2 degrees of freedom; (D4) $t(1)$, standard Cauchy distribution.

\medskip

{\noindent \bf Example 5.2}. We next consider the following model:
\begin{equation*}
  y_t=\gamma_1 \phi(x_{t}^\top {\bf\theta_1})+ \gamma_2(z_{t}^\top {\bf\theta_3})+e_t,
\end{equation*}
where $\phi(\cdot)$ is the standard normal density function, ${\bf\theta}^\top_1=(\theta_{11},\theta_{12})=(1,1)/\sqrt{2}$,
${\bf\theta}^\top_2=(\theta_{21},\theta_{22})=(1,1)/\sqrt{2}$,
$\gamma_1=2$, $\gamma_2=1$. $e_t$ is generated the same as in Example 5.1.

For demonstration, three loss functions are entertained in both examples: (L1) Huber's loss $\rho_c (e)=\frac{1}{2}e^2\cdot 1\{|e|\leq c\}+(\delta (|e|-\frac{1}{2} c^2) \cdot 1\{|e|> c\}$ with $c=1.25$; (L2) absolute errors loss $\rho(e)=|e|$, and (L3) quantile loss $\rho_\tau(e)=e\cdot (\tau-I\{e<0\}$ with $\tau=0.3$. The simulations are conducted for $n=50, 100, 200$ with 5,000 replications. For the quantile regression, the regression residual is normalized such that its $\tau$-th quantile is set as 0.
To measure the estimation accuracy of the nonlinear least-squares estimates for the index parameters, we compute the bias, estimated standard deviations and mean squared errors for each element of the indices. For space limitation, we only report the mean squared errors in Tables \ref{rmse1}-\ref{rmse2} for Examples 5.1-5.2, respectively. Each table contains the estimation results under the three loss functions L1-L3, and four types of error distributions D1-D4.

The main findings from Table \ref{rmse1} are summarized as follows.  First, the mean squared errors are decreasing as the sample size $n$ increases, which indicate the consistency of the robust M estimators. Second, the index parameter estimates associated with the nonstationary variables enjoy super-rate of convergence, as shown in all the cases, in comparison with the standard $\sqrt{n}$ rate of the estimators associated with the stationary variable. In particular, the estimators of $\gamma_1$, $\theta_{11}$ and $\theta_{12}$ are close to $n^2$-consistent, while those of $\gamma_2$, $\theta_{21}$ and $\theta_{22}$ are close to $n^3$-consistent. These results are consistent with our theoretical developments. Finally, the three estimates considered are quite robust in terms of involving the different error distributions. It is found that the Huber's estimator is the most efficient, except for the case with the Cauchy ($t(1)$) errors. The least absolute error estimate becomes the most efficient one when the errors are Cauchy, which is consistent with the theory.

Turning to Table \ref{rmse2} where the cointegrating function is integrable, there are a few new findings which are summarized as follows. First, the estimator of $\gamma_1$ is consistent but converges at a relatively slow rate. This is primarily due to the integrable nature of the cointegrating function. Second, the normalized estimators of $\theta_{11}$ and $\theta_{12}$ are converging at a relatively faster rate than that of $\gamma_1$ does. This finding corroborates the comment below Corollary 4.2 and that discovered in \citet{dgd2016}.

{\small
\begin{table}[tbp]
\caption{Mean Squared Error of M-estimator under Example 5.1 ($\times 10^2$). }
\label{rmse1}
\centering
\resizebox{0.9\textwidth}{!}{
\begin{tabular}{rcccccccccccccccc}
\toprule
&& \multicolumn{3}{c}{D1: Normal}&& \multicolumn{3}{c}{D2: Mixed normal}
&& \multicolumn{3}{c}{D3: t(2)}&& \multicolumn{3}{c}{D4: Cauchy}\\
\cline{3-5}\cline{7-9}\cline{11-13}\cline{15-17}
 &&L1&L2&L3&&L1&L2&L3&&L1&L2&L3&&L1&L2&L3\\
\midrule
$n=50$\\
$\gamma_{1}$&&10.33&12.08&13.23&&12.23&14.61&14.86&&19.07&18.45&22.59&&38.99&32.25&53.94\\
$\theta_{11}$&&1.340&1.303&1.110&&1.478&1.420&1.275&&2.549&1.876&1.769&&4.422&2.893&3.042\\
$\theta_{12}$&&0.865&0.966&0.934&&0.959&1.043&1.037&&1.454&1.309&1.426&&2.395&1.911&2.400\\
$\gamma_{2}$&&4.209&5.194&5.990&&5.012&6.250&6.926&&8.615&9.06&11.38&&18.21&16.81&22.42\\
$\theta_{21}$&&0.152&0.188&0.189&&0.210&0.240&0.243&&0.327&0.323&0.385&&0.684&0.554&0.887\\
$\theta_{22}$&&0.139&0.175&0.178&&0.173&0.208&0.218&&0.279&0.278&0.344&&0.541&0.463&0.708\\
$\gamma_{3}$&&0.512&0.744&0.804&&0.608&0.826&0.925&&1.014&1.139&1.515&&1.821&1.787&2.944\\
$\theta_{31}$&&0.276&0.389&0.422&&0.317&0.438&0.476&&0.543&0.609&0.844&&0.965&0.947&1.580\\
$\theta_{32}$&&0.276&0.385&0.424&&0.318&0.442&0.479&&0.538&0.605&0.818&&0.951&0.920&1.585\\
$n=100$\\
$\gamma_{1}$&&2.370&3.204&3.264&&2.755&3.654&3.821&&4.561&4.813&5.796&&7.061&6.593&10.40\\
$\theta_{11}$&&0.271&0.330&0.336&&0.339&0.429&0.389&&0.564&0.533&0.611&&0.902&0.734&1.018\\
$\theta_{12}$&&0.226&0.283&0.295&&0.277&0.355&0.336&&0.398&0.411&0.481&&0.613&0.558&0.797\\
$\gamma_{2}$&&0.497&0.696&0.739&&0.577&0.776&0.853&&0.962&1.012&1.299&&1.590&1.513&2.518\\
$\theta_{21}$&&0.017&0.023&0.025&&0.020&0.027&0.030&&0.032&0.033&0.045&&0.051&0.049&0.092\\
$\theta_{22}$&&0.017&0.023&0.024&&0.020&0.026&0.029&&0.031&0.032&0.044&&0.049&0.048&0.083\\
$\gamma_{3}$&&0.216&0.308&0.338&&0.260&0.361&0.412&&0.408&0.445&0.618&&0.679&0.652&1.134\\
$\theta_{31}$&&0.118&0.169&0.185&&0.143&0.197&0.224&&0.218&0.240&0.331&&0.363&0.351&0.607\\
$\theta_{32}$&&0.118&0.169&0.185&&0.144&0.198&0.221&&0.216&0.240&0.329&&0.370&0.359&0.622\\
$n=200$\\
$\gamma_{1}$&&0.589&0.825&0.873&&0.648&0.856&0.956&&1.135&1.211&1.472&&1.712&1.524&2.679\\
$\theta_{11}$&&0.060&0.083&0.090&&0.068&0.092&0.099&&0.119&0.124&0.155&&0.193&0.158&0.285\\
$\theta_{12}$&&0.057&0.078&0.084&&0.065&0.087&0.093&&0.109&0.114&0.138&&0.158&0.135&0.239\\
$\gamma_{2}$&&0.062&0.089&0.095&&0.068&0.094&0.102&&0.111&0.119&0.159&&0.180&0.168&0.310\\
$\theta_{21}$&&0.002&0.003&0.003&&0.002&0.003&0.003&&0.004&0.004&0.005&&0.006&0.005&0.010\\
$\theta_{22}$&&0.002&0.003&0.003&&0.002&0.003&0.003&&0.004&0.004&0.005&&0.006&0.005&0.010\\
$\gamma_{3}$&&0.103&0.149&0.161&&0.118&0.165&0.184&&0.182&0.200&0.271&&0.286&0.274&0.485\\
$\theta_{31}$&&0.053&0.078&0.087&&0.063&0.088&0.094&&0.096&0.106&0.148&&0.150&0.137&0.260\\
$\theta_{32}$&&0.053&0.079&0.088&&0.063&0.088&0.095&&0.095&0.105&0.148&&0.150&0.137&0.259\\
\bottomrule
\end{tabular}
}
\end{table}

\begin{table}[tbp]
\caption{Mean Squared Error of M-estimator under Example 5.2  ($\times 10$). }
\label{rmse2}
\centering
\resizebox{0.9\textwidth}{!}{
\begin{tabular}{rcccccccccccccccc}
\toprule
&& \multicolumn{3}{c}{D1: Normal}&& \multicolumn{3}{c}{D2: Mixed normal}
&& \multicolumn{3}{c}{D3: t(2)}&& \multicolumn{3}{c}{D4: Cauchy}\\
\cline{3-5}\cline{7-9}\cline{11-13}\cline{15-17}
 &&L1&L2&L3&&L1&L2&L3&&L1&L2&L3&&L1&L2&L3\\
\midrule
$n=50$\\
$\gamma_{1}$&&0.658&0.928&0.670&&0.780&1.054&0.734&&1.150&1.270&0.967&&1.734&1.612&1.843\\
$\theta_{11}$&&0.123&0.136&0.121&&0.145&0.158&0.137&&0.213&0.185&0.184&&0.311&0.232&0.247\\
$\theta_{12}$&&0.124&0.135&0.124&&0.149&0.159&0.135&&0.222&0.189&0.186&&0.327&0.230&0.248\\
$\gamma_{2}$&&0.017&0.025&0.024&&0.019&0.026&0.026&&0.030&0.032&0.038&&0.050&0.044&0.073\\
$\theta_{21}$&&0.011&0.017&0.016&&0.014&0.018&0.018&&0.020&0.020&0.025&&0.032&0.028&0.047\\
$\theta_{22}$&&0.011&0.017&0.016&&0.014&0.018&0.018&&0.020&0.021&0.025&&0.032&0.028&0.047\\
$n=100$\\
$\gamma_{1}$&&0.406&0.582&0.470&&0.466&0.620&0.501&&0.677&0.717&0.698&&1.060&0.953&1.124\\
$\theta_{11}$&&0.048&0.060&0.056&&0.056&0.071&0.062&&0.084&0.082&0.098&&0.123&0.097&0.133\\
$\theta_{12}$&&0.047&0.060&0.057&&0.056&0.073&0.062&&0.085&0.082&0.101&&0.124&0.100&0.133\\
$\gamma_{2}$&&0.007&0.011&0.011&&0.009&0.012&0.011&&0.014&0.015&0.017&&0.021&0.020&0.029\\
$\theta_{21}$&&0.005&0.007&0.008&&0.006&0.008&0.009&&0.009&0.009&0.012&&0.013&0.012&0.021\\
$\theta_{22}$&&0.005&0.007&0.008&&0.006&0.008&0.008&&0.009&0.009&0.012&&0.014&0.012&0.021\\
$n=200$\\
$\gamma_{1}$&&0.237&0.364&0.297&&0.290&0.385&0.333&&0.455&0.482&0.495&&0.708&0.603&0.830\\
$\theta_{11}$&&0.018&0.026&0.026&&0.023&0.030&0.030&&0.032&0.034&0.039&&0.048&0.041&0.064\\
$\theta_{12}$&&0.018&0.026&0.025&&0.023&0.029&0.030&&0.032&0.034&0.040&&0.050&0.042&0.064\\
$\gamma_{2}$&&0.004&0.005&0.005&&0.004&0.005&0.005&&0.006&0.007&0.008&&0.009&0.008&0.013\\
$\theta_{21}$&&0.002&0.003&0.004&&0.003&0.004&0.004&&0.004&0.004&0.006&&0.006&0.005&0.010\\
$\theta_{22}$&&0.002&0.003&0.004&&0.003&0.004&0.004&&0.004&0.005&0.006&&0.006&0.005&0.010\\
\bottomrule
\end{tabular}}
\end{table}
}

\subsection{Return predictability}

To demonstrate the practical relevance of our proposed model over some natural competitors, we investigate its applicability in stock return prediction. It is common to use a linear predictive mean regression in the literature, which has led to considerable disagreements in the empirical findings as to whether stock returns are predictable or not (\citealp{CT2008}; \citealp{WG2008}). Recently, \citet{KAP2015a} use a nonparametric mean regression, \citet{Lee2016}, and \citet{FL2019} consider a quantile linear regression, and \citet{TLW2021} adopt a nonparametric quantile framework to re-investigate this important issue. We shall demonstrate the nonlinearity in return predictability through the use of the proposed multiple-index model with both stationary and nonstationary predictors.

The data sets to be used for return prediction are obtained from \citet{WG2008}. They have been extensively used in the predictive literature, including the recent balanced predictive mean regression model by Ren et al. (2019), the linear quantile predictive regression by \citet{Lee2016}, and \citet{FL2019}, the linear prediction with cointegrated variables by \citet{kasy2020}, the LASSO predictive regression of
\citet{LSG2021}, and the nonparametric quantile predictive regression of \citet{TLW2021}, among others. The monthly data spans from January 1927 to December 2005. The dependent variable, excess stock return, is defined as the difference between the S\&P 500 index return, including dividends and the one month Treasury bill rate. The nonstationary (persistent) predictors include dividend-price ($dp$), dividend-payout ratio ($de$), long term yield ($lty$), book to market ($bm$) ratios, T-bill rate ($tbl$), default yield spread ($dfy$), net equity expansion ($ntis$), earnings-price ($ep$), term spread ($tms$), while the stationary variables include default return spread ($dfr$), long term rate of return ($ltr$), stock variance ($svar$) and inflation ($infl$). The first-order serial correlations of these variables and their time series plot are available from, for example, \citet{LSG2021}. For detailed description on each series and the data construction, please refer to \citet{WG2008}.

To measure the performance of the predictive regression models under discussion, we use the pseudo out-of-sample $R^2$, defined as
$$PR^2=\frac{\sum_{t=T_1+1}^{T_2} \rho(\widehat e_t)}{\sum_{t=T_1+1}^{T_2} \rho(\overline{e}_t)},$$
where $\widehat{e}_t$ and $\overline{e}_t$ are the respective out-of-sample prediction errors at time $t$ for a given model and that for the base model with only a constant predictor, for the forecasting sample spanning from $T_1$ to $T_2$. Four loss functions are entertained for $\rho(\cdot)$: (SE) squared errors loss $\rho(e)=e^2$; (AE) absolute errors loss $\rho(e)=|e|$,  (HL) Huber's loss $\rho_c (e)=\frac{1}{2}e^2\cdot 1\{|e|\leq c\}+c\cdot (e-\frac{1}{2} c) \cdot 1\{|e|> c\}$ with $c=1.25$; and (QL) quantile loss $\rho_\tau (e)=e\cdot (\tau-I(e<0))$. Positive $PR^2$ indicates the better performance of the given model over the base model, and the larger the value is, the better the performance is.

For space limitation, we follow \citet{KAP2015a}, \citet{Lee2016}, and \citet{TLW2021} to consider two subsamples: (i) the period spanning from January 1927 to December 2005, where significant predictability has been discovered, and (ii) the tranquil period starting from January 1952 to December 2005, where mixed evidences are found for predictability. We consider a rolling-window scheme for the out-of-sample return prediction and let the rolling window in-sample size (RWS) be 120 (10 years), 240 (20 years), and 360 (30 years), respectively. For example, the forecast starts from January 1927 and ends at December 2005 for the first subsample when RWS equals 120.

In contrast to \citet{TLW2021}, our proposed model allows for the use of multivariate nonstationary predictors together with those stationary ones to capture nonlinearity in the return predictability. There are numerous choices of combinations of the nonstationary and stationary predictors. For demonstration purposes, we use the stationary predictor $z=\{inf, svar\}$, and choose the nonstationary predictors as either Case A: $x=\{bm,lty\}$ or Case B: $x=\{dp,tms\}$.  For the nonstationary index, we consider either a linear transformation $g_1(u)=\gamma u$ or an integrable transformation $g_1(u)=\ell(u)=\gamma_1 \cdot exp(-u^2)+\gamma_2 \cdot u \cdot exp(-u^2)$. Similarly, we use either $g_2(v)=\delta v$ or $g_2(v)=\ell(v)=\delta_1 \cdot exp(-v^2)+\delta_2 \cdot v \cdot exp(-v^2)$ for the stationary index. The choice for the nonlinear transformation is made based on the (normalized) Hermite polynomials in the sieve literature where Hermite polynomials form a basis in the related function space.
This amounts to four possible combinations. 
The extensive analysis with a comprehensive set of predictors noted above and all possible specifications of the nonlinear link functions remains an interesting but a challenging work, which is left as future study.

{\small
\begin{table}[h!]
\centering
\caption{Pseudo Out-of-sample $R^2$ for excess return prediction with Case A: $x=\{bm,lty\}$ and $z=\{inf, svar\}$, and Case B: $x=\{dp,tms\}$ and $z=\{inf, svar\}$.}\label{PR2_1}
\resizebox{0.9\textwidth}{!}{
\begin{tabular}[t]{lc l rrr l rrr l rrr l rrr}
\toprule
\multicolumn{2}{l}{Case A}&&\multicolumn{7}{c}{Panel A: January 1927-December 2005}&&\multicolumn{7}{c}{Panel B: January 1952-December 2005}\\
\cline{1-2}\cline{4-10}\cline{12-18}
        &&&\multicolumn{3}{c}{$g_2(v)=v$}       &&  \multicolumn{3}{c}{$\ell(v)$}   &&\multicolumn{3}{c}{$g_2(v)=v$}       &&  \multicolumn{3}{c}{$\ell(v)$}   \\
                       \cline{4-6}\cline{8-10}\cline{12-14}\cline{16-18}
$g_1(u)$       &RWS&& SE         &  AE        &  HL        && SE         &  AE        &  HL      && SE         &  AE        &  HL&& SE         &  AE        &  HL\\
                      \hline
$u$                & 120 && \bf{3.06}  &  \bf{3.19}  &  \bf{2.58} &&  0.80  &  0.36 &  -3.62&&-2.74  &  0.14  & -2.13 && -2.48  &  0.33  & -9.12\\
                      & 240 && \bf{6.58}  &  \bf{3.26}  &  \bf{7.63} &&  3.48 &   1.02  &  2.35&& 1.62 &   \bf{0.63} &   \bf{3.94} &&  \bf{3.07} &  -0.72  &  1.83\\
                      & 360 && \bf{6.57}  &  \bf{0.76}  &  \bf{7.84} &&  3.21 &   0.48  &  5.08&& 1.77 &  -3.31 &   3.43 &&  3.30  &  1.18  &  \bf{5.58}\\

$\ell(u)$&
                       120 &&0.69 &   0.14  &  1.09   &&  0.53 &   0.35 &   1.56 && -1.67 &   0.08 &  -1.60 && \bf{-0.46} &  \bf{0.37} &  \bf{-0.31}\\
                    &240 &&2.34 &   1.13 &   1.91    &&  0.47  &  0.28  &  0.44&& 1.99 &  -0.24 &   1.54  &&0.07 &   0.12  &  0.01\\
                    &360 &&5.52  &  0.64  &  3.43    &&  0.74  &  0.36  &  0.59 &&\bf{4.75} &   \bf{1.35}  &  4.04  &&0.51 &   0.23  &  0.73\\

\multicolumn{2}{l}{Case B}\\
\cline{1-2}\cline{4-10}\cline{12-18}
$u$    & 120 && -4.73&   -3.60&   -5.35 &&  4.53  &  1.05  &  5.23  &&  -3.06 &  -2.63 &  -3.35 && 2.28  &  1.54  & \bf{ 4.39}\\
                      & 240 && -7.58&   -4.51&   -7.74 &&  9.67   & 2.26  &  {\bf 11.2}&&-5.80 &  -3.51 &  -5.79 &&  6.70  &  1.68  &  \bf{8.82}\\
                      & 360 && -6.65&   -3.78&   -7.16 &&  10.7  &  {\bf 4.35}  &  11.1&& -7.48  & -3.79 &  -7.42 &&  8.91  &  \bf{2.99}  &  8.95\\

$\ell(u)$&
                       120 &&2.15  &   {\bf 2.39}  &   4.30   &&  {\bf 5.53}  &  1.35  &  {\bf 5.37} &&   -0.41  &  \bf{2.58}  &  3.04 && \bf{2.63} &   0.94  &  3.85\\
                    &240 &&8.75  &  {\bf 3.40}  &  10.3    &&   {\bf 10.0} &   2.56  &  10.9&&\bf{7.56}  &  \bf{1.99}  &  8.59  &&6.12  &  0.69  &  6.97\\
                    &360 &&8.48 &   3.53  &  10.5    &&   {\bf 10.8} &   3.91  &  {\bf 11.2} &&6.18  &  1.58   & 8.79  && \bf{10.2}  &  1.97  &  \bf{9.10}\\

\bottomrule
\end{tabular}}
\end{table}
}

{\small

\begin{table}[h!]
\centering
\caption{Pseudo Out-of-sample $R^2$ for quantile excess return prediction with $x=\{dp,tms\}$ and $z=\{inf, svar\}$.}\label{PR2_3}
\resizebox{0.9\textwidth}{!}{
\begin{tabular}[t]{llc l rrrrr r rrrrr}
\toprule
&&&\multicolumn{12}{c}{Panel A: January 1927-December 2005}\\
\cline{5-15}

$g_2(v)$&$g_1(u)$       &RWS&&$\tau=0.05$ &0.1&0.2&0.3&0.4&0.5&0.6&0.7&0.8&0.9&0.95\\
                      \hline
$v$ &$u$    & 120 &&-1.87 &   0.32 &  -1.09 &  -0.99 &  -2.41 &  -3.60 &  -4.08 & -4.51 &  -6.05 &  8.86 &  -7.96\\
        &  & 240 &&\bf{3.89}  &  1.33  &  0.11  & -1.08  & -2.66  & -4.51 &  -4.69 &  -6.77 &  -9.21 &  -11.2  & -15.8\\
        &  & 360 &&0.30  &  3.05  &  0.93 &  -1.00 &  -1.87 &  -3.82 &  -4.55  & -7.14  & -8.95 &  -13.7 &  -18.3\\

&$\ell(u)$&
                       120 &&\bf{2.83} &  -2.64  &  1.31 &   1.97  &  0.89  &  \bf{2.39}  &  0.96 & 0.61 &   4.74 &   0.87  &  6.05\\
             &     & 240 && 1.95 &   2.93  &  4.06  &  4.39  &  3.69  &  \bf{3.40}  &  1.67 &  3.65  &  5.51 &   \bf{6.71}  & -1.14\\
             &     & 360 && 4.40 &   4.40  &  \bf{5.08} &   6.10  &  3.15  &  3.53  &  3.85  & 2.71 &   \bf{7.89}  &  \bf{6.50}   & \bf{5.82}  \\
\hline
$\ell(v)$&$u$    & 120 &&1.20  &  2.17  &  1.57  &  1.64  &  \bf{1.45}  &  1.05 &   \bf{2.28}  &  \bf{2.61}  &  4.08  &  \bf{2.39}  &  4.33\\
     &     & 240 &&3.19 &   4.02  &  4.06  &  5.08 &   \bf{4.14}  &  2.26 &   \bf{2.01}  &  2.13  &  6.36  &  7.16  &  \bf{2.99}\\
      &    & 360 &&\bf{4.74}  &  4.94 &   4.75  &  \bf{6.40} &   \bf{4.67}  &  \bf{4.35}  &  \bf{4.81} &   4.33  &  7.36 &   5.64 &   4.49\\

&$\ell(u)$&
                       120 &&1.65 &   \bf{2.22} &   \bf{2.06}  &  \bf{2.52} &   1.42  &  1.35  &  1.46 &   2.16 &   \bf{5.45} &  2.16 &   \bf{5.17}\\
              &    & 240 &&3.40 &   \bf{4.20}  &  \bf{4.28} &   \bf{5.51}  &  3.06  &  2.56  &  1.68  &  \bf{4.80} &   \bf{7.29}  &  6.33 &   0.45\\
              &    & 360 &&4.48 &   \bf{5.14}  &  4.91 &   5.58  &  3.96  &  3.73  &  3.48 &   \bf{4.98} &   7.29 &   5.50 &   3.61\\

&&&\multicolumn{12}{c}{Panel B: January 1952-December 2005}\\
\midrule
$v$&$u$    & 120 &&-1.69 &  -0.11 &  -0.80 &  -0.91 &  -1.05 &  -2.62 &  -2.51& -3.16 &  -4.26 &  -5.52 &  -6.29\\
         & & 240 &&\bf{4.59} &   1.45  &  0.64  & -0.93  & -1.49  & -3.50 &  -3.46 & -5.18 &  -7.27 &  -10.7 &  -12.5\\
         & & 360 &&4.06  &  3.20 &   1.25  & -1.47  & -2.97 &  -3.80 &  -4.54 & -4.92 &  -7.09 &  -13.5 &  -18.1\\

&$\ell(u)$&
                       120 &&\bf{2.01} &  -3.89 &   2.09 &   2.51  &  0.72 &   \bf{2.58} &   0.33 &  -1.51 &   2.99  &  3.88  &  \bf{7.13}\\
              &     & 240 &&2.91 &   2.91 &   3.47 &   4.23 &   2.52 &   \bf{1.99} &  -1.00 &  0.36  &  0.68  &  \bf{4.00}  &  3.00 \\
              &     & 360 &&7.65 &   7.37 &   \bf{7.18} &   7.12 &   3.31 &   1.58  &  1.71 &  -1.17  & \bf{-0.70}  & \bf{-4.28} &  \bf{-4.34}  \\

\hline
$\ell(v)$&$u$    & 120 &&    0.69 &   1.80 &   1.58 &   2.56  &  \bf{1.70} &   1.54 &   \bf{2.13} &   \bf{0.46} &   2.10  &  4.88  &  4.19\\
     &     & 240 &&    3.43 &   4.59 &   3.69 &   5.31 &   \bf{3.11}&    1.68 &   \bf{0.83}&   -1.94 &   1.97  &  3.25 &   \bf{4.90}\\
      &    & 360 &&    9.30  &  7.43 &   6.69 &   \bf{8.13} &   \bf{4.48} &   \bf{2.99} &   \bf{2.29}  &  \bf{0.46} &  -1.14  & -4.84  & -4.73\\

&$\ell(u)$&
                       120 &&    1.61 &   \bf{2.01} &   \bf{2.43}  &  \bf{2.90} &   1.00   & 0.94 &   1.90 &  -0.66  &  \bf{3.69}  &  \bf{5.02}   & 6.10\\
              &     & 240 &&    4.28 &   \bf{5.35} &   \bf{4.14} &   \bf{5.82} &   1.87 &   0.69 &  -0.21 &   \bf{0.93}  &  \bf{2.92} &   3.07 &   2.60\\
              &     & 360 &&    \bf{9.64} &   \bf{8.53} &   7.15 &   7.18  &  3.36  &  1.87 &   0.79 &  -0.33  & -2.26  & -5.65&   -8.88\\

\bottomrule
\end{tabular}}
\end{table}
}

Table \ref{PR2_1} collects the pseudo out-of-sample $R^2$ for excess return prediction with the above two sets of selected predictors and specifications of the link functions, respectively, and for the loss functions SE, AE and HL. For each loss function and each RWS, the combination of the link functions that produces the largest $PR^2$ is bolded. There are several interesting findings.
First, the proposed models demonstrate significant return predictability compared to the baseline model with only an intercept. For Case A with $x=\{bm,lty\}$, the linear predictive regression outperform other specifications with nonlinearities for all three loss functions for the first sample period January 1927 to December 2005 (Panel A). When it comes to the second sample period in Panel B, nonlinear predictability is discovered, even though the predictability is seen to decline, as that discovered by \citet{CY2006}. For Case B with $x=\{dp,tms\}$, we find no predictability from the linear predictive model for both sample periods. The pseudo out-of-sample $R^2$ can reach 10.8\%, 4.35\%, 11.2\% for the first sample period under the squared error loss, the least absolute error loss, and the Huber loss, respectively. The nonlinear predictability is also seen to decline in the ``tranquil'' period, but remains significant with the pseudo out-of-sample $R^2$ being as large as 10.2\%, when both indices enter the model in the nonlinear manner. This finding is in contrast to the weak nonlinear predictability discovered by \citet{KAP2015a} with only a single nonstationary predictor. To summarize, significant return predictability has been revealed in the nonlinear predictive regression with both stationary and nonstationary indices we introduced.


As the recent literature has witnessed a growing interest in investigating the return prediction in quantiles (Lee, 2016; Fan and Lee, 2019; Tu, et al., 2021), we next study the performance of our proposed model in predicting the return quantiles. We consider the quantile level $\tau=0.05,0.1,0.2,\ldots,0.9,0.95$. The analysis will only focus on the nonlinear effect of the nonstationary index as discovered from Tables \ref{PR2_1}. Therefore, we only present the results for $x=\{dp,tms\}$ and $z=\{inf, svar\}$ in Table \ref{PR2_3} at all the quantile levels specified above, for the two subsamples mentioned earlier. The specifications for transformation on the stationary and nonstationary components are the same as above. From Table \ref{PR2_3}, it is observed that linear quantile predictability seems to only exist for the lower quantile levels, such as those at $\tau=0.05,0.1$ and $0.2$. However, nonlinear quantile predictability prevails at all quantile levels. Even at the lower quantiles, the proposed nonlinear index model outperforms the linear model except for a very few cases. 
In addition, it seems that the nonlinear predictability becomes stronger as the quantile level moves to the two ends, especially for the first subsample.

\section{Conclusion}

The paper has studied a class of additive single index models with diverse regressors including I(1) process, time trend and $\alpha$-mixing stationary process. To deliver the asymptotic theory of robust estimation, a generalized function approach has been proposed that has independent interest, as shown in Section two with a simple linear model. This approach associated with Theorem 2.1 has answered a call from the literature to establish the convergence rate of regular sequences to nonsmooth loss functions. Thus, we therefore believe that Theorem 2.1 fills a gap in the literature.

The generalized function approach is then applied in general robust estimation for a class of additive single--index cointegrating time series models where the objective function is constructed with possible nonsmooth loss, such as LAD, quantile loss and Huber's loss; the corresponding asymptotic theory has been established, respectively, according to H-regular and I-regular classes of the regression functional forms involving I(1) processes. Monte Carlo simulations are conducted to verify the performance of estimators proposed in finite sample situations; and an empirical study on stock returns has also been implemented to exhibit both the relevance and applicability of the proposed model and estimation method developed in the paper.

\section*{Acknowledgements}

Dong would like to thank the financial support from National Natural Science Foundation of China (Grant 72073143) and the Fundamental Research Funds
for the Central Universities, Zhongnan University of Economics and Law (2722022EG001); Gao acknowledges financial support from the Australian Research Council Discovery Grants Program under Grant Number: DP200102769; Peng acknowledges the Australian Research Council Discovery Grants Program for its financial support under Grant Number DP210100476; and Tu (Corresponding Author) would like to thank support from National Natural Science Foundation of China (Grant 72073002, 12026607, 92046021), the Center for Statistical Science at Peking University, and Key Laboratory of Mathematical Economics and Quantitative Finance (Peking University), Ministry of Education.

{\footnotesize

\bibliography{123}

\begin{thebibliography}{}

\bibitem[Badu, 1989]{badu1989}
Badu, G.~J. (1989).
\newblock {Strong representations for LAD estimators in linear models}.
\newblock {\em Probability theory and related fields}, 83:547--558.

\bibitem[Bahadur, 1966]{bahadur1966}
Bahadur, R.~R. (1966).
\newblock {A note on quantiles in large samples}.
\newblock {\em Annals of Mathematical Statistics}, 37:577--581.

\bibitem[Bai et~al., 1992]{bai1992}
Bai, Z., Rao, C.~R., and Wu, Y. (1992).
\newblock {M estimation of multivariate linear regression parameters under a
  convex discrepancy function}.
\newblock {\em Statistica Sinica}, 2:237--254.

\bibitem[Bauer and Kohler, 2019]{bk2019}
Bauer, B. and Kohler, M. (2019).
\newblock On deep learning as a remedy for the course of dimensionality in
  nonparametric regression.
\newblock {\em Annals of Statistics}, 47(4):2261--2285.

\bibitem[Bickel, 1974]{bickel1974}
Bickel, P.~J. (1974).
\newblock {Edgeworth expansions in nonparametric statistics}.
\newblock {\em Annals of Statistics}, 2:1--20.

\bibitem[Bosq, 1996]{bosq1996}
Bosq, D. (1996).
\newblock {\em {Nonparametric Statistics for Stochastic Processes: Estimation
  and Prediction}}.
\newblock Springer, New York.

\bibitem[Bravo et~al., 2021]{degui2021}
Bravo, F., Li, D., and Tj{\o}stheim, D. (2021).
\newblock {Robust nonlinear regression estimation in null recurrent time
  series}.
\newblock {\em Journal of Econometrics}, 224:416--438.

\bibitem[Campbell and Thompson, 2008]{CT2008}
Campbell, J.~Y. and Thompson, S.~B. (2008).
\newblock Predicting excess stock returns out of sample: Can anything beat the
  historical average?
\newblock {\em Review of Financial Studies}, 21(4):1509--1531.

\bibitem[Campbell and Yogo, 2006]{CY2006}
Campbell, J.~Y. and Yogo, M. (2006).
\newblock Efficient tests of stock return predictability.
\newblock {\em Journal of Financial Economics}, 81(1):27--60.

\bibitem[Chen, 2007]{chen2007}
Chen, X. (2007).
\newblock {Large Sample Sieve Estimation of Semi--Nonparametric Models, Chapter
  76 edited by James J. Heckman and Edward E. Leamer}.
\newblock {\em Handbook of Econometrics}, 6B:5549--5632.

\bibitem[Chen et~al., 2019]{degui2019}
Chen, X., Li, D., Li, Q., and Li, Z. (2019).
\newblock {Nonparametric estimation of conditional quantile functions in the
  presence of irrelevant covariates}.
\newblock {\em Journal of Econometrics}, 212:433--450.

\bibitem[Chen et~al., 2001]{crs2001}
Chen, X., Racine, J., and Swanson, N. (2001).
\newblock Semiparametric arx neural network models with an application to
  forecasting inflation.
\newblock {\em IEEE Transactions on Neural Networks}, 12(6):674--683.

\bibitem[Chen and Shen, 1998]{cs1998}
Chen, X. and Shen, X. (1998).
\newblock Sieve extremum estimates for weakly dependent data.
\newblock {\em Econometrica}, 66(2):298--314.

\bibitem[Chen and White, 1999]{cw1999}
Chen, X. and White, H. (1999).
\newblock Improved rates and asymptotic normality for nonparametric neural
  network estimators.
\newblock {\em IEEE Transactions on Information Theory}, 45(6):682--691.

\bibitem[Cybenko, 1989]{cg1989}
Cybenko, G. (1989).
\newblock Approximation by superpositions of a sigmoidal function.
\newblock {\em Mathematics of Control, Signals and Systems}, 2(2):303--314.

\bibitem[Davies et~al., 1992]{davis1992}
Davies, R.~A., Knight, K., and Liu, J. (1992).
\newblock {M-estimation for autoregressions with infinite variance}.
\newblock {\em Stochastic Processes and Their Application}, 40:145--180.

\bibitem[Dong and Gao, 2018]{donggao2018}
Dong, C. and Gao, J. (2018).
\newblock Specification testing driven by orthogonal series for nonlinear
  cointegration with endogeneity.
\newblock {\em Econometric Theory}, 34:754--789.

\bibitem[Dong and Gao, 2019]{dg2019}
Dong, C. and Gao, J. (2019).
\newblock Expansion and estimation of levy process functionals in nonlinear and
  nonstationary time series regression.
\newblock {\em Econometric Reviews}, 38:125--150.

\bibitem[Dong et~al., 2015]{dgp2015}
Dong, C., Gao, J., and Peng, B. (2015).
\newblock {Semiparametric single-index panel data models with cross-sectional
  dependence}.
\newblock {\em Journal of Econometrics}, 188:301--312.

\bibitem[Dong et~al., 2016]{dgd2016}
Dong, C., Gao, J., and Tj{\o}stheim, D. (2016).
\newblock Estimation for single-index and partially linear single-index
  integrated models.
\newblock {\em Annals of Statistics}, 44:425--453.

\bibitem[Dong et~al., 2017]{dgdy2017}
Dong, C., Gao, J., Tj{\o}stheim, D., and Yin, J. (2017).
\newblock Specification testing for nonlinear multivariate cointegrating
  regressions.
\newblock {\em Journal of Econometrics}, 200:104--117.

\bibitem[Dong and Linton, 2018]{DL2018}
Dong, C. and Linton, O. (2018).
\newblock Additive nonparametric models with time variable and both stationary
  and nonstationary regressors.
\newblock {\em Journal of Econometrics}, 207:212--236.

\bibitem[Fan et~al., 1994]{fanjq1994}
Fan, J., Hu, T.-C., and Truong, Y.~K. (1994).
\newblock {Robust nonparametric function estimation}.
\newblock {\em Scandinavian Journal of Statistics}, 21:433--446.

\bibitem[Fan et~al., 2003]{fan2003}
Fan, J., Jiang, J., Zhang, C., and Zhou, Z. (2003).
\newblock {Time-dependent diffusion models for term structure dynamics}.
\newblock {\em Statistica Sinica}, 13:965--992.

\bibitem[Fan and Lee, 2019]{FL2019}
Fan, R. and Lee, J.~H. (2019).
\newblock Predictive quantile regressions under persistence and conditional
  heteroskedasticity.
\newblock {\em Journal of Econometrics}, 213(1):261--280.

\bibitem[Gao et~al., 2009]{gao2009}
Gao, J., Li, D., and Lin, Z. (2009).
\newblock Robust estimation in parametric time series models under long-- and
  short--range dependent structures.
\newblock {\em Australian and New Zealand Journal of Statistics},
  51(2):161--181.

\bibitem[Gel'fand and Shilov, 1964]{gelfand1964}
Gel'fand, I.~M. and Shilov, G.~E. (1964).
\newblock {\em {Generalized Functions}}.
\newblock Academic Press, New York.

\bibitem[Geyer, 1994]{charles1994}
Geyer, C.~J. (1994).
\newblock {On the asymptotics of constrained $M$-estimation}.
\newblock {\em The Annals of Statistics}, 22:1993--2010.

\bibitem[He and Shao, 2000]{hexuming2000}
He, X. and Shao, Q. (2000).
\newblock {On parameters of increasing dimensions}.
\newblock {\em Journal of Multivariate Analysis}, 73:120--135.

\bibitem[Huber, 1964]{huber1964}
Huber, P.~J. (1964).
\newblock {Robust estimation of a location parameter}.
\newblock {\em Annals of Mathematical Statistics}, 35:73--101.

\bibitem[Kanwal, 1983]{kanwal1983}
Kanwal, R.~P. (1983).
\newblock {\em {Generalized Fuctions: Theory and Technique}}.
\newblock Academic Press, New York.

\bibitem[Kasparis et~al., 2015]{KAP2015a}
Kasparis, I., Andreou, E., and Phillips, P. (2015).
\newblock Nonparametric predictive regression.
\newblock {\em Journal of Econometrics}, 185:468--494.

\bibitem[Knight, 1989]{knight1989}
Knight, K. (1989).
\newblock {Limit theory for autoregressive-parameter estimates in an
  infinite-variance random walk}.
\newblock {\em Canadian Journal of Statistics}, 17:261--278.

\bibitem[Koenker and Bassett, 1978]{koenker1978}
Koenker, R. and Bassett, G. (1978).
\newblock {Regression quantiles}.
\newblock {\em Econometrica}, 46:33--50.

\bibitem[Kohler and Krzyz\'{a}k, 2017]{kk2017}
Kohler, M. and Krzyz\'{a}k, A. (2017).
\newblock Nonparametric regression based on hierarchical interaction models.
\newblock {\em IEEE Transactions on Information Theory}, 63(3):1620--1630.

\bibitem[Koo et~al., 2020]{kasy2020}
Koo, B., Anderson, H.~M., Seo, M.~H., and Yao, W. (2020).
\newblock High dimensional predictive regression in the presence of
  cointegration.
\newblock {\em Journal of Econometrics}, 219:456--477.

\bibitem[Lee et~al., 2022]{LSG2021}
Lee, J., Shi, Z., and Gao, Z. (2022).
\newblock On lasso for predictive regression.
\newblock {\em Journal of Econometrics}, 229:322--349.

\bibitem[Lee, 2016]{Lee2016}
Lee, J.~H. (2016).
\newblock Predictive quantile regression with persistent covariates: Ivx-qr
  approach.
\newblock {\em Journal of Econometrics}, 192:105--118.

\bibitem[Li et~al., 2016]{degui2016}
Li, D., Tj{\o}stheim, D., and Gao, J. (2016).
\newblock {Estimation in nonlinear regression with Harris recurrent {M}arkov
  chains}.
\newblock {\em Annals of Statistics}, 44:1957--1987.

\bibitem[Milne, 1980]{milne1980}
Milne, R.~D. (1980).
\newblock {\em {Applied Functional Analysis: An Introductory Treatment}}.
\newblock Pitman Publishing Limited, London.

\bibitem[Park and Phillips, 1999]{phillips1999}
Park, J.~Y. and Phillips, P. C.~B. (1999).
\newblock {Asymptotics for nonlinear transformations of integrated time
  series}.
\newblock {\em Econometric Theory}, 15:269--298.

\bibitem[Park and Phillips, 2000]{phillips2000}
Park, J.~Y. and Phillips, P. C.~B. (2000).
\newblock {Nonstationary binary choice}.
\newblock {\em Econometrica}, 68:1249--1280.

\bibitem[Park and Phillips, 2001]{phillips2001}
Park, J.~Y. and Phillips, P. C.~B. (2001).
\newblock {Nonlinear regression with integreted time series}.
\newblock {\em Econometrica}, 69(1):117--161.

\bibitem[Phillips, 1991]{phillips1991}
Phillips, P. C.~B. (1991).
\newblock {A shortcut to LAD estimator asymptotics}.
\newblock {\em Econometric Theory}, 7:450--463.

\bibitem[Phillips, 1995]{phillips1995}
Phillips, P. C.~B. (1995).
\newblock {Robust nonstationary regression}.
\newblock {\em Econometric Theory}, 11:912--951.

\bibitem[Phillips and Solo, 1992]{phillips1992}
Phillips, P. C.~B. and Solo, V. (1992).
\newblock {Asymptotics for linear processes}.
\newblock {\em Annals of Statistics}, 20(2):971--1001.

\bibitem[Pollard, 1991]{pollard1991}
Pollard, D. (1991).
\newblock {Asymptotics for the least absolute deviation regression estimators}.
\newblock {\em Econometric Theory}, 7:186--199.

\bibitem[Rockafellar, 1970]{rockafellar1970}
Rockafellar, R.~T. (1970).
\newblock {\em {Convex Analysis}}.
\newblock Princeton University Press, Princeton.

\bibitem[Rudin, 2004]{rudin2004}
Rudin, W. (2004).
\newblock {\em {Principles of Mathematical Analysis}}.
\newblock McGraw-Hill Companies, Inc., New York.

\bibitem[Schmidt-Hieber, 2020]{sh2020}
Schmidt-Hieber, J. (2020).
\newblock Nonparametric regression using deep neural networks with {ReLu}
  activation function.
\newblock {\em Annals of Statistics}, 48(4):1875--1897.

\bibitem[Stein and Shakarchi, 2003]{stein2003}
Stein, E.~M. and Shakarchi, R. (2003).
\newblock {\em {Fourier Analysis An Introduction}}.
\newblock Princeton University Press, Princeton.

\bibitem[Susarla and Walter, 1981]{walter1981}
Susarla, V. and Walter, G. (1981).
\newblock {Estimation of a multivariate density function using delta
  sequences}.
\newblock {\em Annals of Statistics}, 9:347--355.

\bibitem[Tu et~al., 2022]{TLW2021}
Tu, Y., Liang, H.~Y., and Wang, Q. (2022).
\newblock Nonparametric inference for quantile cointegrations with stationary
  covariates.
\newblock {\em Journal of Econometrics}, 230:453--482.

\bibitem[van~der Vaart, 1996]{vaart1996}
van~der Vaart, A.~W. (1996).
\newblock {\em {Weak Convergence and Empirical Processes with Applications to
  Statistics}}.
\newblock Springer Series in Statistics. Springer, New York.

\bibitem[Walter and Blum, 1979]{walter1979}
Walter, G. and Blum, J. (1979).
\newblock {Probability density estimation using delta sequences}.
\newblock {\em Annals of Statistics}, 7:328--340.

\bibitem[Wang et~al., 2018]{qiying2018}
Wang, Q., Wu, D., and Zhu, K. (2018).
\newblock {Model checks for nonlinear cointegrating regression}.
\newblock {\em Journal of Econometrics}, 207:261--284.

\bibitem[Welch and Goyal, 2008]{WG2008}
Welch, I. and Goyal, A. (2008).
\newblock A comprehensive look at the empirical performance of equity premium
  prediction.
\newblock {\em Review of Financial Studies}, 21:455--508.

\bibitem[Xiao, 2009]{xiao2009a}
Xiao, Z. (2009).
\newblock Quantile cointegrating regression.
\newblock {\em Journal of Econometrics}, 150:248--260.

\bibitem[Zhou et~al., 2018]{fan2018}
Zhou, W., Bose, K., Fan, J., and Liu, H. (2018).
\newblock {A new perspective on robust M-estimation: finite sample theory and
  applications to dependence-adjusted multiple testing}.
\newblock {\em Annals of Statistics}, 46:1904--1931.

\end{thebibliography}

}

\appendix

{\small

\section{Lemmas}

Under Assumption 3.1 and w.l.o.g. letting $x_0=0$ a.s., similar to (A.5) and (A.6) of \citet{dgdy2017}, we have
\begin{align}\label{xt}
x_t=\sum_{i=1}^tw_i=\sum_{j=-\infty}^tB_{t,j}\eta_{j}, \ \ \text{where}\ B_{t,j}=\sum_{i=\max(1,j)}^tA_{i-j},
\end{align}
and for $t>s>0$,
\begin{align}\label{xts}
x_t=&\sum_{i=1}^tw_i=\sum_{i=s+1}^tw_i+x_s=x_{ts}+x_{ts}^*,
\end{align}
where $x_{ts}=\sum_{j=s+1}^tB_{t,j}\eta_{j}$, $x_{ts}^*=x_s+ \bar{x}_{ts}$,  and $\bar{x}_{ts}=\sum_{j=-\infty}^s \left(\sum_{i=s+1}^tA_{i-j}\right)\eta_{j}$. As a result, $x_{ts}$ is independent of $x_{ts}^*$. Denote $d_{ts}^2:= \mathbb{E}(x_{ts} x_{ts}^\top)$ and from the B-N decomposition (\citealp{phillips1992}) we have $d_{ts}^2\sim t-s$ when $t-s$ is large. The representations of \eqref{xt} and \eqref{xts}, along with the following lemma, will be used for asymptotic analysis.

\begin{lemma}\label{lemma1}
Let Assumption 3.1 hold. Let $x_{j,t}=\theta_{1j}^{0\top} x_t$ be unit root process and define $x_{j,ts}=\theta_{1j}^{0\top} x_{ts}$ for $t>s$ where $x_{ts}$ is given by \eqref{xts}, $j=1, \cdots, p_1$; and define $d_{j,t}^2=\mathbb{E}(x_{j,t}^2)$ and $d_{j,ts}^2=\mathbb{E}(x_{j,ts}^2)$.
\begin{enumerate}[(1)]
\item For large $t$, $d_{j,t}^{-1}x_{j,t}$ have densities $f_{jt}(u)$ which are uniformly bounded over $u\in \mathbb{R}$ and $t$. Meanwhile, the partial derivatives of $f_{jt}(u)$ are uniformly bounded as well. Consequently, $f_{jt}(u)$ satisfy a uniform Lipschitz condition
\begin{equation}\label{lemma01}
\sup_{u\in \mathbb{R}}|f_{jt}(u+\triangle u)-f_{jt}(u)|\le C \,|\triangle u|
\end{equation}
for some $C>0$ and any $\triangle u\in \mathbb{R}$.

\item For large $t-s$, $d_{j,ts}^{-1}x_{j,ts}$ have uniformly bounded densities $f_{j,ts}(u)$ over all $u\in \mathbb{R}$ and $(t,s)$. Additionally, $f_{j, ts}(u)$ have bounded partial derivatives and satisfy Lipshitz condition similar to \eqref{lemma01} as well.
\end{enumerate}
\end{lemma}

\medskip

\begin{lemma}\label{lemma2}
Consider H-regular regression functions. Recall $x_{nt}=n^{-1/2}x_t$, and $Z_{1}(x_{nt})$ and $Z_2(z_t)$ are defined in \eqref{41a}.
Let $\psi(u)$ be the subgradient of $\rho(u)$. Under Assumptions 2.1, 3.1-3.3 and 3.4(i), as $n\to\infty$,
\begin{align}\label{lemma2a}
&\frac{1}{\sqrt{n}}\sum_{t=1}^n\psi(e_t)\begin{pmatrix}Z_{1}(x_{nt})\\
Z_{2}(z_t)
\end{pmatrix}
\to_D\begin{pmatrix}\int_0^1Z_{1}(B(r))dU(r)\\
N(0, a_1\Sigma)
\end{pmatrix},
\end{align}
where $\Sigma=\int_0^1[\mathbb{E}Z_2 (h(r,v_1) ) \, Z_2(h(r,v_1))^\top ]dr$ and $a_1>0$ given in Assumption 3.3.

Moreover, as $n\to\infty$,
\begin{align}\label{lemma2b}
&\frac{1}{n}\sum_{t=1}^nZ_2(z_t)Z_2(z_t)^\top\to_P\Sigma,
\end{align}
and
\begin{align}\label{lemma2c}
&\frac{1}{n}\sum_{t=1}^n\begin{pmatrix}
Z_1(x_{nt})Z_1(x_{nt})^\top\\
Z_2(z_t)Z_1(x_{nt})^\top
\end{pmatrix}
\to_D \begin{pmatrix}
\int_0^1 Z_1(B(r))Z_1(B(r))^\top dr\\
\int_0^1[\mathbb{E}Z_2(h(r,v_1))]Z_1(B(r))^\top dr
\end{pmatrix}.
\end{align}
\end{lemma}

\begin{lemma}\label{lemma3}
Let $Z_t=(Z_1(x_{nt})^\top, Z_2(z_t)^\top)^\top$ where $Z_1$ and $Z_2$ are given in Lemma \ref{lemma2}. Under Assumptions 2.1, 3.1-3.3 and 3.4(i), as $n\to\infty$,
\begin{align}
 &\frac{1}{\sqrt{n}}\sum_{t=1}^n[\rho'_m(e_t)-\psi(e_t)]
Z_{t}=O_P(m^{-1/2}n^{1/2}), \label{lemma3a}\\
&\frac{1}{n}\sum_{t=1}^n[\rho''_m(e_t)-\e\rho''_m(e_t)]Z_tZ_t'=O_P(n^{-1/2}). \label{lemma3b}
\end{align}
\end{lemma}

\begin{lemma}\label{lemma4}
Consider I-regular regression functions. Recall vector $Z_t=(Z_1(x_{nt})^\top, Z_2(z_t)^\top)^\top$ defined in \eqref{42c1}.
Under Assumptions 2.1, 3.1-3.3 and 3.4(ii), we have
\begin{align}
 &\sum_{t=1}^n[\rho_m'(e_t)-\psi(e_t)]Z_1(x_{nt})=O_P(n^{1/4}m^{-1/2}), \label{lemma4a1} \\
 & \sum_{t=1}^n[\rho_m'(e_t)-\psi(e_t)]Z_2(z_{t})=O_P(n^{1/2}m^{-1/2}), \label{lemma4a2}
\end{align}
and
\begin{align}
 &\sum_{t=1}^n[\rho_m''(e_t)-\e\rho''_m(e_t)]Z_1(x_{nt})Z_1(x_{nt})^\top=
 O_P(n^{-1/4}),\label{lemma4b1}\\
 &\sum_{t=1}^n[\rho_m''(e_t)-\e\rho''_m(e_t)]Z_2(z_{t})Z_2(z_{t})^\top=
 O_P(n^{-1/2}),\label{lemma4b2}\\
 &\sum_{t=1}^n[\rho_m''(e_t)-\e\rho''_m(e_t)]Z_1(x_{nt})Z_2(z_{t})^\top
 =O_P(n^{-1/2}),\label{lemma4b3}
\end{align}
as $n\to\infty$ where $m=[n^{2+\varepsilon}]$.
\end{lemma}

\begin{lemma}\label{lemma5}
For $Z_t$ defined in Lemma \ref{lemma4}, Under Assumptions 2.1, 3.1-3.3 and 3.4(ii), we have jointly
\begin{align}
&\mathcal{R}_n\equiv\sum_{t=1}^nZ_tZ_t^\top\to_D\mathcal{R}, \label{lemma5a} \\
&\sum_{t=1}^n\psi(e_t)Z_t\to_D\mathcal{R}^{1/2}N(0, a_1 I), \label{lemma5b}
\end{align}
as $n\to\infty$, where $a_1$ is given in Assumption 3.3 and the identity matrix has dimension $d_1+p_1+d_2p_2+p_2$. Consequently, as $n\to\infty$,
\begin{equation}\label{lemma5c}
\mathcal{R}^{-1/2}_n\sum_{t=1}^n\psi(e_t)Z_t\to_DN(0, a_1 I).
\end{equation}
\end{lemma}

\section{Proofs of the main results}

{\bf Proof of Theorem \ref{th21}} ~~(1) By Theorem 9.42 of \citet{rudin2004}, we may take derivative under the integral,
\begin{align*}
\rho_m'(u)=&-\int \rho(x)\phi_m'(x-u)dx,
\end{align*}
and further as $m\to\infty$, $\phi_m'(x-u)\to \delta'(x-u)$ in generalized function sense, and hence $\rho_m'(u)\to \rho'(u)$ whenever the latter exists. For the second derivative, the same argument applies.

(2) Note that
\begin{align*}
\sup_u|\rho_m(u)-\rho(u)|=&\sup_u \left|\int \rho(x)\phi_m(x-u)dx-\rho(u)\right|\\
\le&\sup_u\int |\rho(x+u)-\rho(u)|\phi_m(x)dx\\
\le&C\int |x|\phi_m(x)dx =\frac{C}{\sqrt{2m}}\int |x|\phi(x)dx\\
 =&Cm^{-1/2},
\end{align*}
where $\phi(\cdot)$ is the density of standard normal variable and $C$ may not be the same at each appearance.

(3) Consequently, for these $\rho(u)$, $\rho_m(e_t)-\rho(e_t)=O(m^{-1/2})$ almost surely. Moreover,
\begin{align*}
&\e|\rho_m'(e_t)-\psi(e_t)|=\int |\rho_m'(u)-\psi(u)|f(u)du\\
=&\int \left|-\int \rho(x)\phi_m'(x-u)dx-\rho'(u)\right|f(u)du\\
=&\int \left|-\int \rho(x+u)\phi_m'(x)dx-\rho'(u)\right|f(u)du\\
=&\int \left|\int \rho'(x+u)\phi_m(x)dx-\rho'(u)\right|f(u)du\\
\le&\iint |\rho'(x+u)-\rho'(u)|\phi_m(x)dxf(u)du\\
=&\int \left[\int |\rho'(x+u)-\rho'(u)|f(u)du\right]\phi_m(x)dx\\
=&\int_0^\infty \left[\int (\rho'(x+u)-\rho'(u))f(u)du\right]\phi_m(x)dx\\
&+\int_{-\infty}^0 \left[\int (\rho'(u)-\rho'(x+u))f(u)du\right]\phi_m(x)dx\\
=&-\int_0^\infty \left[\int (\rho(x+u)-\rho(u))f'(u)du\right]\phi_m(x)dx\\
&-\int_{-\infty}^0 \left[\int (\rho(u)-\rho(x+u))f'(u)du\right]\phi_m(x)dx\\
\le&\int \left[\int |\rho(u)-\rho(x+u)||f'(u)|du\right]\phi_m(x)dx,
\end{align*}
by integration by parts and noting that $\rho'(u)=\psi(u)$ is an increasing function. It follows that
\begin{align*}
\e|\rho_m'(e_t)-\psi(e_t)|
\le C\int |x|\phi_m(x)dx \int |f'(u)|du=Cm^{-1/2}.
\end{align*}

(4) Note that
\begin{align*}
&\e[\rho_m''(e_t)-\rho''(e_t)]=\int [\rho_m''(u)-\rho''(u)]f(u)du\\
=&\iint [\rho''(x+u)-\rho''(u)]\phi_m(x)dxf(u)du\\
=&\int \left[\int [\rho''(x+u)-\rho''(u)]f(u)du\right]\phi_m(x)dx\\
=&\int \left[\int [\rho(x+u)-\rho(u)]f''(u)du\right]\phi_m(x)dx,
\end{align*}
which implies $|\e[\rho_m''(e_t)-\rho''(e_t)]|\le C m^{-1/2}$ immediately.

(5) Note that
\begin{align*}
& |\e[\rho''(e_t+\epsilon)-\rho''(e_t)]|
=\left|\int[\rho''(u+\epsilon)-\rho''(u)]f(u)du\right|\\
=&\left|\int[\rho(u+\epsilon)-\rho(u)]f''(u)du\right|
\le\int|\rho(u+\epsilon)-\rho(u)||f''(u)|du\\
\le& |\epsilon| \int |f''(u)|du=C|\epsilon|.
\end{align*}
Therefore,
\begin{align*}
|\e[\rho_m''(e_t+\epsilon)-\rho''_m(e_t)]|\le& |\e[\rho''(e_t+\epsilon)-\rho''(e_t)]|+|\e[\rho_m''(e_t+\epsilon)-
\rho''(e_t+\epsilon)]|\\
&+|\e[\rho_m''(e_t)-\rho''(e_t)]|\\
=&O(m^{-1/2}+\epsilon).
\end{align*}
\qed

{\bf Proof of Theorem \ref{th22}}.\ \ Define
\begin{equation*}
\tilde{\Pi}_{mn}(\beta)\equiv \sum_{t=1}^n[\rho_m(e_t- n^{-1/2} x_t'\beta)-\rho_m(e_t)].
\end{equation*}

By virtue of Theorem \ref{th21}, $\sup_u|\rho_m(u)-\rho(u)|\le C m^{-1/2}$, we have
\begin{equation}\label{2c}
\sup_{\beta\in \mathbb{R}^d} \left|\tilde{\Pi}_n(\beta)-\tilde{\Pi}_{mn}(\beta)\right|
=O_{a.s.}(nm^{-1/2}).
\end{equation}

In fact,
\begin{align*}
&\sup_{\beta\in \mathbb{R}^d} \left|\tilde{\Pi}_n(\beta)-\tilde{\Pi}_{mn}(\beta)\right|\\
\le& \sum_{t=1}^n\sup_{\beta\in \mathbb{R}^d} |\rho(e_t-n^{-1/2}x_t'\beta)-\rho_m(e_t-n^{-1/2}x_t'\beta)|
+\sum_{t=1}^n|\rho_m(e_t)-\rho(e_t)|\\
=&O_{a.s.}(nm^{-1/2}).
\end{align*}

Thus, as long as we take $m=[n^{2+\varepsilon}]$ with $\varepsilon>0$, it is legitimate to analyze $\tilde{\Pi}_{mn}(\beta)$ that is sufficiently smooth.

Now, using Taylor expansion to second order we have
\begin{align*}
 &\rho_m(e_t-n^{-1/2}x_t'\beta)-\rho_m(e_t)
 =-\DF{1}{\sqrt{n}}\rho'_m(e_t)x_t'\beta+\DF{1}{2n}\rho_m''(e_t)[x_t'\beta]^2.
\end{align*}
Accordingly, we may write
\begin{align*}
\tilde{\Pi}_{mn}(\beta)=&\sum_{t=1}^n\left(-\DF{1}{\sqrt{n}}\rho'_m(e_t)x_t'\beta
+\DF{1}{2n}\rho_m''(e_t)[x_t'\beta]^2\right) \\
=&\left(-\DF{1}{\sqrt{n}}\sum_{t=1}^n\rho'_m(e_t)x_t'\right)\beta+\DF{1}{2}\beta' \left(\DF{1}{n} \sum_{t=1}^n\rho''_m(e_t)x_tx_t'\right)\beta \\
=&\left(-\DF{1}{\sqrt{n}}\sum_{t=1}^n\psi(e_t)x_t'\right)\beta+\DF{1}{2} \beta' \left(\DF{1}{n} \sum_{t=1}^n\e[\rho''_m(e_t)]x_tx_t'\right)\beta \\
&+\left(\DF{1}{\sqrt{n}}\sum_{t=1}^n[\rho'_m(e_t)-\psi(e_t)]x_t'\right)\beta \\
&+\DF{1}{2}\beta' \left(\DF{1}{n} \sum_{t=1}^n\{\rho''_m(e_t)-\e[\rho''_m(e_t)]\} x_tx_t'\right)\beta  \\
=&\left(-\DF{1}{\sqrt{n}}\sum_{t=1}^n\psi(e_t)x_t'\right)\beta+\DF{a_2}{2}
\beta' \left(\DF{1}{n} \sum_{t=1}^nx_tx_t'\right)\beta \\
&+\left(\DF{1}{\sqrt{n}}\sum_{t=1}^n[\rho'_m(e_t)-\psi(e_t)]x_t'\right)\beta \\
&+\DF{1}{2}\beta' \left(\DF{1}{n} \sum_{t=1}^n\{\rho''_m(e_t)-\e[\rho''_m(e_t)]\} x_tx_t'\right)\beta \\
&+\DF{1}{2}(\e[\rho''_m(e_t)]-a_2)
\beta' \left(\DF{1}{n} \sum_{t=1}^nx_tx_t'\right)\beta
\end{align*}
where $a_2=\e[\rho''(e_t)]>0$, and due to Theorem 2.1, $\rho'_m(e_t)-\psi(e_t)=O_P(m^{-1/2})$ uniformly in $t$, $\e[\rho''_m(e_t)-a_2]=O(m^{-1/2})$, and $\DF{1}{n} \sum_{t=1}^n\{\rho''_m(e_t)-\e[\rho''_m(e_t)]\} x_tx_t'=O_P(n^{-1/2})$. Indeed,
\begin{align*}
&\e\left\|\DF{1}{n} \sum_{t=1}^n\{\rho''_m(e_t)-\e[\rho''_m(e_t)]\} x_tx_t'\right\|^2\\
=&\DF{1}{n^2} \sum_{t=1}^n\e\{\rho''_m(e_t)-\e[\rho''_m(e_t)]\}^2\e \|x_t\|^4\\
=&C\DF{1}{n}\{\e[\rho''_m(e_t)]^2-(\e[\rho''_m(e_t)])^2\},
\end{align*}
and moreover,
\begin{align*}
\e[\rho''_m(e_t)]^2=&\int [\rho''_m(u)]^2f(u)du
= \int \left[\int \rho(x)\phi''_m(x-u)dx\right]^2f(u)du\\
=&\int \left[\int \rho(x+u)\phi''_m(x)dx\right]^2f(u)du\\
=&\int \left[\int \rho(x+u)m\sqrt{\DF{m}{\pi}}(2mx^2-1)e^{-mx^2}dx\right]^2f(u)du\\
=&\int \left[\int \rho\left(\DF{x}{\sqrt{m}}+u\right)m \DF{1}{\sqrt{\pi}}(2x^2-1)e^{-x^2}dx\right]^2f(u)du\\
=&\int \left[\int \left(\rho(u)+\psi(u)\DF{x}{\sqrt{m}}+\rho''(u)\DF{x^2}{m}+O(m^{-3/2})\right)m \DF{1}{\sqrt{\pi}}(2x^2-1)e^{-x^2}dx\right]^2f(u)du\\
=&\int \left[\int \rho''(u)\DF{1}{\sqrt{\pi}} x^2(2x^2-1)e^{-x^2}dx\right]^2f(u)du+O(m^{-1/2})\\
=&\int[\rho''(u)]^2f(u)du+O(m^{-1/2}),
\end{align*}
by Taylor expansion, and similarly
\begin{align*}
\e\rho''_m(e_t)=&\int \rho''_m(u)f(u)du=\iint \rho(x+u)\phi''_m(x)dxf(u)du \\
 =& \iint \rho(x+u)m\sqrt{\DF{m}{\pi}}(2mx^2-1)e^{-mx^2}dx f(u)du\\
 =&\iint \rho\left(\DF{x}{\sqrt{m}}+u\right)m \DF{1}{\sqrt{\pi}}(2x^2-1)e^{-x^2}dxf(u)du\\
 =&\iint \left(\rho(u)+\psi(u)\DF{x}{\sqrt{m}}+\rho''(u)\DF{x^2}{m}+O(m^{-3/2})\right)m \DF{1}{\sqrt{\pi}}(2x^2-1)e^{-x^2}dxf(u)du\\
 =&\int \rho''(u)f(u)du+O(m^{-1/2}).
\end{align*}
Thus,
\begin{equation*}
\e[\rho''_m(e_t)]^2-(\e[\rho''_m(e_t)])^2=\int[\rho''(u)]^2f(u)du-\left(\int \rho''(u)f(u)du\right)^2+O(m^{-1/2}).
\end{equation*}

It is readily seen that $|\tilde{\Pi}_{mn}(\beta)-Q_n(\beta)|=O_P(n^{-1/2})$, that together with \eqref{2c} implies \eqref{th22b}.

Moreover,
\begin{align}\label{2f}
&\sup_{\|\beta\|\le c}|\tilde{\Pi}_{mn}(\beta)-Q_n(\beta)|\notag\\
\le&\sup_{\|\beta\|\le c}
\left|\left(\DF{1}{\sqrt{n}}\sum_{t=1}^n[\rho'_m(e_t)-\psi(e_t)]x_t'\right)\beta \right|\notag\\
&+\sup_{\|\beta\|\le c}
\left|\DF{1}{2}\beta' \left(\DF{1}{n} \sum_{t=1}^n\{\rho''_m(e_t)-\e[\rho''_m(e_t)]\} x_tx_t'\right)\beta \right|\notag \\
&+\sup_{\|\beta\|\le c}
\left|\DF{1}{2}(\e[\rho''_m(e_t)]-a_2)
\beta' \left(\DF{1}{n} \sum_{t=1}^nx_tx_t'\right)\beta\right|\notag\\
\le&c\left\|\DF{1}{\sqrt{n}}\sum_{t=1}^n[\rho'_m(e_t)-\psi(e_t)]x_t \right\|
+\DF{c^2}{2}\left\|\DF{1}{n} \sum_{t=1}^n\{\rho''_m(e_t)-\e[\rho''_m(e_t)]\} x_tx_t' \right\|\notag \\
&+\DF{c^2}{2}|\e[\rho''_m(e_t)]-a_2|\left\|\DF{1}{n} \sum_{t=1}^n x_tx_t' \right\|\notag \\
=&O_P(m^{-1/2}n^{1/2}+n^{-1/2}+m^{-1/2})=O_P(n^{-1/2}),
\end{align}
for any $c>0$.

It is also clear that
\begin{equation}\label{2g}
\sup_{\|\beta\|\le c_n}|\tilde{\Pi}_{mn}(\beta)-Q_n(\beta)|=o_P(1),
\end{equation}
for $c_n=o(n^{1/4})$.

In view of \eqref{2c}, along with \eqref{2f} and \eqref{2g}, assertions \eqref{th22c} and \eqref{th22d} follow immediately.

We next shall show
\begin{equation}\label{2h}
\widehat{\beta}-\widehat{\beta}_Q=o_P(n^{-1/2+\lambda}), \ \ \ n\to\infty,
\end{equation}
where $\widehat{\beta}$ is given by \eqref{2b}.

First of all, $\widehat{\beta}-\widehat{\beta}_Q=o_P(1)$, similar to the Convexity Lemma in \citet{pollard1991}.

Moreover, by Lipschitz condition of $\rho(\cdot)$,
\begin{align*}
&|\tilde{\Pi}_n(\widehat{\beta})-\tilde{\Pi}_n(\widehat{\beta}_Q)|  \\
\le&\sum_{t=1}^n|\rho(e_t-n^{-1/2}x_t'\widehat{\beta})- \rho(e_t-n^{-1/2}x_t'\widehat{\beta}_Q)|\\
\le&C\frac{1}{\sqrt{n}}\sum_{t=1}^n\|x_t\|\|\widehat{\beta}-\widehat{\beta}_Q\|\\
=&\|\widehat{\beta}-\widehat{\beta}_Q\|\sqrt{n} \e(\|x_t\|)(1+o_P(1)).
\end{align*}
Accordingly, for any $\epsilon>0$, and some $M>0$ chosen later,
\begin{align*}
&\p(n^{1/2-\lambda}\|\widehat{\beta}-\widehat{\beta}_Q\|> \epsilon)\le\p( |\tilde{\Pi}_n(\widehat{\beta})-\tilde{\Pi}_n(\widehat{\beta}_Q)|>\epsilon n^{\lambda} \e(\|x_t\|))\\
=&\p( |\tilde{\Pi}_n(\widehat{\beta})-\tilde{\Pi}_n(\widehat{\beta}_Q)|>\epsilon n^{\lambda} \e(\|x_t\|), |\tilde{\Pi}_n(\widehat{\beta}_Q)-Q_n(\widehat{\beta}_Q)|>Mn^{-1/2})\\
&+\p( |\tilde{\Pi}_n(\widehat{\beta})-\tilde{\Pi}_n(\widehat{\beta}_Q)|>\epsilon n^{\lambda} \e(\|x_t\|), |\tilde{\Pi}_n(\widehat{\beta}_Q)-Q_n(\widehat{\beta}_Q)|\le Mn^{-1/2})\\
\le&\p( |\tilde{\Pi}_n(\widehat{\beta}_Q)-Q_n(\widehat{\beta}_Q)|>Mn^{-1/2})\\
&+\p( |\tilde{\Pi}_n(\widehat{\beta})-Q_n(\widehat{\beta}_Q)|>\epsilon n^{\lambda} \e(\|x_t\|)/2)\\
=&\p(|\tilde{\Pi}_n(\widehat{\beta}_Q)-Q_n(\widehat{\beta}_Q)|>Mn^{-1/2})\\
&+\p(|\tilde{\Pi}_n(\widehat{\beta})-Q_n(\widehat{\beta})|>\epsilon n^{\lambda} \e(\|x_t\|)/4)\\
&+\p(|Q_n(\widehat{\beta})-Q_n(\widehat{\beta}_Q)|>\epsilon n^{\lambda} \e(\|x_t\|)/4).
\end{align*}

For any $\eta>0$, one may choose sufficient large $M>0$ such that the first term is less than $\eta/2$, because $|\tilde{\Pi}_n(\beta)-Q_n(\beta)| =O_P(n^{-1/2})$. Due to the same reason the second term is $o(1)$ for $n$ large; and straightforward algebra yields $Q_n(\widehat{\beta})-Q_n(\widehat{\beta}_Q)=o_P(1)$ because of $\widehat{\beta}-\widehat{\beta}_Q=o_P(1)$, and then the third term is $o(1)$ too. This proves \eqref{2h}. \qed

\medskip

{\bf Proof of Theorem \ref{th41}}\ \ Let $\lambda_{ij}, q_{ij}\to\infty$ ($i=1,2$, $j=1,\ldots, p_i$) with $n$ that we shall specify later and let $\beta_{1j}=q_{1j}(\gamma_{1j}-\gamma_{1j}^0)$, $\beta_{2j}=q_{2j}(\gamma_{2j}-\gamma_{2j}^0)$, $\alpha_{1j}=\lambda_{1j}(\theta_{1j}-\theta_{1j}^0)$, $\alpha_{2j}=\lambda_{2j}(\theta_{2j}-\theta_{2j}^0)$. We shift down our objective function by a constant that certainly does not affect the minimization, and reparameterize
\begin{align*}
&L_n(\theta,\gamma)=\sum_{t=1}^n\left[\rho\left(y_t- \sum_{j=1}^{p_1} \gamma_{1j}g_{1j}(x_{t}^\top\theta_{1j})- \sum_{j=1}^{p_2} \gamma_{2j} g_{2j}(z_{t}^\top\theta_{2j})\right)-\rho(e_t)\right]\\
=&\sum_{t=1}^n\Big[\rho\Big(e_t+\sum_{j=1}^{p_1}
[\gamma_{1j}^{0}g_{1j}(x_{t}^\top\theta_{1j}^{0})- \gamma_{1j}g_{1j}(x_{t}^\top\theta_{1j})]+\sum_{j=1}^{p_2}[
\gamma_{2j}^{0}g_{2j}(z_{t}^\top\theta_{2j}^{0})-\gamma_{2j} g_{2j}(z_{t}^\top\theta_{2j})]\Big)
-\rho(e_t)\Big]\\
=&\sum_{t=1}^n\Big[\rho\Big(e_t+\sum_{j=1}^{p_1}
[\gamma_{1j}^{0}g_{1j}(x_{t}^\top\theta_{1j}^{0})- \gamma_{1j}g_{1j}(x_{t}^\top\theta_{1j}^{0}+x_{t}^\top \lambda_{1j}^{-1}\alpha_{1j})]\\
&\qquad +\sum_{j=1}^{p_2}[
\gamma_{2j}^{0}g_{2j}(z_{t}^\top\theta_{2j}^{0})-\gamma_{2j} g_{2j}(z_{t}^\top\theta_{2j}^{0}+z_{t}^\top \lambda_{2j}^{-1}\alpha_{2j})]\Big)
 -\rho(e_t)\Big]\\
=&\sum_{t=1}^n\Big[\rho\Big(e_t+\sum_{j=1}^{p_1}
[\gamma_{1j}^{0}(g_{1j}(x_{t}^\top\theta_{1j}^{0})-
g_{1j}(x_{t}^\top\theta_{1j}^{0}+x_{t}^\top \lambda_{1j}^{-1}\alpha_{1j}))+(\gamma_{1j}^{0}- \gamma_{1j})g_{1j}(x_{t}^\top\theta_{1j}^{0}+x_{t}^\top \lambda_{1j}^{-1}\alpha_{1j})]\\
&+\sum_{j=1}^{p_2}[
\gamma_{2j}^{0}(g_{2j}(z_{t}^\top\theta_{2j}^{0})-
g_{2j}(z_{t}^\top\theta_{2j}^{0}+z_{t}^\top \lambda_{2j}^{-1}\alpha_{2j}))+(\gamma_{2j}^{0}-\gamma_{2j}) g_{2j}(z_{t}^\top\theta_{2j}^{0}+z_{t}^\top \lambda_{2j}^{-1}\alpha_{2j})]\Big)
 -\rho(e_t)\Big]\\
=&\sum_{t=1}^n\Big[\rho\Big(e_t+\sum_{j=1}^{p_1}
[\gamma_{1j}^{0}(g_{1j}(x_{t}^\top\theta_{1j}^{0})-
g_{1j}(x_{t}^\top\theta_{1j}^{0}+x_{t}^\top \lambda_{1j}^{-1}\alpha_{1j}))+q_{1j}^{-1} \beta_{1j}g_{1j}(x_{t}^\top\theta_{1j}^{0}+x_{t}^\top \lambda_{1j}^{-1}\alpha_{1j})]\\
&+\sum_{j=1}^{p_2}[
\gamma_{2j}^{0}(g_{2j}(z_{t}^\top\theta_{2j}^{0})-
g_{2j}(z_{t}^\top\theta_{2j}^{0}+z_{t}^\top \lambda_{2j}^{-1}\alpha_{2j}))+q_{2j}^{-1} \beta_{2j} g_{2j}(z_{t}^\top\theta_{2j}^{0}+z_{t}^\top \lambda_{2j}^{-1}\alpha_{2j})]\Big)
 -\rho(e_t)\Big]\\
\equiv&\tilde{L}_n(\alpha, \beta).
\end{align*}
As argued in Section 2, this reparameterization is adopted in the literature quite often, for example, \citet{bickel1974}, \citet{badu1989}, \citet{davis1992} and \citet{phillips1995}. We then consider the optimization of $\tilde{L}_n(\alpha, \beta)$ over $(\alpha, \beta)$. It is clear that if $(\widehat\alpha,\widehat\beta)$ minimizes $\tilde{L}_n(\alpha, \beta)$, $(\widehat\theta,\widehat\gamma)$ minimizes $L_n(\theta, \gamma)$ where $\widehat\beta_{1j}=q_{1j}(\widehat\gamma_{1j}-\gamma_{1j}^0)$, $\widehat\beta_{2j}=q_{2j}(\widehat\gamma_{2j}-\gamma_{2j}^0)$, $\widehat\alpha_{1j}=\lambda_{1j}(\widehat\theta_{1j}-\theta_{1j}^0)$, and $\widehat\alpha_{2j}=\lambda_{2j}(\widehat\theta_{2j}-\theta_{2j}^0)$.

For simplicity, denote
\begin{align*}
\Delta_{1t}(\alpha_{1j}, \beta_{1j})= &\gamma_{1j}^{0}(g_{1j}(x_{t}^\top\theta_{1j}^{0})-
g_{1j}(x_{t}^\top\theta_{1j}^{0}+x_{t}^\top \lambda_{1j}^{-1}\alpha_{1j}))+q_{1j}^{-1} \beta_{1j}g_{1j}(x_{t}^\top\theta_{1j}^{0}+x_{t}^\top \lambda_{1j}^{-1}\alpha_{1j}),  \\ \intertext{and}
\Delta_{2t}(\alpha_{2j}, \beta_{2j})=& \gamma_{2j}^{0}(g_{2j}(z_{t}^\top\theta_{2j}^{0})-
g_{2j}(z_{t}^\top\theta_{2j}^{0}+z_{t}^\top \lambda_{2j}^{-1}\alpha_{2j}))+q_{2j}^{-1} \beta_{2j} g_{2j}(z_{t}^\top\theta_{2j}^{0}+z_{t}^\top \lambda_{2j}^{-1}\alpha_{2j}),
\end{align*}
and thus $\tilde{L}_n(\alpha, \beta)$ is simplified as
\begin{equation}\label{Qn}
\tilde{L}_n(\alpha, \beta)= \sum_{t=1}^n\Big[\rho\Big(e_t+ \sum_{j=1}^{p_1}\Delta_{1t}(\alpha_{1j}, \beta_{1j})+\sum_{j=1}^{p_2}\Delta_{2t}(\alpha_{2j}, \beta_{2j})\Big)
 -\rho(e_t)\Big].
\end{equation}

On the other hand, using Taylor expansion,
\begin{align*}
\Delta_{1t}(\alpha_{1j}, \beta_{1j})= &\gamma_{1j}^{0}(g_{1j}(x_{t}^\top\theta_{1j}^{0})-
g_{1j}(x_{t}^\top\theta_{1j}^{0}+x_{t}^\top \lambda_{1j}^{-1}\alpha_{1j}))+q_{1j}^{-1} \beta_{1j}g_{1j}(x_{t}^\top\theta_{1j}^{0})\\
&+q_{1j}^{-1} \beta_{1j}[g_{1j}(x_{t}^\top\theta_{1j}^{0}+x_{t}^\top \lambda_{1j}^{-1}\alpha_{1j})-g_{1j}(x_{t}^\top\theta_{1j}^{0})]\\
=&-\gamma_{1j}^{0}[\dot{g}_{1j}(x_{t}^\top\theta_{1j}^{0})x_{t}^\top \lambda_{1j}^{-1}\alpha_{1j}+\frac{1}{2}\ddot{g}_{1j}(x_{t}^\top
\theta_{1j}^{0*})(x_{t}^\top \lambda_{1j}^{-1}\alpha_{1j})^2]\\
&+q_{1j}^{-1} \beta_{1j}g_{1j}(x_{t}^\top\theta_{1j}^{0})\\
&+q_{1j}^{-1} \beta_{1j}[\dot{g}_{1j}(x_{t}^\top\theta_{1j}^{0})x_{t}^\top \lambda_{1j}^{-1}\alpha_{1j}+\frac{1}{2}\ddot{g}_{1j}(x_{t}^\top
\theta_{1j}^{0*})(x_{t}^\top \lambda_{1j}^{-1}\alpha_{1j})^2],
\end{align*}
where $x_{t}^\top\theta_{1j}^{0*}$ is on the segment joining $x_{t}^\top\theta_{1j}^{0}$ and $x_{t}^\top\theta_{1j}^{0}+x_{t}^\top \lambda_{1j}^{-1}\alpha_{1j}$.

Suppose that Assumption 3.4(i) holds. Let $\lambda_{1j}=\dot{\nu}_j(\sqrt{n})n$ and $q_{1j}=\nu_j(\sqrt{n})\sqrt{n}$, $j=1, \cdots, p_1$. Denote $x_{nt}=x_t/\sqrt{n}$ for simplicity. Observe that in $\Delta_{1t}(\alpha_{1j}, \beta_{1j})$,
\begin{align*}
 &\ddot{g}_{1j}(x_{t}^\top\theta_{1j}^{0*})(x_{t}^\top \lambda_{1j}^{-1}\alpha_{1j})^2=\ddot{g}_{1j}(x_{nt}^\top\theta_{1j}^{0*})
 (x_{nt}^\top \alpha_{1j})^2 \frac{\ddot{\nu}_j(\sqrt{n})}{\dot{\nu}_j(\sqrt{n})^2n}=o_P\left(
 \frac{1}{\dot{\nu}_j(\sqrt{n})n}\right), \\
&q_{1j}^{-1}\dot{g}_{1j}(x_{t}^\top\theta_{1j}^{0})x_{t}^\top \lambda_{1j}^{-1}\alpha_{1j}=\frac{1}{\nu_j(\sqrt{n})n}
  \dot{g}_{1j}(x_{nt}^\top\theta_{1j}^{0})x_{nt}^\top \alpha_{1j}=O_P\left(
 \frac{1}{\nu_j(\sqrt{n})n}\right),\\
&q_{1j}^{-1}\ddot{g}_{1j}(x_{t}^\top\theta_{1j}^{0*})(x_{t}^\top \lambda_{1j}^{-1}\alpha_{1j})^2=\ddot{g}_{1j}(x_{nt}^\top\theta_{1j}^{0*})
(x_{nt}^\top \alpha_{1j})^2 \frac{\ddot{\nu}_j(\sqrt{n})}{\nu_j(\sqrt{n})\dot{\nu}_j(\sqrt{n})^2n\sqrt{n}}\\
&\ \ \ \ =o_P\left(\frac{1}{\nu_j(\sqrt{n})\dot{\nu}_j(\sqrt{n})n\sqrt{n}}\right),
\end{align*}
from which we conclude
\begin{equation}\label{delta1}
\Delta_{1t}(\alpha_{1j}, \beta_{1j})=\frac{1}{\sqrt{n}}\left[-\gamma_{1j}^{0} \dot{g}_{1j}(x_{nt}^\top\theta_{1j}^{0})x_{nt}^\top \alpha_{1j}+
\beta_{1j}g_{1j}(x_{nt}^\top\theta_{1j}^{0})\right](1+o_P(1)).
\end{equation}

Moreover,
\begin{align*}
\Delta_{2t}(\alpha_{2j}, \beta_{2j})=& \gamma_{2j}^{0}(g_{2j}(z_{t}^\top\theta_{2j}^{0})-
g_{2j}(z_{t}^\top\theta_{2j}^{0}+z_{t}^\top \lambda_{2j}^{-1}\alpha_{2j}))+q_{2j}^{-1} \beta_{2j} g_{2j}(z_{t}^\top\theta_{2j}^{0})\\
&+q_{2j}^{-1} \beta_{2j}[ g_{2j}(z_{t}^\top\theta_{2j}^{0}+z_{t}^\top \lambda_{2j}^{-1}\alpha_{2j})-g_{2j}(z_{t}^\top\theta_{2j}^{0})]\\
=&-\gamma_{2j}^{0}[\dot{g}_{2j}(z_{t}^\top\theta_{2j}^{0})z_{t}^\top \lambda_{2j}^{-1}\alpha_{2j}+\frac{1}{2}
\ddot{g}_{2j}(z_{t}^\top\theta_{2j}^{0*})(z_{t}^\top \lambda_{2j}^{-1}\alpha_{2j})^2]\\
&+q_{2j}^{-1} \beta_{2j} g_{2j}(z_{t}^\top\theta_{2j}^{0})\\
&+q_{2j}^{-1} \beta_{2j}[\dot{g}_{2j}(z_{t}^\top\theta_{2j}^{0})z_{t}^\top \lambda_{2j}^{-1}\alpha_{2j}+\frac{1}{2}
\ddot{g}_{2j}(z_{t}^\top\theta_{2j}^{0*})(z_{t}^\top \lambda_{2j}^{-1}\alpha_{2j})^2],
\end{align*}
where $z_{t}^\top\theta_{2j}^{0*}$ is on the segment joining $z_{t}^\top\theta_{2j}^{0}$ and $z_{t}^\top\theta_{2j}^{0}+z_{t}^\top \lambda_{2j}^{-1}\alpha_{2j}$.

Now, let $\lambda_{2j}=\sqrt{n}$ and $q_{2j}=\sqrt{n}$, $j=1, \cdots, p_2$. Accordingly, in $\Delta_{2t}(\alpha_{2j}, \beta_{2j})$,
\begin{align*}
 &\ddot{g}_{2j}(z_{t}^\top\theta_{2j}^{0*})(z_{t}^\top \lambda_{2j}^{-1}\alpha_{2j})^2=\frac{1}{n}\ddot{g}_{2j}
 (z_{t}^\top\theta_{2j}^{0*})(z_{t}^\top \alpha_{2j})^2=O_P(n^{-1}),  \\
 &q_{2j}^{-1} \beta_{2j}\dot{g}_{2j}(z_{t}^\top\theta_{2j}^{0})z_{t}^\top \lambda_{2j}^{-1}\alpha_{2j}=O_P(n^{-1}),\\
 &q_{2j}^{-1} \beta_{2j}\ddot{g}_{2j}(z_{t}^\top\theta_{2j}^{0*})(z_{t}^\top \lambda_{2j}^{-1}\alpha_{2j})^2=O_P(n^{-3/2}),
\end{align*}
from which we have
\begin{equation}\label{delta2}
\Delta_{2t}(\alpha_{2j}, \beta_{2j})= \frac{1}{\sqrt{n}}\left[-\gamma_{2j}^{0}\dot{g}_{2j}(z_{t}^\top
\theta_{2j}^{0})z_{t}^\top \alpha_{2j}+\beta_{2j} g_{2j}(z_{t}^\top\theta_{2j}^{0})\right](1+o_P(1)).
\end{equation}

Thus, equations \eqref{delta1} and \eqref{delta2} deliver the leading terms of $\Delta_{1t}(\alpha_{1j}, \beta_{1j})$ and $\Delta_{2t}(\alpha_{2j}, \beta_{2j})$, respectively, while they are all of $O_P(n^{-1/2})$. Hence,
\begin{equation}\label{delta}
\Delta_{t}(\alpha, \beta)\equiv \sum_{j=1}^{p_1}\Delta_{1t}(\alpha_{1j}, \beta_{1j})+\sum_{j=1}^{p_2}\Delta_{2t}(\alpha_{2j}, \beta_{2j}),
\end{equation}
that is of order $O_P(n^{-1/2})$, validating the use of Taylor expansion in the sequel.

Next, since $\rho(\cdot)$ satisfies Assumption 2.1, invoking the regular sequence $\rho_m(u)$ defined in Theorem \ref{th21}, we have $
\sup_{u\in \mathbb{R}}|\rho_m(u)-\rho(u)|\le Cm^{-1/2}.$

Hence, $\sup_{\alpha, \beta}|\tilde{L}_n(\alpha, \beta)-\tilde{L}_{mn}(\alpha, \beta)|=O(nm^{-1/2})$ almost surely, where
\begin{equation}\label{Qmn}
\tilde{L}_{mn}(\alpha, \beta)= \sum_{t=1}^n\Big[\rho_m\Big(e_t+ \Delta_{t}(\alpha, \beta)\Big)-\rho_m(e_t)\Big].
\end{equation}
Notice that if we let $m$ sufficiently large, for example, $m=n^{2+\varepsilon}$ for some $\varepsilon>0$, $\tilde{L}_n(\alpha, \beta)$ is approximated by $\tilde{L}_{mn}$ with a negligible difference. The benefit of $\tilde{L}_{mn}$ is that $\rho_m(\cdot)$ is sufficiently smooth.

Accordingly, taking Taylor expansion to second order we have
\begin{align*}
\tilde{L}_{mn}(\alpha, \beta)=& \sum_{t=1}^n\Big\{\rho'_m(e_t)\Delta_{t}(\alpha, \beta)+\frac{1}{2}\rho''_m(e_t)\Big[\Delta_{t}(\alpha, \beta) \Big]^2\Big\}.
\end{align*}

For expositional purposes, denote
\begin{align*}
\Lambda_1= &(\alpha_{11}, \cdots, \alpha_{1p_1}, \beta_{11}, \cdots, \beta_{1p_1})^\top, \\
\Lambda_2= &(\alpha_{21}, \cdots, \alpha_{2p_2}, \beta_{21}, \cdots, \beta_{2p_2})^\top,\\
\intertext{and further denote}
\Lambda=&(\Lambda_1^\top, \Lambda_2^\top)^\top, \ \ \ \text{and}\ \ \
Z_t=(Z_1(x_{nt})^\top, Z_2(z_t)^\top)^\top.
\end{align*}
This enables us to write
\begin{equation*}
\Delta_{t}(\alpha, \beta)=\frac{1}{\sqrt{n}}[Z_1(x_{nt})^\top \Lambda_1+Z_2(z_t)^\top \Lambda_2]=\frac{1}{\sqrt{n}}Z_t^\top \Lambda.
\end{equation*}
We then rewrite
\begin{align*}
\tilde{L}_{mn}(\alpha, \beta)=& \sum_{t=1}^n\Big\{\rho'_m(e_t)\Delta_{t}(\alpha, \beta)+\frac{1}{2}\rho''_m(e_t)\Big[\Delta_{t}(\alpha, \beta)\Big]^2\Big\}\\
=&\left(\frac{1}{\sqrt{n}}\sum_{t=1}^n\psi(e_t)Z_{t}'\right)\Lambda+ \frac{1}{2}\Lambda'\left(\frac{1}{n}\sum_{t=1}^n\rho''_m(e_t)
Z_tZ_t'\right)\Lambda\\
&+\left(\frac{1}{\sqrt{n}}\sum_{t=1}^n[\rho'_m(e_t)-\psi(e_t)]Z_{t}'
\right)\Lambda\\
=&\left(\frac{1}{\sqrt{n}}\sum_{t=1}^n\psi(e_t)Z_{t}'\right)\Lambda+ \frac{1}{2}\e[\rho''_m(e_t)]\Lambda'\left(\frac{1}{n}\sum_{t=1}^n
Z_tZ_t'\right)\Lambda\\
&+\left(\frac{1}{\sqrt{n}}\sum_{t=1}^n[\rho'_m(e_t)-\psi(e_t)]Z_{t}'
\right)\Lambda\\
&+\frac{1}{2}\Lambda'\left(\frac{1}{n}\sum_{t=1}^n
[\rho''_m(e_t)-\e\rho''_m(e_t)]Z_tZ_t'\right)\Lambda\\
=&\left(\frac{1}{\sqrt{n}}\sum_{t=1}^n\psi(e_t)Z_{t}'\right)\Lambda+ \frac{1}{2}a_2\Lambda'\left(\frac{1}{n}\sum_{t=1}^n
Z_tZ_t'\right)\Lambda \\ &+\frac{1}{2}\{\e[\rho''_m(e_t)]-a_2\}\Lambda'\left(\frac{1}{n}\sum_{t=1}^n
Z_tZ_t'\right)\Lambda\\
&+\left(\frac{1}{\sqrt{n}}\sum_{t=1}^n[\rho'_m(e_t)-\psi(e_t)]Z_{t}'
\right)\Lambda\\
&+\frac{1}{2}\Lambda'\left(\frac{1}{n}\sum_{t=1}^n
[\rho''_m(e_t)-\e\rho''_m(e_t)]Z_tZ_t'\right)\Lambda.
\end{align*}

Denoting
\begin{equation}\label{quadratic}
Q_n(\alpha, \beta)\equiv \left(\frac{1}{\sqrt{n}}\sum_{t=1}^n\psi(e_t)Z_t\right)^\top \Lambda +\frac{a_2}{2}\Lambda^\top \left(\frac{1}{n}\sum_{t=1}^nZ_tZ_t^\top \right)\Lambda,
\end{equation}
we then have $\tilde{L}_{mn}(\alpha,\beta)-Q_n(\alpha,\beta)\to_P 0$ for all $(\alpha, \beta)$ due to Lemma A.3; and for any $c>0$,
\begin{align*}
&\sup_{\|\alpha\|^2+\|\beta\|^2\le c}| \tilde{L}_{mn}(\alpha,\beta)-Q_n(\alpha,\beta)| \\
\le &\frac{c}{2}
|\e[\rho''_m(e_t)]-a_2|\left\|\frac{1}{n}\sum_{t=1}^n
Z_tZ_t'\right\|+c\left\|\frac{1}{\sqrt{n}}\sum_{t=1}^n[\rho'_m(e_t)-\psi(e_t)]
Z_{t}\right\|\\
&+\frac{c}{2}\left\|\frac{1}{n}\sum_{t=1}^n
[\rho''_m(e_t)-\e\rho''_m(e_t)]Z_tZ_t'\right\|\\
=&O_P(m^{-1/2}n^{1/2}+n^{-1/2}),
\end{align*}
where $\e[\rho''_m(e_t)]-a_2=\e[\rho''_m(e_t)-\rho''(e_t)]=O(m^{-1/2})$ due to Theorem \ref{th21} and Lemma \ref{lemma3}.

Because $a_2>0$, the quadratic form \eqref{quadratic} has unique minimum,
\begin{equation}\label{minimizer}
\widehat{\Lambda}_Q=a_2^{-1}\left(\frac{1}{n}\sum_{t=1}^nZ_tZ_t^\top \right)^{-1} \frac{1}{\sqrt{n}}\sum_{t=1}^n\psi(e_t)Z_t.
\end{equation}

Note that $\widehat{\Lambda}$ is a vector containing all estimators of interest, that is, $\widehat{\Lambda}=(\widehat{\Lambda}_1^\top, \ \ \widehat{\Lambda}_2^\top)^\top$, where $\widehat{\Lambda}_1$ and $ \widehat{\Lambda}_2$ are defined in \eqref{41b1}.
We may show $\widehat{\Lambda}=\widehat{\Lambda}_Q+o_P(1)$ as what we did in Theorem 2.2. Hence, using the joint convergence in Lemma \ref{lemma2}, as $n\to \infty$, $\widehat{\Lambda}=O_P(1)$ and then
\begin{equation*}
\widehat{\Lambda}\to_DB^{-1}B_1,
\end{equation*}
where the matrix $B$ and vector $B_1$ are given in \eqref{41b}.
This finishes the proof.\qed

\medskip

{\bf Proof of Corollary \ref{cor1}}\ \ Observe that
\begin{align*}
\widehat{e}_t=&y_t-\sum_{j=1}^{p_1}\widehat\gamma_{1j}g_{1j}
(\widehat{\theta}_{1j}^{\,\top} x_t) -\sum_{j=1}^{p_2} \widehat\gamma_{2j}g_{2j}(\widehat{\theta}_{2j}^{\,\top} z_t)\\
=&e_t+\sum_{j=1}^{p_1}[\gamma_{1j}^0g_{1j}
(\theta_{1j}^{0\,\top} x_t)-\widehat\gamma_{1j}g_{1j}
(\widehat{\theta}_{1j}^{\,\top} x_t)]+\sum_{j=1}^{p_1}[\gamma_{2j}^0g_{2j}( \theta_{2j}^{0\,\top} z_t) - \widehat\gamma_{2j}g_{2j} (\widehat{\theta}_{2j}^{\,\top} z_t)]\\
=&e_t+\sum_{j=1}^{p_1}\Delta_{1t}(\widehat\alpha_{1j}, \widehat\beta_{1j}) +\sum_{j=1}^{p_2}\Delta_{2t}(\widehat\alpha_{2j}, \widehat\beta_{2j})=e_t+
\Delta_{t}(\widehat\alpha, \widehat\beta),
\end{align*}
and it is shown in the proof of the theorem that $\Delta_{t}(\widehat\alpha, \widehat\beta)=n^{-1/2}Z_t^\top \widehat{\Lambda}=O_P(n^{-1/2+\epsilon})$ uniformly in $t$ for any $\epsilon\in (0,1/2)$.

Notice that
\begin{align*}
\widehat{a}_1-a_1=&\frac{1}{n}\sum_{t=1}^n \{[\psi(\widehat{e}_t)]^2-\e [\psi(e_t)]^2\}\\
=&\frac{1}{n}\sum_{t=1}^n\{[\psi(\widehat{e}_t)]^2- [\psi(e_t)]^2\} +\frac{1}{n}\sum_{t=1}^n \{[\psi(e_t)]^2-\e [\psi(e_t)]^2\}\\
=&\frac{1}{n}\sum_{t=1}^n\{\psi(\widehat{e}_t)-\psi(e_t)\} \{\psi(\widehat{e}_t)+\psi(e_t)\} +\frac{1}{n}\sum_{t=1}^n \{[\psi(e_t)]^2-\e [\psi(e_t)]^2\}\\
\equiv & I_1+I_2, \ \ \text{say}.
\end{align*}
The second term $I_2=o_P(1)$ since $\{\psi(e_t)\}$ is a martingale difference sequence and hence is ergodic, so is $\{\psi(e_t)^2\}$. In the first term, $|\psi(\widehat{e}_t)+\psi(e_t)|\le c$ because $\rho(\cdot)$ is Lipschitz; and
\begin{align*}
|I_1|\le&\frac{c}{n}\sum_{t=1}^n|\psi(\widehat{e}_t)-\psi(e_t)|\\
\le & \frac{c}{n}\sum_{t=1}^n|\psi(\widehat{e}_t)-\rho_m'(\widehat{e}_t)|
+\frac{c}{n}\sum_{t=1}^n|\rho_m'(\widehat{e}_t)-
\rho_m'(e_t)|+\frac{c}{n}\sum_{t=1}^n|\rho_m'(e_t) -\psi(e_t)|\\
=&I_{11}+I_{12}+I_{13}, \ \ \text{say}.
\end{align*}

$I_{13}=o_P(1)$ because by Theorem \ref{th21}, $\e|\rho_m'(e_t) -\psi(e_t)|=O(m^{-1/2})$; similarly $I_{11}=o_P(1)$ by Theorem \ref{th21}, $\widehat{e}_t=e_t+\Delta_{t}(\widehat\alpha, \widehat\beta)$ and conditional argument on $\Delta_{t}(\widehat\alpha, \widehat\beta)$. To show $I_{12}=o_P(1)$, notice that $\rho_m'(\widehat{e}_t)-
\rho_m'(e_t)=\rho_m''(e_t^*)\Delta_{t}(\widehat\alpha, \widehat\beta)$ where $e_t^*$ is on the segment joining $\widehat{e}_t$ and $e_t$. Here, for large $m$, $\rho_m''(\cdot)$ is nonnegative due to convexity of $\rho(\cdot)$. By Theorem \ref{th21} again, $\e\rho_m''(e_t^*)\to \e\rho''(e_t)$ as $m, n\to\infty$. Then,
\begin{equation*}
I_{12}=O_P(n^{-1/2+\epsilon})\frac{c}{n}\sum_{t=1}^n\rho_m''(e_t^*)=
O_P(n^{-1/2+\epsilon}),
\end{equation*}
which gives $I_{12}=o_P(1)$. Accordingly, $I_1=o_P(1)$, so we have $\widehat{a}_1\to_Pa_1$.

Notice further that
\begin{align*}
\widehat{a}_2-a_2=&\frac{1}{n}\sum_{t=1}^n \{\rho''_m(\widehat{e}_t)-\e [\rho''(e_t)]\}\\
=& \frac{1}{n}\sum_{t=1}^n[\rho''_m(\widehat{e}_t)-\rho''_m(e_t)] +\frac{1}{n}\sum_{t=1}^n \{\rho''_m(e_t)-\e [\rho''_m(e_t)]\} + \e\{\rho''_m(e_t)-\rho''(e_t)\} \\
\equiv&I_3+I_4+O(m^{-1/2}), \ \ \text{say}.
\end{align*}
Here $I_3=o_P(1)$ because of the differentiability of $\rho''_m$ and $\widehat{e}_t-e_t=\Delta_{t}(\widehat\alpha, \widehat\beta)$; $I_4=o_P(1)$ due to ergodicity of $\{e_t\}$. Accordingly, $\widehat{a}_2\to_Pa_2$.
\qed

\medskip

{\bf Proof of Theorem \ref{th42}}. \ \ Define $D_n=\text{diag}(\sqrt[4]{n}, \sqrt[4]{n}^3I_{d_1-1})$, $p_n=\sqrt{n}$ and $q_n=\sqrt[4]{n}$ for later use. Define
\begin{align}\label{th32a}
\begin{split}
\alpha_{1}=&D_nP^\top (\theta_{1}-\theta_{1}^0),\ \beta_{1j}=q_n (\gamma_{1j}-\gamma_{1j}^0), \ \ j=1,\cdots, p_1, \\ \alpha_{2j}=&p_n(\theta_{2j}-\theta_{2j}^0),\ \ \ \beta_{2j}=p_n (\gamma_{2j}-\gamma_{2j}^0),\ j=1,\cdots, p_2.
\end{split}
\end{align}
In what follows we may write
\begin{align*}
x_t^\top \theta_{1}=&x_t^\top \theta_{1}^0+x_t^\top (\theta_{1}-\theta_{1}^0)=x_t^\top \theta_{1}^0+x_{n,t}^\top \alpha_{1}, \
\ \text{where}\ x_{n,t}\equiv D_n^{-1}P^\top x_t,\\
z_t^\top \theta_{2j}=&z_t^\top \theta_{2j}^0+z_t^\top (\theta_{2j}-\theta_{2j}^0)=z_t^\top \theta_{2j}^0+p_n^{-1}z_t^\top \alpha_{2j}.
\end{align*}

We then have
\begin{align*}
&L_n(\theta,\gamma)=\sum_{t=1}^n\left[\rho\left(y_t- \sum_{j=1}^{p_1} \gamma_{1j}g_{1j}(x_{t}^\top\theta_{1})- \sum_{j=1}^{p_2} \gamma_{2j} g_{2j}(z_{t}^\top\theta_{2j})\right)-\rho(e_t)\right]\\
=&\sum_{t=1}^n\Big[\rho\Big(e_t+\sum_{j=1}^{p_1}
[\gamma_{1j}^{0}g_{1j}(x_{t}^\top\theta_{1}^{0})- \gamma_{1j}g_{1j}(x_{t}^\top\theta_{1})]+\sum_{j=1}^{p_2}[
\gamma_{2j}^{0}g_{2j}(z_{t}^\top\theta_{2j}^{0})-\gamma_{2j} g_{2j}(z_{t}^\top\theta_{2j})]\Big)
-\rho(e_t)\Big]\\
=&\sum_{t=1}^n\Big[\rho\Big(e_t+\sum_{j=1}^{p_1}
[\gamma_{1j}^{0}g_{1j}(x_{t}^\top\theta_{1}^{0})- \gamma_{1j}g_{1j}(x_{t}^\top\theta_{1}^{0}+x_{n,t}^\top \alpha_{1})]\\
&\qquad +\sum_{j=1}^{p_2}[
\gamma_{2j}^{0}g_{2j}(z_{t}^\top\theta_{2j}^{0})-\gamma_{2j} g_{2j}(z_{t}^\top\theta_{2j}^{0}+z_{t}^\top p_{n}^{-1}\alpha_{2j})]\Big)
 -\rho(e_t)\Big]\\
=&\sum_{t=1}^n\Big[\rho\Big(e_t+\sum_{j=1}^{p_1}
[\gamma_{1j}^{0}(g_{1j}(x_{t}^\top\theta_{1}^{0})-
g_{1j}(x_{t}^\top\theta_{1}^{0}+x_{n,t}^\top \alpha_{1}))+(\gamma_{1j}^{0}- \gamma_{1j})g_{1j}(x_{t}^\top\theta_{1}^{0}+x_{n,t}^\top \alpha_{1})]\\
&+\sum_{j=1}^{p_2}[
\gamma_{2j}^{0}(g_{2j}(z_{t}^\top\theta_{2j}^{0})-
g_{2j}(z_{t}^\top\theta_{2j}^{0}+z_{t}^\top p_{n}^{-1}\alpha_{2j}))+(\gamma_{2j}^{0}-\gamma_{2j}) g_{2j}(z_{t}^\top\theta_{2j}^{0}+z_{t}^\top p_{n}^{-1}\alpha_{2j})]\Big)
 -\rho(e_t)\Big]\\
=&\sum_{t=1}^n\Big[\rho\Big(e_t+\sum_{j=1}^{p_1}
[\gamma_{1j}^{0}(g_{1j}(x_{t}^\top\theta_{1}^{0})-
g_{1j}(x_{t}^\top\theta_{1}^{0}+x_{n,t}^\top \alpha_{1}))+q_{n}^{-1} \beta_{1j}g_{1j}(x_{t}^\top\theta_{1}^{0}+x_{n,t}^\top \alpha_{1})]\\
&+\sum_{j=1}^{p_2}[
\gamma_{2j}^{0}(g_{2j}(z_{t}^\top\theta_{2j}^{0})-
g_{2j}(z_{t}^\top\theta_{2j}^{0}+z_{t}^\top p_{n}^{-1}\alpha_{2j}))+p_{n}^{-1} \beta_{2j} g_{2j}(z_{t}^\top\theta_{2j}^{0}+z_{t}^\top p_{n}^{-1}\alpha_{2j})]\Big)
 -\rho(e_t)\Big]\\
\equiv&\tilde{L}_n(\alpha, \beta).
\end{align*}
The reason and benefit of reparameterization of $\tilde{L}_n(\alpha, \beta)$ are the same as stated in Theorem 4.1.

We then consider the optimization of $\tilde{L}_n(\alpha, \beta)$ over $(\alpha, \beta)$. It is clear that if $(\widehat\alpha,\widehat\beta)$ minimizes $\tilde{L}_n(\alpha, \beta)$, $(\widehat\theta,\widehat\gamma)$ minimizes $L_n(\theta, \gamma)$ where $\widehat\beta_{1j}=q_{n}(\widehat\gamma_{1j}-\gamma_{1j}^0)$, $\widehat\beta_{2j}=p_{n}(\widehat\gamma_{2j}-\gamma_{2j}^0)$, $\widehat\alpha_{1}=D_nP^\top(\widehat\theta_{1}-\theta_{1}^0)$, and $\widehat\alpha_{2j}=p_{n}(\widehat\theta_{2j}-\theta_{2j}^0)$.

Observe that
\begin{align*}
\Delta_{1t}(\alpha_{1}, \beta_{1j})\equiv&
\gamma_{1j}^{0}(g_{1j}(x_{t}^\top\theta_{1}^{0})-
g_{1j}(x_{t}^\top\theta_{1}^{0}+x_{n,t}^\top \alpha_{1}))+q_{n}^{-1} \beta_{1j}g_{1j}(x_{t}^\top\theta_{1}^{0}+x_{n,t}^\top \alpha_{1}) \\
=&\gamma_{1j}^{0}(g_{1j}(x_{t}^\top\theta_{1}^{0})-
g_{1j}(x_{t}^\top\theta_{1}^{0}+x_{n,t}^\top \alpha_{1}))+q_{n}^{-1} \beta_{1j}g_{1j}(x_{t}^\top\theta_{1}^{0}) \\
&+q_{n}^{-1} \beta_{1j}[g_{1j}(x_{t}^\top\theta_{1}^{0}+x_{n,t}^\top \alpha_{1})-g_{1j}(x_{t}^\top\theta_{1}^{0})], \\
\Delta_{2t}(\alpha_{2j}, \beta_{2j})\equiv &\gamma_{2j}^{0}(g_{2j}(z_{t}^\top\theta_{2j}^{0})-
g_{2j}(z_{t}^\top\theta_{2j}^{0}+z_{t}^\top p_{n}^{-1}\alpha_{2j}))+p_{n}^{-1} \beta_{2j} g_{2j}(z_{t}^\top\theta_{2j}^{0}+z_{t}^\top p_{n}^{-1}\alpha_{2j})\\
=&\gamma_{2j}^{0}(g_{2j}(z_{t}^\top\theta_{2j}^{0})-
g_{2j}(z_{t}^\top\theta_{2j}^{0}+z_{t}^\top p_{n}^{-1}\alpha_{2j}))+p_{n}^{-1} \beta_{2j} g_{2j}(z_{t}^\top\theta_{2j}^{0})\\
&+p_{n}^{-1} \beta_{2j}[ g_{2j}(z_{t}^\top\theta_{2j}^{0}+z_{t}^\top p_{n}^{-1}\alpha_{2j})-g_{2j}(z_{t}^\top\theta_{2j}^{0})].
\end{align*}

Further, invoking Taylor expansion,
\begin{align*}
\Delta_{1t}(\alpha_{1}, \beta_{1j})=& -\gamma_{1j}^{0}\dot{g}_{1j}(x_{t}^\top\theta_{1}^{0})x_{n,t}^\top \alpha_{1}+\frac{1}{2}\ddot{g}_{1j}(x_{t}^\top\theta_{1}^{*}) (x_{n,t}^\top \alpha_{1})^2 +q_{n}^{-1}\beta_{1j} g_{1j}(x_{t}^\top\theta_{1}^{0})\\
&+q_{n}^{-1} \beta_{1j}[\dot{g}_{1j}(x_{t}^\top\theta_{1}^{0})x_{n,t}^\top \alpha_{1}+\frac{1}{2}\ddot{g}_{1j}(x_{t}^\top\theta_{1}^{*})(x_{n,t}^\top \alpha_{1})^2],
\end{align*}
where $\theta_{1}^{*}$ is on the segment joining $\theta_{1}^{0}$ and $\alpha_{1}$.

Notice that $x_{n,t}=D_n^{-1}P^\top x_t=[n^{-1/4}x_{t}^\top \theta_{1}^{0}, n^{-3/4}x_t^\top P_{2}]^\top$, and thus
\begin{align*}
&q_{n}^{-1}\dot{g}_{1j}(x_{t}^\top\theta_{1}^{0})x_{n,t}^\top
= [n^{-1/2}\dot{g}_{1j}(x_{t}^\top\theta_{1}^{0})
x_{t}^\top\theta_{1}^{0}, n^{-1}\dot{g}_{1j}(x_{t}^\top\theta_{1}^{0})x_t^\top P_{2}];
\end{align*}
and also
\begin{align*}
&\ddot{g}_{1j}(x_{t}^\top\theta_{1}^{*}) (x_{n,t}^\top \alpha_{1})^2=
\ddot{g}_{1j}(x_{t}^\top\theta_{1}^{*}) (x_{t}^\top (\theta_{1}-\theta_{1}^0))^2\\
=&\ddot{g}_{1j}(x_{t}^\top\theta_{1}^{*}) (x_{t}^\top P^*D_n^{-1}D_nP^{*\top} (\theta_{1j}-\theta_{1j}^0))^2
=\ddot{g}_{1j}(x_{t}^\top\theta_{1}^{*}) (x_{t}^\top P^*D_n^{-1}\alpha_{1}^*))^2\\
=&n^{-1/2}\ddot{g}_{1j}(x_{t}^\top\theta_{1}^{*})(x_{t}^\top\theta_{1}^{*})^2
+n^{-3/2}\ddot{g}_{1j}(x_{t}^\top\theta_{1}^{*})(x_{t}^\top P^*\alpha_{1,2}^*)^2\\
&+2n^{-1}\ddot{g}_{1j}(x_{t}^\top\theta_{1}^{*})(x_{t}^\top\theta_{1}^{*}) (x_{t}^\top P_{2}^*\alpha_{1,2}^*)
\end{align*}
where $P^*=(\theta_{1}^{*}/\|\theta_{1}^{*}\|, P_{2}^*)$ is an orthogonal matrix, $\alpha_1^*=D_nP^{*\top} (\theta_{1j}-\theta_{1j}^0)$ and $\alpha_{1,2}^*=\sqrt[4]{n}^3P_2^{*\top} (\theta_{1j}-\theta_{1j}^0)$. Thus, due to Lemmas 2 and 3 in \citet[p1256]{phillips2000} and the $I$-regular property of $u\dot{g}_{1j}(u)$, $u\ddot{g}_{1j}(u)$ and $u^2\ddot{g}_{1j}(u)$, $\Delta_{1t}(\alpha_{1j}, \beta_{1j})$ has two leading terms, that is,
\begin{align}\label{th32b}
\Delta_{1t}(\alpha_{1}, \beta_{1j})=&
[-\gamma_{1j}^{0}\dot{g}_{1j}(x_{t}^\top\theta_{1}^{0})x_{n,t}^\top \alpha_{1}+q_{n}^{-1}\beta_{1j} g_{1j}(x_{t}^\top\theta_{1j}^{0})](1+o_P(1)).
\end{align}

Similarly,
\begin{align*}
\Delta_{2t}(\alpha_{2j}, \beta_{2j})
=&-\gamma_{2j}^{0}\dot{g}_{2j}(z_{t}^\top\theta_{2j}^{0})z_{t}^\top p_{n}^{-1}\alpha_{2j}+\frac{1}{2}\gamma_{2j}^{0}\ddot{g}_{2j}
(z_{t}^\top\theta_{2j}^{*})(z_{t}^\top p_{n}^{-1}\alpha_{2j})^2
+p_{n}^{-1} \beta_{2j} g_{2j}(z_{t}^\top\theta_{2j}^{0})\\
&+p_{n}^{-1} \beta_{2j}[\dot{g}_{2j}(z_{t}^\top\theta_{2j}^{0})z_{t}^\top p_{n}^{-1}\alpha_{2j}+\frac{1}{2}\ddot{g}_{2j}
(z_{t}^\top\theta_{2j}^{*})(z_{t}^\top p_{n}^{-1}\alpha_{2j})^2]\\
=&p_n^{-1}[-\gamma_{2j}^{0}\dot{g}_{2j}(z_{t}^\top\theta_{2j}^{0})z_{t}^\top \alpha_{2j}+\beta_{2j} g_{2j}(z_{t}^\top\theta_{2j}^{0})]\\
&+\frac{1}{2}p_{n}^{-2}\gamma_{2j}^{0}\ddot{g}_{2j}
(z_{t}^\top\theta_{2j}^{*})(z_{t}^\top \alpha_{2j})^2+p_{n}^{-2} \beta_{2j}[\dot{g}_{2j}(z_{t}^\top\theta_{2j}^{0})z_{t}^\top \alpha_{2j}+\frac{1}{2}p_{n}^{-1}\ddot{g}_{2j}
(z_{t}^\top\theta_{2j}^{*})(z_{t}^\top \alpha_{2j})^2]\\
=&p_n^{-1}[-\gamma_{2j}^{0}\dot{g}_{2j}(z_{t}^\top\theta_{2j}^{0})z_{t}^\top \alpha_{2j}+\beta_{2j} g_{2j}(z_{t}^\top\theta_{2j}^{0})](1+o_P(1)).
\end{align*}

For convenience, define
\begin{equation}\label{th32c}
\Delta_t(\alpha, \beta)=\sum_{j=1}^{p_1}\Delta_{1t}(\alpha_{1}, \beta_{1j})+ \sum_{j=1}^{p_2} \Delta_{2t}(\alpha_{2j}, \beta_{2j}),
\end{equation}
and due to the I-regularity of the functions involved, it is readily seen that $\Delta_t(\alpha, \beta)\sim n^{-1/4}$. This validates the use of Taylor expansion in the sequel.

We then write
\begin{align*}
\tilde{L}_n(\alpha, \beta)=&\sum_{t=1}^n\Big[\rho\Big(e_t+\Delta_t(\alpha, \beta)\Big) -\rho(e_t)\Big],\\
\tilde{L}_{mn}(\alpha, \beta)\equiv&\sum_{t=1}^n\Big[\rho_m\Big(e_t+\Delta_t(\alpha, \beta)\Big) -\rho_m(e_t)\Big],
\end{align*}
where $\rho_m(\cdot)$ is the regular sequence defined in Theorem \ref{th21}, and therefore,
\begin{equation}\label{th32d}
\sup_{\alpha, \beta}|\tilde{L}_{n}(\alpha, \beta)-\tilde{L}_{mn}(\alpha, \beta)|\le Cm^{-1/2}n,
\end{equation}
where $C$ is an absolute constant independent of both $m$ and $n$.

One may choose sufficiently large $m$, say $m=n^{2+\varepsilon}$, such that the difference between $\tilde{L}_{n}(\alpha, \beta)$ and $\tilde{L}_{mn}(\alpha, \beta)$ is negligible. We now focus on $\tilde{L}_{mn}(\alpha, \beta)$ that is differentiable w.r.t. all parameters.

Taking Taylor expansion to second order, we have
\begin{align}\label{th32e}
\begin{split}
\tilde{L}_{mn}(\alpha, \beta)= &\sum_{t=1}^n\Big[\rho_m \Big(e_t+\Delta_t(\alpha, \beta)\Big) -\rho_m(e_t)\Big]\\
=&\sum_{t=1}^n\Big[\rho_m'(e_t)\Delta_t(\alpha, \beta)+\frac{1}{2}\rho_m''(e_t)\Delta_t(\alpha, \beta)^2\Big].
\end{split}
\end{align}

To write $\tilde{L}_{mn}(\alpha, \beta)$ in vector form, recall $\Lambda=(\Lambda_1^\top, \Lambda_2^\top)^\top$,
\begin{align*}
\Lambda_1= &(\alpha_{1}^\top, \beta_{11}, \cdots, \beta_{1p_1})^\top, &
\Lambda_2= &(\alpha_{21}^\top, \cdots, \alpha_{2p_2}^\top, \beta_{21}, \cdots, \beta_{2p_2})^\top,
\end{align*}
and further recall $p_n=\sqrt{n}$ and $q_n=\sqrt[4]{n}$, $\dot{g}(u)=\sum_{j=1}^{p_1}\gamma_{1j}^0\dot{g}_{1j}(u)$, and $Z_t=(Z_1(x_{nt})^\top, Z_2(z_t)^\top)^\top$ given by \eqref{42c1}.
We then write
\begin{equation}\label{th32f}
\Delta_{t}(\alpha, \beta)=Z_1(x_{nt})^\top \Lambda_1+Z_2(z_t)^\top \Lambda_2=Z_t^\top \Lambda,
\end{equation}
so that $\tilde{L}_{mn}(\alpha, \beta)$ in \eqref{th32e} can be written as
\begin{align}\label{th32g}
\begin{split}
\tilde{L}_{mn}(\alpha, \beta)= &\sum_{t=1}^n\Big[\rho_m'(e_t)Z_t^\top \Lambda +\frac{1}{2}\rho_m''(e_t)(Z_t^\top \Lambda)^2\Big]\\
=&\left(\sum_{t=1}^n\rho_m'(e_t)Z_t\right)^\top \Lambda +\frac{1}{2}\Lambda^\top\left(\sum_{t=1}^n\rho_m''(e_t)Z_tZ_t^\top\right) \Lambda\\
=&\left(\sum_{t=1}^n\psi(e_t)Z_t\right)^\top \Lambda+\frac{1}{2}a_2\Lambda^\top \left(\sum_{t=1}^nZ_tZ_t^\top\right) \Lambda\\
&+\left(\sum_{t=1}^n[\rho_m'(e_t)-\psi(e_t)]Z_t\right)^\top \Lambda+\frac{1}{2}\Lambda^\top \left(\sum_{t=1}^n[\e(\rho''_m(e_t)|\ra_{t-1})-a_2]Z_tZ_t^\top\right) \Lambda\\
&+\frac{1}{2}\Lambda^\top \left(\sum_{t=1}^n[\rho_m''(e_t)-\e(\rho''_m(e_t)|\ra_{t-1})]Z_tZ_t^\top\right) \Lambda.
\end{split}
\end{align}

Let
\begin{equation}\label{th32h}
Q_n(\alpha, \beta)=\left(\sum_{t=1}^n\psi(e_t)Z_t\right)^\top \Lambda+\frac{1}{2}a_2\Lambda^\top \left(\sum_{t=1}^nZ_tZ_t^\top\right) \Lambda.
\end{equation}
Then, for any $(\alpha, \beta)$, due to Lemmas \ref{lemma4},
\begin{align*}
 &|\tilde{L}_{mn}(\alpha, \beta)-Q_n(\alpha, \beta)| \\
 \le &\left\|\sum_{t=1}^n[\rho_m'(e_t)-\psi(e_t)]Z_t\right\|\|\Lambda\|+ \frac{1}{2}\|\Lambda\|^2 \left\|\sum_{t=1}^n[\e(\rho''_m(e_t)|\ra_{t-1})-a_2]Z_tZ_t^\top\right\|\\
 &+\frac{1}{2}\|\Lambda\|^2 \left\|\sum_{t=1}^n[\rho_m''(e_t)-\e\rho''_m(e_t)]Z_tZ_t^\top\right\|\\
=&O_P(n^{1/2}m^{-1/2})+O_P(m^{-1/2})+O_P(n^{-1/4})\\
=&O_P(n^{-1/4}),
\end{align*}
for $m=\left[n^{2+\varepsilon}\right]$. This also holds uniformly on any compact region for $(\alpha, \beta)$. Moreover, $Q_n(\alpha, \beta)$ has unique minimizer
\begin{equation*}
\widehat{\Lambda}_Q=-\left(a_2\sum_{t=1}^nZ_tZ_t^\top\right)^{-1}
\sum_{t=1}^n\psi(e_t)Z_t,
\end{equation*}
for that we can show $\widehat{\Lambda}=\widehat{\Lambda}_Q+o_P(1)$ as what we did in Theorem 2.2.

Then, $\widehat{\Lambda}=O_P(1)$ is obtained from that of $\widehat{\Lambda}_Q$ by virtue of the joint convergence in Lemma \ref{lemma5}, and moreover,
\begin{equation*}
\widehat{\Lambda}\to_D\DF{\sqrt{a_1}}{a_2}\mathcal{R}^{-1/2}N(0, I),
\end{equation*}
as $n\to\infty$, where $I$ is identity of order $d_1+p_1+p_2d_2+p_2$. \qed

\medskip

{\bf Proof of Corollary \ref{cor2}}.\ \ We now consider the convergence of $\widehat\theta_1-\theta_1^0$. Observe that
\begin{align*}
&\sqrt[4]{n}(\widehat\theta_1-\theta_1^0)=\sqrt[4]{n} (D_nP^\top)^{-1}(D_nP^\top)(\widehat\theta-\theta_0)\\
\to_D&\DF{\sqrt{a_1}}{a_2}P\; \text{diag}(1, {\bf 0}_{d_1-1})N(0,r_{11})
=_D\DF{\sqrt{a_1}}{a_2}N(0, r_{11}^{11}\theta_1^0(\theta_1^0)^\top).
\end{align*}
\qed

\newpage

\section*{Appendix C.\ Preliminaries on generalized functions}

\renewcommand{\theequation}{C.\arabic{equation}}
\renewcommand{\thetable}{C.\arabic{table}}
\setcounter{equation}{0} \setcounter{lemma}{0} \setcounter{subsection}{0} %
\setcounter{figure}{0}

Since we shall adopt a generalized function approach in our paper, some preliminaries are given in this section. Note that in mathematic context generalized functions are called distributions or tempered distributions according to the spaces of test functions.

\medskip

\noindent {\bf Definition} (Space $D$) \ \emph{The space of all functions $\phi(x)$ defined on the real line satisfying the following properties is called Space $D$}.
\begin{itemize}
  \item \emph{$\phi(x)$ is an infinitely differentiable function defined at every point on $\mathbb{R}$. That is, $\phi^{(k)}(x)$ exists for any positive integer $k$};
  \item \emph{There is a constant $A>0$ such that $\phi(x)\equiv 0$ for $|x|>A$, or equivalently it has a compact support.}
\end{itemize}

Note that $D$ is a linear space over real set. A function $\phi(x)\in D$ is of $C^\infty$, also known as a test function. There exist many different types of functions in $D$, and it is noteworthy that for any continuous function $f(x)$ with compact support, there is a function $\phi(x)\in D$ such that $|f(x)-\phi(x)|< \varepsilon$ for all $x$ and any given $\varepsilon>0$. This implies the denseness of $D$ in any $C[a,b]$. See \citet[p. 3]{gelfand1964}.
\medskip

\noindent {\bf Definition} (Convergence in $D$) \ \emph{A sequence $\{\phi_m\}$ in $D$ is said to converge to a function $\phi_0$ if the following conditions are satisfied:}
\begin{itemize}
  \item \emph{All $\phi_m$ as well as $\phi_0$ vanish outside a common region};
  \item \emph{$\phi_m^{(k)}\to\phi_0^{(k)}$ uniformly over $\mathbb{R}$ as $m\to\infty$ for all $k\ge 0$}.
\end{itemize}

It can be shown that $\phi_0\in D$, and therefore $D$ is closed under this definition.

\medskip

\noindent {\bf Definition} (Distribution) \ \emph{A continuous linear functional on the space $D$ is called a distribution. The space of all distributions on $D$ is denoted by $D'$}.
\medskip

Distribution is another name of generalized function. For the definitions of continuity and linearity of a functional, please see \citet[p. 25]{kanwal1983}. The space $D'$ is called the dual space of $D$, is itself a linear space.

The set of distributions that are mostly useful are those generated by locally integrable functions. Indeed, if $f(x)$ is locally integrable on $\mathbb{R}$, it generates a distribution $f:\ \phi \mapsto \mathbb{R}$ through
\begin{equation*}
\langle f, \phi\rangle=\int f(x)\phi(x)dx, \ \ \ \forall \phi\in D.
\end{equation*}

Such defined distribution is called a \emph{regular distribution}. Remarkably, it is proved in \citet[p. 27]{kanwal1983} that two continuous functions that generate the same regular distributions are identical. Moreover, if two locally integrable functions produce the same regular distributions, they are identical almost everywhere. These enable one to identify functions from the distributions they generate.

All distributions other than regular ones are called \emph{singular}. Thus, $D'$ is larger than $D$, because all $\phi\in D$ are distributions so that $D\subset D'$, and there do exist singular distributions, in particular Dirac delta $\delta(\cdot)$: $\phi\mapsto \phi(x_0)$, $\forall \, \phi\in D$, with a fixed $x_0\in \mathbb{R}$, is a singular distribution as shown in \citet[p. 4]{gelfand1964}.

By definition, generalized functions cannot be assigned values at isolated points, while statements about a generalized function on a neighbourhood of points can be given in a well-defined way. This means that generalized functions are determined by its local property. Detailed discussions can be found in \citet[p. 5]{gelfand1964}.

As is well known, not all ordinary functions are differentiable. In contradiction, generalized functions always have derivatives that are generalized functions too, and consequently have derivatives of any order. See \citet[p. 18]{gelfand1964}.

\medskip

\noindent {\bf Definition} (Derivative of distributions) \emph{Let $f\in D'$. A functional $g$ defined on $D$ given by}
\begin{equation*}
\langle g, \phi\rangle =-\langle f, \phi'\,\rangle, \ \ \ \forall\, \phi\in D
\end{equation*}
\emph{is called the derivative of $f$, denoted as $f'$ or $df/dx$}.
\medskip

The definition is motivated by integration by parts. To be an eligible member of $D'$, note that such defined $f'$ is linear and continuous. Accordingly, all generalized functions have derivatives of all orders. Moreover, the generalized derivative of a regular distribution agrees with the conventional one whenever the latter exists.
\medskip

\noindent {\bf Definition} (Convergence in $D'$) \ \emph{A sequence of distributions $f_m\in D'$, $m=1,2,\cdots$, is said to converge to a distribution $f\in D'$ if}
\begin{equation*}
\lim_{m\to\infty}\langle f_m,\phi\rangle=\langle f, \phi\rangle, \ \ \ \forall\, \phi\in D.
\end{equation*}
\emph{A set of distributions $\{f_\epsilon\}$ indexed by real $\epsilon$ is said converging to $f$ when $\epsilon\to\epsilon_0$, if for $ \forall\, \phi\in D$, $\lim_{\epsilon\to \epsilon_0} \langle f_\epsilon,\phi\rangle=\langle f, \phi\rangle$}.

\emph{A series of distributions $\sum_{m=1}^\infty f_m$ converges to a distribution $f\in D$ if the sequence of partial sum $s_M=\sum_{m=1}^M f_m$ converges to $f$ as $M\to\infty$.}

These definitions contain the convergence of ordinary functions as a special case. Indeed, suppose that all members of distribution sequence $\{f_m\}$ are regular, and $f_m(x)$ converge to $f(x)$ uniformly on any compact interval, then
\begin{equation*}
\lim_{m\to\infty}\langle f_m,\phi\rangle=\lim_{m\to\infty}\int f_m(x)\phi(x)dx=\int f(x)\phi(x)dx=\langle f, \phi\rangle, \ \ \forall\, \phi\in D,
\end{equation*}
by uniform convergence theorem.

A consequence of the definition is that if $f_m\to f$ then $f_m^{(k)}\to f^{(k)}$ for any $k>0$; if $\sum_{m=1}^\infty f_m$ converges to $f$, then the series can be differentiated term by term as many times as required. See \citet[p. 59]{kanwal1983}.

The most important sequence of distributions is a sequence of regular distributions $\{f_m\}$, so-called delta-convergent sequence, that converges to $\delta$ distribution. This may be regarded as a bridge between regular and singular distributions.

\begin{lemma}
Let $f(x)$ be a nonnegative function defined on $\mathbb{R}$ such that $\int f(x)dx=1$. Put $f_\epsilon(x)=\epsilon^{-1}f(x/\epsilon)$, $\epsilon>0$. Then $\lim_{\epsilon\to 0}f_\epsilon(x)=\delta(x)$.
\end{lemma}

This is exactly the theorem in \citet[p. 62]{kanwal1983} for univariate functions. This result enables us to construct delta sequence. Observe that $\epsilon$ can be replaced by $1/m$ to have sequence $f_m(x)$ that convergence to $\delta(x)$ as $m\to\infty$.

In some cases we should consider a space that is larger than $D$ as test function space, in order to extend the compact support of test functions in $D$ to the entire real line.

\medskip

\noindent {\bf Definition} (Space $S$)\ The space $S$ of test functions of rapid decay contains all functions $\phi$ defined on $\mathbb{R}$ that satisfy
\begin{itemize}
  \item $\phi(x)$ is infinitely differentiable, i.e. $\phi(x)\in C^\infty$;
  \item $\phi(x)$, as well as its derivatives of all orders, vanishes at infinity faster than any power of $1/|x|$, i.e. for any $p,k\ge 0$, $|x^p\phi^{(k)}(x)|\le C_{pk}$ where the constant $C_{pk}$ only depends on $p$, $k$ and $\phi$.
\end{itemize}

The space $S$ is linear and clearly $D\subset S$. Accordingly $S'\subset D'$ because a continuous linear functional on $S$ is also a continuous linear functional on $D$. Similarly to $D$ and $D'$, we may define the convergence of sequence and the derivative of distributions in $S$ and $S'$. One may find these in \citet[p.17]{gelfand1964} and \citet[p.138]{kanwal1983}.

\section*{D.\ Proofs of Lemmas}

\renewcommand{\theequation}{D.\arabic{equation}}
\renewcommand{\thetable}{C.\arabic{table}}

{\bf Proof of Lemma A.1}\ \ This is exactly Lemma A.2 in \citet{donggao2018}. \qed

\medskip

{\bf Proof of Lemma A.2}\ \ Recall that $z_t=h(\tau_t,v_t)$, and $\tau_t=t/n$ and $v_t=q(\eta_t,\cdots, \eta_{t-d_0+1})+\tilde{v_t}$, where $\tilde{v}_t$ is independent of $\{\eta_j, j\in \mathbb{Z}\}$. Thus, $z_t$ and $x_t$ are correlated through these $\eta$'s. In view of (A.1) we may write $x_t=\sum_{i=t-d_0+1}^t w_i+ \sum_{i=1}^{t-d_0}w_i\equiv w_{t,d_0}+x_{t-d_0}$. Then, because $n^{-1/2}w_{t,d_0}=o_P(1)$, $x_{t}$ and $x_{t-d_0}$ have the same asymptotic distribution. Meanwhile, $x_{t-d_0}$ and $z_t$ are mutually independent.

By the continuity of $\ddot{g}_{1j}(\cdot)$, $\dot{g}_{1j}(x_{nt}^\top \theta_{1j}^0)- \dot{g}_{1j}(x_{n,t-d_0}^\top \theta_{1j}^0)=O_P(n^{-1/2})$. We therefore may replace $\dot{g}_{1j}(x_{nt}^\top \theta_{1j}^0)$ by $\dot{g}_{1j}(x_{n, t-d_0}^\top \theta_{1j}^0)$ in the derivation in the sequel but we avoid doing so and treat them independent for simplicity.

It follows from \citet{phillips2001} that
\begin{equation*}
\frac{1}{\sqrt{n}}\sum_{t=1}^n\psi(e_t)Z_1(x_{nt})\to_D \int_0^1 Z_1(B(r)) dU(r),
\end{equation*}
while since $z_t$ is a strictly stationary and $\alpha$-mixing stationary, and due to the martingale difference structure imposed in Assumption 3.3 it is evidently that $\frac{1}{\sqrt{n}}\sum_{t=1}^n\psi(e_t)Z_2(z_t)\to_D N(0,a_1\Sigma)$, where the conditional covariance matrix converging to $\Sigma$ in probability is shown below. The joint convergence then follows by the independence between $x_{n,t-d_0}$ and $z_t$. This finishes (A.4).

Now, consider the convergence in probability in (A.5). Notice that $z_t$ is $\alpha$-mixing, so that by Assumption 3.2, $\frac{1}{n} \sum_{t=1}^n[Z_2(z_t)Z_2(z_t)^\top- \mathbb{E}Z_2(z_t)Z_2(z_t)^\top]=o_P(1)$. Indeed,
\begin{align*}
&\mathbb{E}\left\| \frac{1}{n} \sum_{t=1}^n[Z_2(z_t)Z_2(z_t)^\top- \mathbb{E}Z_2(z_t)Z_2(z_t)^\top]\right\|^2 = \frac{1}{n^2}\mathbb{E}\sum_{t=1}^n\left\|Z_2(z_t)Z_2(z_t)^\top-\mathbb{E} Z_2(z_t)Z_2(z_t)^\top]\right\|^2\\
&+\frac{2}{n^2}\mathbb{E}\sum_{t=2}^n\sum_{s=1}^{t-1}\text{tr}\left([ Z_2(z_t)Z_2(z_t)^\top- \mathbb{E}Z_2(z_t)Z_2(z_t)^\top] [ Z_2(z_s)Z_2(z_s)^\top- \mathbb{E}Z_2(z_s)Z_2(z_s)^\top]^\top\right)\\
\le&\frac{1}{n^2}\sum_{t=1}^n\mathbb{E}[\|Z_2(z_t)\|^4] +\frac{2}{n^2}\sum_{t=2}^n\sum_{s=1}^{t-1}\alpha^{1/2}(t-s)
[\mathbb{E}(\|Z_2(z_t)\|^4)]^{1/4}[\mathbb{E}(\|Z_2(z_t)\|^4)]^{1/4}\\
=&O(n^{-1})=o(1)
\end{align*}
where we use the Davydov's inequality for $\alpha$-mixing processes with $p=2$ and $q=r=4$ \citep[p 19]{bosq1996} and the condition in Assumption 3.2.

It then suffices to show that, as $n\to\infty$,
\begin{align*}
\frac{1}{n} \sum_{t=1}^n\mathbb{E}Z_2(z_t)Z_2(z_t)^\top =&\frac{1}{n} \sum_{t=1}^n\mathbb{E}Z_2(h(\tau_t,v_t))Z_2(h(\tau_t,v_t))^\top\\
=&\sum_{t=1}^{n-1}\int_{\tau_{t}}^{\tau_{t+1}}
\mathbb{E}Z_2(h(r,v_t))Z_2(h(r,v_t))^\top dr+o(1)\\
=&\int_0^1\mathbb{E}Z_2(h(r,v_1))Z_2(h(r,v_1))^\top+o(1)\\
\to& \Sigma.
\end{align*}

Finally we consider (A.6). Invoking the martingale structure of the error term, it suffices to consider the convergence of
\begin{align*}
&\frac{1}{n}\sum_{t=1}^n\begin{pmatrix}
Z_1(x_{nt})Z_1(x_{nt})^\top\\
[\mathbb{E}Z_2(z_t)]Z_1(x_{nt})^\top
\end{pmatrix}
=\frac{1}{n}\sum_{t=1}^n\begin{pmatrix}
Z_1(x_{nt})Z_1(x_{nt})^\top\\
[\mathbb{E}Z_2(h(\tau_t,z_1))]Z_1(x_{nt})^\top
\end{pmatrix}\\
&\to_D \begin{pmatrix}
\int_0^1 Z_1(B(r))Z_1(B(r))^\top dr\\
\int_0^1[\mathbb{E}Z_2(h(r,v_1))]Z_1(B(r))^\top dr
\end{pmatrix},
\end{align*}
due to Theorem 3.2 of \citet[p. 131]{dg2019}.\qed

{\bf Proof of Lemma A.3}\ \ Note by Theorem 2.1 that $\rho'_m(e_t)-\psi(e_t)=O_P(m^{-1/2})$ uniformly in $t$, while $x_{nt}$ and $z_t$ are all bounded in probability. It follows from the continuities of functions $g_{1j}$, $g_{2j}$ as well as their derivatives, (A.7) is fulfilled.

To show (A.8), notice that
\begin{align*}
&\e\left\|\frac{1}{n}\sum_{t=1}^n[\rho''_m(e_t)-\mathbb{E}\rho''_m(e_t)]
Z_tZ_t^\top\right\|^2\\
=&\frac{1}{n^2}\sum_{t=1}^n\e\{\e([\rho''_m(e_t)-\mathbb{E}\rho''_m(e_t)]^2|
\ra_{t-1})\|Z_tZ_t^\top\|^2\}.
\end{align*}

Similar to Section 2, we have
\begin{equation*}
\e([\rho''_m(e_t)-\mathbb{E}\rho''_m(e_t)]^2|\ra_{t-1}) =\int[\rho''(u)]^2f_{t,e}(u)du-\left(\int \rho''(u)f_{t,e}(u)du\right)^2+O(m^{-1/2}).
\end{equation*}
Here, $f_{t,e}(u)$ is the density of $e_t$ given $\ra_{t-1}$. Under assumption $\max_t f_{t,e}(u)\le f_e(u)$ that is a density and belongs to $S$, $\e([\rho''_m(e_t)-\mathbb{E}\rho''_m(e_t)]^2|\ra_{t-1})$ is bounded by $\int[\rho''(u)]^2f_e(u)du$. The assertion follows immediately due to the assumption on $z_t$ and Lemma A.1.\qed

\noindent{\bf Proof of Lemma A.4}\ \ In $Z_1(x_{nt})$, note that in the sub-vector $\sum_{j=1}^{p_1}\gamma_{1j}^0\dot{g}_{1j}(x_t^\top \theta_{1}^0)x_{n,t}^\top$, $x_{n,t}=(n^{-1/4}x_t^\top \theta_{1j}^0, n^{-3/4}x_t^\top P_{2j})$. We show (A.9) by three typical variables in $Z_1(x_{nt})$, that is,
\begin{align*}
I_1\equiv & n^{-1/4}\sum_{t=1}^n[\rho_m'(e_t)-\psi(e_t)]
\dot{g}_{11}(x_t^\top \theta_{1}^0)x_t^\top \theta_{1}^0=O_P(n^{1/4}m^{-1/2}),\\
I_2\equiv &n^{-3/4}\sum_{t=1}^n[\rho_m'(e_t)-\psi(e_t)]
\dot{g}_{11}(x_t^\top \theta_{1}^0)x_t^\top P_{2}=O_P(n^{1/4}m^{-1/2}),\\
I_3\equiv &n^{-1/4}\sum_{t=1}^n[\rho_m'(e_t)-\psi(e_t)]g_{11}(x_t^\top \theta_{11}^0)=O_P(n^{1/4}m^{-1/2}).
\end{align*}

Because of similarity, we show the second one only. Notice that
\begin{align*}
\e|I_2|= &n^{-3/4}\e\left|\sum_{t=1}^n[\rho_m'(e_t)-\psi(e_t)]\dot{g}_{11}(x_t^\top \theta_{1}^0)x_t^\top P_{2}\right|\\
\le&n^{-3/4}\sum_{t=1}^n\e\left|[\rho_m'(e_t)-\psi(e_t)]\dot{g}_{11}(x_t^\top \theta_{1}^0)x_t^\top P_{2}\right|\\
=&n^{-3/4}\sum_{t=1}^n\e(\e[|\rho_m'(e_t)-\psi(e_t)||\ra_{t-1}]
\left|\dot{g}_{11}(x_t^\top \theta_{1}^0)x_t^\top P_{2}\right|).
\end{align*}

Moreover, similar to the proof of Theorem 2.1,
\begin{equation*}
\e[|\rho_m'(e_t)-\psi(e_t)||\ra_{t-1}]\le C m^{-1/2}\int |f_{t,e}'(u)|du.
\end{equation*}
Thus, by the upper bound of $f_{t,e}$ in Assumption 3.3,
\begin{align*}
\e|I_1|\le&Cn^{-3/4}m^{-1/2}\sum_{t=1}^n\e
\left|\dot{g}_{11}(x_t^\top \theta_{1}^0)x_t^\top P_{2}\right| =Cn^{1/4}m^{-1/2},
\end{align*}
invoking the joint density of $x_t^\top \theta_{1}^0$ and $x_t^\top P_{2}$ in Lemma A.1 of \citet[p 446]{dgd2016}.

To show (A.10), note that all elements in $Z_2(z_t)$ are stationary and $\alpha$-mixing. Accordingly, taking the first element as example,
\begin{align*}
 &\frac{1}{\sqrt{n}}\e\left\|\sum_{t=1}^n[\rho_m'(e_t)-\psi(e_t)]
 \dot{g}_{21}(z_t^\top \theta_{21}^0)z_{t}^\top\right\| \\
\le &\frac{1}{\sqrt{n}}\sum_{t=1}^n\e(\e[|\rho_m'(e_t)-\psi(e_t)||\ra_{t-1}]
 \dot{g}_{21}(z_t^\top \theta_{21}^0)\|z_{t}\|)\\
 \le &Cm^{-1/2}\frac{1}{\sqrt{n}}\sum_{t=1}^n\e\|
 \dot{g}_{21}(z_t^\top \theta_{21}^0)z_{t}\|=Cm^{-1/2}n^{1/2}.
\end{align*}

Next, we show (A.11)-(A.13). For (A.11), we take the left-top element as example. Note that
\begin{align*}
\e|I_4|^2\equiv &\e\left(\DF{1}{\sqrt{n}}\sum_{t=1}^n[\rho_m''(e_t)-\e\rho''_m(e_t)]
 (\dot{g}_{11}(x_t^\top \theta_{1}^0)x_t^\top \theta_{1}^0)^2\right)^2\\
=&\DF{1}{n}\sum_{t=1}^n\e(\e[(\rho_m''(e_t)-\e\rho''_m(e_t))^2|\ra_{t-1}]
 (\dot{g}_{11}(x_t^\top \theta_{1}^0)x_t^\top \theta_{1}^0)^4),
\end{align*}
and the main step is to calculate the conditional expectation $\e[(\rho_m''(e_t)-\e\rho_m''(e_t))^2|\ra_{t-1}]$ that, however, similar to the proof of Theorem 2.2, equals
\begin{equation*}
\e[(\rho_m''(e_t)-\e\rho_m''(e_t))^2|\ra_{t-1}]=\int [\rho''(u)]^2f_{t,e}(u)du -\left[\int \rho''(u)f_{t,e}(u)du\right]^2 +O(m^{-1/2}),
\end{equation*}
that in turn is bounded by $\int [\rho''(u)]^2f_{e}(u)du+O(m^{-1/2})$ due to Assumption 3.3. This yields $\e|I_4|^2=O(n^{-1/2})$, hence $I_4=O_P(n^{-1/4})$.

(A.12) follows evidently, so we omit the proof.

We also show (A.13) using one element as an exemplar,
\begin{align*}
I_5\equiv &\DF{1}{\sqrt{n}n^{3/4}}\sum_{t=1}^n[\rho_m''(e_t)-\e\rho''_m(e_t)]
 g_{21}(z_t^\top \theta_{21}^0)z_t\dot{g}_{11}(x_t^\top \theta_{1}^0)x_t^\top P_{2}.
\end{align*}
Observe that
\begin{align*}
\e\|I_5\|^2= &\DF{1}{n^{5/2}}\sum_{t=1}^n\e(\e[(\rho_m''(e_t)-\e\rho''_m(e_t))^2|\ra_{t-1}]
 \|g_{21}(z_t^\top \theta_{21}^0)z_t\|^2\|\dot{g}_{11}(x_t^\top \theta_{1}^0)x_t^\top P_{2}\|^2)\\
 \le&C\DF{1}{n^{5/2}}\sum_{t=1}^n\e
 \|g_{21}(z_t^\top \theta_{21}^0)z_t\|^2\e\|\dot{g}_{11}(x_t^\top \theta_{1}^0)x_t^\top P_{2}\|^2\\
 =&C\DF{1}{n^{5/2}}\sum_{t=1}^n
 \e\|\dot{g}_{11}(x_t^\top \theta_{1}^0)x_t^\top P_{2}\|^2\\
 =&C\DF{1}{n},
\end{align*}
where again we use the joint density in Lemma A.1 of \citet{dgd2016}. \qed

\medskip

In the following lemma, recall $Z_t$ defined in Lemma A.4, and denote $\dot{g}(u)=\sum_{j=1}^{p_1}\gamma_{1j}^0\dot{g}_{1j}(u)$ for succinctness. To show the limits of $\mathcal{R}_n\equiv \sum_{t=1}^nZ_tZ_t^\top$ and $\sum_{t=1}^n\psi(e_t)Z_t$ we need to define a complicated matrix. Let $\mathcal{R}$ be a square matrix of order $(d_1+p_1)+p_2(d_2+1)$, standing for the limit of $\mathcal{R}_n$, that we divide into blocks $\mathcal{R}=(\mathcal{R}_{ij})_{2\times 2}$, conformably with the blocks in $Z_tZ_t^\top$, i.e. $Z_1Z_1^\top$, $Z_1Z_2^\top$, $Z_2Z_1^\top$ and $Z_2Z_2^\top$.  $\mathcal{R}_{11}=(R_{ij}^{11})$ is a symmetric matrix of order $d_1+p_1$ that has elements,
\begin{align*}
R_{11}^{11}=& L_1(1,0)\int  [\dot{g}(u)]^2u^2du,  &
R_{1,2:d_1}^{11}=&\int_0^1B_2(r)^\top dL_1(r,0)\int [\dot{g}(u)]^2udu, \\
R_{1, d_1+1}^{11}=&L_1(1,0)\int ug_{11}(u)\dot{g}(u)du, \cdots&
R_{1, d_1+p_1}^{11}=&L_1(1,0)\int ug_{1p_1}(u)\dot{g}(u)du \\
R_{2:d_1, 2:d_1}^{11}=& \int_0^1B_2(r)B_2(r)^\top dL_1(r,0)\int [\dot{g}(u)]^2du, & \\
R_{2:d_1, d_1+1}^{11}=& \int_0^1B_2(r) dL_1(r,0)\int g_{11}(u)\dot{g}(u)du, \cdots &R_{2:d_1, d_1+p_1}^{11}=& \int_0^1B_2(r) dL_1(r,0)\int g_{1p_1}(u)\dot{g}(u)du,\\
R_{d_1+1,d_1+1}^{11}=&L_1(1,0)\int [g_{11}(u)]^2du, \cdots & R_{d_1+1,d_1+p_1}^{11}=&L_1(1,0)\int g_{11}(u)g_{1p_1}(u)du\\
\vdots & & &\vdots \\
R_{d_1+p_1,d_1+1}^{11}=&L_1(1,0)\int g_{11}(u)g_{1p_1}(u)du, \cdots & R_{d_1+p_1,d_1+p_1}^{11}=&L_1(1,0)\int [g_{1p_1}(u)]^2du.
\end{align*}

Notice that $\mathcal{R}_{22}=\Sigma$, the same as given in Theorem 4.1, while $\mathcal{R}_{12}=\mathcal{R}_{12}^\top=0$ a zero matrix of $(d_1+p_1)\times (d_2p_2+p_2)$. Thus, $\mathcal{R}$ is a diagonal block matrix.

\medskip

\noindent{\bf Proof of Lemma A.5}\ \ The joint convergence is implied from the joint convergence of underlying processes given in Assumption 3.3.

Recall
\begin{align*}
Z_1(x_{nt})\equiv &(-\dot{g}(x_t^\top \theta_{1}^0)x_{n,t}^\top,\ n^{-1/4}g_{11}(x_t^\top \theta_{1}^0), \cdots, n^{-1/4}g_{1p_1}(x_t^\top \theta_{1}^0))^\top, \\
Z_2(z_t)\equiv &\frac{1}{\sqrt{n}}(-\gamma_{21}^0\dot{g}_{21}(z_t^\top \theta_{21}^0)z_{t}^\top, \cdots, -\gamma_{2p_2}^0\dot{g}_{2p_2}(z_t^\top \theta_{2p_2}^0)z_{t}^\top, g_{21}(z_t^\top \theta_{21}^0), \cdots, g_{2p_2}(z_t^\top \theta_{2p_2}^0))^\top.
\end{align*}

Notice that
\begin{align*}
\sum_{t=1}^nZ_tZ_t^\top=&\sum_{t=1}^n\begin{pmatrix}Z_{1}(x_{nt})
Z_{1}(x_{nt})^\top & Z_{1}(x_{nt})Z_{2}(z_t)^\top \\
Z_{2}(z_t)Z_{1}(x_{nt})^\top & Z_{2}(z_t)Z_{2}(z_t)^\top \end{pmatrix},
\end{align*}
and $\sum_{t=1}^nZ_{1}(x_{nt})Z_{1}(x_{nt})^\top$ has elements that converge jointly due to \citet{phillips2000} and \citet{phillips2001},
\begin{align*}
 \sum_{t=1}^n
[\dot{g}(x_t^\top \theta_{1}^0)]^2x_{n,t}x_{n,t}^\top &= \sum_{t=1}^n
[\dot{g}(x_t^\top \theta_{1}^0)]^2 \begin{pmatrix} n^{-1/2}(x_t^\top \theta_{1}^0)^2 & n^{-1} (x_t^\top \theta_{1}^0) (P_2x_t)^\top\\
n^{-1} (x_t^\top \theta_{1}^0) (P_2x_t)& n^{-3/2} (P_2x_t)(P_2x_t)^\top \end{pmatrix} \\
\to_D & \begin{pmatrix}L_1(1,0)\int  [\dot{g}(u)]^2u^2du  & \int_0^1B_2(r)^\top dL_1(r,0)\int [\dot{g}(u)]^2udu\\
 \int_0^1B_2(r) dL_1(r,0)\int [\dot{g}(u)]^2udu & \int_0^1B_2(r)B_2(r)^\top dL_1(r,0)\int [\dot{g}(u)]^2du  \end{pmatrix},
\end{align*}
\begin{align*}
-\sum_{t=1}^nn^{-1/4}g_{11}(x_t^\top \theta_{1}^0)\dot{g}(x_t^\top \theta_{1}^0)x_{n,t}^\top &=-\sum_{t=1}^ng_{11}(x_t^\top \theta_{1}^0) \dot{g}(x_t^\top \theta_{1}^0)\begin{pmatrix} n^{-1/2}x_t^\top \theta_{1}^0 \\ n^{-1} (P_2x_t)^\top \end{pmatrix}\\
\to_D&-\begin{pmatrix} L_1(1,0)\int ug_{11}(u)\dot{g}(u)du \\ \int_0^1 B_2(r) dL_1(r,0)\int g_{11}(u)\dot{g}(u)du  \end{pmatrix},\\
&\vdots \\
-\sum_{t=1}^n n^{-1/4}g_{1p_1}(x_t^\top \theta_{1}^0) \dot{g}(x_t^\top \theta_{1}^0)x_{n,t}^\top
=&-\sum_{t=1}^ng_{1p_1}(x_t^\top \theta_{1}^0) \dot{g}(x_t^\top \theta_{1}^0)\begin{pmatrix} n^{-1/2}x_t^\top \theta_{1}^0 \\ n^{-1} (P_2x_t)^\top \end{pmatrix}\\
\to_D&-\begin{pmatrix} L_1(1,0)\int ug_{1p_1}(u)\dot{g}(u)du \\ \int_0^1 B_2(r) dL_1(r,0)\int g_{1p_1}(u)\dot{g}(u)du  \end{pmatrix},\\
n^{-1/2}\sum_{t=1}^n[g_{11}(x_t^\top \theta_{1}^0)]^2\to_D&L_1(1,0) \int [g_{11}(u)]^2du,\\
&\vdots\\
n^{-1/2}\sum_{t=1}^ng_{11}(x_t^\top \theta_{1}^0)g_{1p_1}(x_t^\top \theta_{1}^0)\to_D&L_1(1,0)\int g_{11}(u) g_{1p_1}(u)du\\
&\vdots\\
n^{-1/2}\sum_{t=1}^n[g_{1p_1}(x_t^\top \theta_{1}^0)]^2\to_D& L_1(1,0) \int [g_{1p_1}(u)]^2du.
\end{align*}

Moreover, $\sum_{t=1}^nZ_{2}(z_{t})Z_{2}(z_{t})^\top\to_P \Sigma$ as we shown in Lemma A.2.

Next, we consider $\sum_{t=1}^nZ_{1}(x_{n,t})Z_{2}(z_{t})^\top$ that has elements:
\begin{align*}
 \DF{1}{\sqrt{n}}\gamma_{21}^0&\sum_{t=1}^n \dot{g}(x_t^\top \theta_{1}^0)x_{n,t}\dot{g}_{21}(z_t^\top \theta_{21}^0)z_t^\top, \cdots
 &\cdots,
 \DF{1}{\sqrt{n}}\gamma_{2p_2}^0&\sum_{t=1}^n \dot{g}(x_t^\top \theta_{1}^0)x_{n,t}\dot{g}_{2p_2}(z_t^\top \theta_{2p_2}^0)z_t^\top,\\
 \DF{-1}{\sqrt{n}}&\sum_{t=1}^n \dot{g}(x_t^\top \theta_{1}^0)x_{n,t}g_{21}(z_t^\top \theta_{21}^0),\cdots
 &\cdots,
 \DF{-1}{\sqrt{n}}&\sum_{t=1}^n \dot{g}(x_t^\top \theta_{1}^0) x_{n,t}g_{2p_2}(z_t^\top \theta_{2p_2}^0),\\
 \DF{-1}{\sqrt[4]{n}^3}\gamma_{21}^0&\sum_{t=1}^n g_{11}(x_t^\top \theta_{1}^0)\dot{g}_{21}(z_t^\top \theta_{21}^0)z_t^\top, \cdots
 &\cdots,
 \DF{-1}{\sqrt[4]{n}^3}\gamma_{2p_2}^0&\sum_{t=1}^n g_{11}(x_t^\top \theta_{1}^0)\dot{g}_{2p_2}(z_t^\top \theta_{2p_2}^0)z_t^\top,\\
  \DF{1}{\sqrt[4]{n}^3}&\sum_{t=1}^n g_{11}(x_t^\top \theta_{1}^0)g_{21}(z_t^\top \theta_{21}^0), \cdots
 &\cdots,
 \DF{1}{\sqrt[4]{n}^3}&\sum_{t=1}^n g_{11}(x_t^\top \theta_{1}^0)g_{2p_2}(z_t^\top \theta_{2p_2}^0)\\
 \vdots & & &\vdots \\
 \DF{-1}{\sqrt[4]{n}^3}\gamma_{21}^0&\sum_{t=1}^n g_{1p_1}(x_t^\top \theta_{1}^0)\dot{g}_{21}(z_t^\top \theta_{21}^0)z_t^\top, \cdots
 &\cdots,
 \DF{-1}{\sqrt[4]{n}^3}\gamma_{2p_2}^0&\sum_{t=1}^n g_{1p_1}(x_t^\top \theta_{1}^0)\dot{g}_{2p_2}(z_t^\top \theta_{2p_2}^0)z_t^\top,\\
 \DF{1}{\sqrt[4]{n}^3}&\sum_{t=1}^n g_{1p_1}(x_t^\top \theta_{1}^0)g_{21}(z_t^\top \theta_{21}^0), \cdots
 &\cdots,
 \DF{1}{\sqrt[4]{n}^3}&\sum_{t=1}^n g_{1p_1}(x_t^\top \theta_{1}^0)g_{2p_2}(z_t^\top \theta_{2p_2}^0).
\end{align*}

All of them have limit zero since they are all over-normalized comparing with elements in $\mathcal{R}_{11}$ and $\mathcal{R}_{22}$. Because of similarity, we only show the following limits
\begin{align*}
I_1=&\DF{1}{\sqrt{n}}\sum_{t=1}^n \dot{g}(x_t^\top \theta_{1}^0)x_{n,t}\dot{g}_{21}(z_t^\top \theta_{21}^0)z_t^\top\to_P0, \\ I_2=&\DF{1}{\sqrt[4]{n}^3}\sum_{t=1}^n g_{11}(x_t^\top \theta_{1}^0)g_{21}(z_t^\top \theta_{21}^0)\to_P0,
\end{align*}
that are two typical elements in the matrix.

Observe that
\begin{align*}
I_1=&\DF{1}{\sqrt{n}}\sum_{t=1}^n \dot{g}(x_t^\top \theta_{1}^0)x_{n,t}\dot{g}_{21}(z_t^\top \theta_{21}^0)z_t^\top\\
=&\DF{1}{\sqrt{n}}\sum_{t=1}^n \dot{g}(x_t^\top \theta_{1}^0)x_{n,t}\e[\dot{g}_{21}(z_t^\top \theta_{21}^0)z_t^\top]\\
&+\DF{1}{\sqrt{n}}\sum_{t=1}^n \dot{g}(x_t^\top \theta_{1}^0)x_{n,t}\{\dot{g}_{21}(z_t^\top \theta_{21}^0)z_t^\top-\e[\dot{g}_{21}(z_t^\top \theta_{21}^0)z_t^\top]\}\\
\equiv& I_{11}+I_{12}, \ \ \text{say}.
\end{align*}

Note that $z_t=h(\tau_t, v_t)$. It follows from Lemma B.1 of \citet[p140]{dg2019} that
\begin{align*}
I_{11}=&\DF{1}{\sqrt{n}}\sum_{t=1}^n \dot{g}(x_t^\top \theta_{1}^0)x_{n,t}\e[\dot{g}_{21}(z_t^\top \theta_{21}^0)z_t^\top]\\
=&\DF{1}{\sqrt[4]{n}}\sum_{t=1}^n \dot{g}(x_t^\top \theta_{1}^0) \begin{pmatrix}n^{-1/2}x_t^\top \theta_{1}^0\\ n^{-1} P_2x_t\end{pmatrix}\e[\dot{g}_{21}(h(\tau_t, v_t)^\top \theta_{21}^0)h(\tau_t, v_t)^\top]\to_P0,
\end{align*}
because there is an extra factor $n^{-1/4}$, while for $I_{12}$ one may calculate
\begin{align*}
\e\|I_{12}\|^2=&\DF{1}{n}\sum_{t=1}^n\e [\dot{g}(x_t^\top \theta_{1}^0)]^2x_{n,t}^\top x_{n,t} \e\|\dot{g}_{21}(z_t^\top \theta_{21}^0)z_t^\top-\e[\dot{g}_{21}(z_t^\top \theta_{21}^0)z_t^\top]\|^2\\
&+\DF{2}{n}\sum_{t=2}^n\sum_{s=1}^{t-1}\e \dot{g}(x_t^\top \theta_{1}^0)\dot{g}(x_s^\top \theta_{1}^0)x_{n,t}^\top x_{n,s} \e tr\{(\dot{g}_{21}(z_t^\top \theta_{21}^0)z_t^\top-\e[\dot{g}_{21}(z_t^\top \theta_{21}^0)z_t^\top])\\
& \qquad \times (\dot{g}_{21}(z_s^\top \theta_{21}^0) z_s^\top -\e[\dot{g}_{21}(z_s^\top \theta_{21}^0)z_s^\top])\}\\
\le&\sup_{r\in[0,1]}\e\|\dot{g}_{21}(h(r,v_t)^\top \theta_{21}^0)h(r,v_t)^\top -\e[\dot{g}_{21}(h(r,v_t)^\top \theta_{21}^0)h(r,v_t)^\top]\|^2\\
&\times\DF{1}{n}\sum_{t=1}^n\e [\dot{g}(x_t^\top \theta_{1}^0)]^2x_{n,t}^\top x_{n,t}\\
&+\sup_{r\in[0,1]}\e\|\dot{g}_{21}(h(r,v_t)^\top \theta_{21}^0)h(r,v_t)^\top -\e[\dot{g}_{21}(h(r,v_t)^\top \theta_{21}^0)h(r,v_t)^\top]\|^4\\
&\times \DF{2}{n}\sum_{t=2}^n\sum_{s=1}^{t-1}\alpha(t-s)\e |\dot{g}(x_t^\top \theta_{1}^0)\dot{g}(x_s^\top \theta_{1}^0)x_{n,t}^\top x_{n,s}|,
\end{align*}
by the argument in Lemma A.2 that $x_t$ and $v_t$ can be regarded as mutually independent. We then further compute the expectations invoking the joint density in Lemma A.1 of \citet[p446]{dgd2016} that shows $\e\|I_{12}\|^2\to 0$ as $n\to\infty$. We omit the details for simplicity while refer the readers who are interested in the derivations to \citet{dgd2016}. Thus, $I_1=o_P(1)$.

Now, let's turn to $I_2$. Observe
\begin{align*}
I_2=&\DF{1}{\sqrt[4]{n}^3}\sum_{t=1}^n g_{11}(x_t^\top \theta_{1}^0)g_{21}(z_t^\top \theta_{21}^0)\\
=&\DF{1}{\sqrt[4]{n}^3}\sum_{t=1}^n g_{11}(x_t^\top \theta_{1}^0)\e[g_{21}(h(\tau_t,v_t)^\top \theta_{21}^0)]\\
&+\DF{1}{\sqrt[4]{n}^3}\sum_{t=1}^n g_{11}(x_t^\top \theta_{1}^0)\{g_{21}(z_t^\top \theta_{21}^0)-\e[g_{21}(z_t^\top \theta_{21}^0)]\}\\
=&I_{21}+I_{22}, \ \ \text{say}.
\end{align*}

According to Lemma B.1 of \citet[p140]{dg2019} again,
\begin{equation*}
\DF{1}{\sqrt{n}}\sum_{t=1}^n g_{11}(x_t^\top \theta_{1}^0)\e[g_{21}(h(\tau_t,v_t)^\top \theta_{21}^0)]\to_D\int_0^1 \e[g_{21}(h(r,v_t)^\top \theta_{21}^0)]dL_1(r,0) \int g_{11}(u)du,
\end{equation*}
so that $I_{21}=O_P(n^{-1/4})$ and due to the same reason as used in $I_{12}$ we have $I_{22}=o_P(1)$. This completes the proof of (A.14).

Notice that $\sum_{t=1}^n\psi(e_t)Z_t$ is a martingale sequence due to Assumption 3.3 and its conditional covariance matrix is $\mathcal{R}_n\equiv \sum_{t=1}^n Z_tZ_t^\top$ that has limit $\mathcal{R}$ a positive definite matrix. A rigorous proof of (A.15) can be achieved following from Lemma 3 of \citet{phillips2000} with a bit modification to contain stationary part $Z_2(z_t)$.

The self-normalized version (A.16) follows immediately from (A.14) and (A.15) due to the joint convergence. \qed

}

\end{document}